\documentclass{emulateapj}
\usepackage{apjfonts}
\usepackage{graphicx}
\usepackage{amsmath}

\newcommand{\be}{\begin{equation}}
\newcommand{\ee}{\end{equation}}
\newcommand{\beqa}{\begin{eqnarray}}
\newcommand{\eeqa}{\end{eqnarray}}

\def\Mo{{\rm M_\odot}}

\newcommand{\Msun}{\rm{M}_{\odot}}

\def\kms{{\ }{\rm km}\,{\rm s}^{-1}}

\newcommand{\massq}{mag arcsec$^{-2}$}
\newcommand{\vphi}{V_{\phi}}

\newcommand{\Lz}{L_z}
\newcommand{\vlos}{$v_{\rm{los}}$ }

\begin{document}
\submitted{The Astrophysical Journal, accepted}
\vspace{1mm}
\slugcomment{{\it The Astrophysical Journal, accepted}}

\shortauthors{VILLALOBOS, KAZANTZIDIS, \& HELMI}
\shorttitle{THICK-DISK EVOLUTION INDUCED BY A GROWING THIN DISK}

\title{Thick-Disk Evolution Induced by the Growth of an Embedded Thin Disk}

\author{\'Alvaro Villalobos,\altaffilmark{1,2}
        Stelios Kazantzidis,\altaffilmark{3}
        and Amina Helmi\altaffilmark{1}}        

\begin{abstract}
  We perform collisionless $N$-body simulations to investigate the
  evolution of the structural and kinematical properties of simulated
  thick disks induced by the growth of an embedded thin disk. The
  thick disks used in the present study originate from
  cosmologically-common $5$:$1$ encounters between initially-thin
  primary disk galaxies and infalling satellites. The growing thin
  disks are modeled as static gravitational potentials and we explore
  a variety of growing-disk parameters that are likely to influence
  the response of thick disks. We find that the final thick-disk
  properties depend strongly on the total mass and radial scale-length
  of the growing thin disk, and much less sensitively on its growth
  timescale and vertical scale-height as well as the initial sense of
  thick-disk rotation.  Overall, the growth of an embedded thin disk
  can cause a substantial contraction in both the radial and vertical
  direction, resulting in a significant decrease in the scale-lengths
  and scale-heights of thick disks. Kinematically, a growing thin disk
  can induce a notable increase in the mean rotation and velocity
  dispersions of thick-disk stars. We conclude that the reformation of
  a thin disk via gas accretion may play a significant role in setting
  the structure and kinematics of thick disks, and thus it is an
  important ingredient in models of thick-disk formation.
\end{abstract}


\altaffiltext{1}{Kapteyn Astronomical Institute, University of
  Groningen, P.O. Box 800, 9700 AV Groningen, The Netherlands; {\tt
  villalobos, ahelmi@astro.rug.nl}.}  
\altaffiltext{2}{Present address: INAF-Osservatorio Astronomico di 
  Trieste, Via Tiepolo 11, I-34143 Trieste, Italy; {\tt 
  villalobos@oats.inaf.it}.}
\altaffiltext{3}{Center for
  Cosmology and Astro-Particle Physics; and Department of Physics; and
  Department of Astronomy, The Ohio State University, 191 West
  Woodruff Avenue, Columbus, OH 43210, USA; {\tt
    stelios@mps.ohio-state.edu}.}

\section{Introduction}
\label{introduction}
\defcitealias{villalobos-helmi2008}{Paper I}
\defcitealias{villalobos-helmi2009}{Paper II}

Hierarchical models of cosmological structure formation, such as the
currently favored cold dark matter (CDM) paradigm
\citep[e.g.,][]{white_rees1978,blumenthal1984}, generically predict
that galaxies are built via the continuous accretion of smaller
systems.  Indeed, a growing body of observational evidence has
recently confirmed this prediction with the discovery of tidal streams
and complex stellar structures in the Milky Way (MW)
\citep[e.g.,][]{ibata1994,helmi-etal1999,yanny2000,ibata2001a,newberg2002,
  majewski2003,martinez2005,belokurov2006}, the Andromeda galaxy
\citep{ibata2001b,ferguson2002,ferguson2005,kalirai2006,ibata2007},
and beyond the Local Group
\citep[e.g.,][]{malin_hadley1997,shang1998,peng2002,forbes2003,
  pohlen2004}.

In the context of CDM, bombardment by infalling satellites may lead to
the heating of thin galactic disks and the subsequent formation of
thick disks (e.g., \citealt{quinn1986}; \citealt{walker1996};
\citealt{velazquez1999}; \citealt{hayashi2006};
\citealt{villalobos-helmi2008}, hereafter
\citetalias{villalobos-helmi2008}; \citealt{kazantzidis2008};
\citealt{read2008}), and can even cause thin-disk destruction
\citep{purcell2009}.  Depending on when the most substantial accretion
event has occurred, a new thin disk may reform, for example, via the
cooling of hot gas in the galactic halo
\citep[e.g.,][]{white_rees1978,mo1998, baugh2006} and be observable at
present.

In this scenario, the dynamical effects of the cooling gas on the
properties of the post-accretion thickened stellar distribution could
be significant. Indeed, as the gas cools down and slowly accumulates
at the center of the system, it can induce concomitant contraction of
the heated stellar component due to its gravity \citep[see
e.g.,][]{elmegreen2006,kazantzidis2009}. This phenomenon is similar in
spirit to the contraction of dark matter halos during the process of
baryonic cooling that leads to the formation of galaxies
\citep[e.g.,][]{zeldovich1980,barnes1984,blumenthal1986}.

Significant theoretical effort, including both semi-analytic modeling
\citep{toth1992,benson2004,hopkins2008} and numerical simulations
\citep{quinn1986,quinn1993,walker1996,huang1997,sellwood1998,velazquez1999,
font2001,ardi2003,gauthier2006,hayashi2006,kazantzidis2008,read2008,
purcell2009,kazantzidis2009,moster2009}
has been devoted to exploring the dynamical heating of galactic disks
via satellite accretion events. However, very few studies have
investigated the response of the resulting thickened stellar component
to the gas accumulation and subsequent reformation of a thin-disk
\citep[see, however,][]{kazantzidis2009,moster2009}. A full model of
satellite-disk interactions in a cosmological context including gas
dynamics and star formation would be ideal to obtain a complete
picture of this physical process.  Unfortunately, due to the fact that
they are highly complex and very costly in terms of computational
power, cosmological simulations that could address in a systematic way
the structural and kinematical evolution of thick disks in response to
growing thin-disks do not yet exist.

Recently, \citet{kazantzidis2009} performed dissipationless $N$-body
simulations to investigate the influence of a slowly growing thin disk
on the properties of an initially-thick galactic disk.  These authors
considered three models of growing disks with masses equal to $10\%$,
$50\%$, and $100\%$ of the mass of the thick disk, and showed that the
latter contracted vertically as well as radially in response to the
growth of the thin-disk component. Not unexpectedly, the magnitude of
the thick-disk structural evolution was found to depend sensitively on
the total mass of the growing disk. However, \citet{kazantzidis2009}
focused on a fairly small region of parameter space and, in addition,
they utilized a fully-formed thick-disk galaxy instead of a galactic
model whose thick disk formed self-consistently by encounters between
an initially-thin disk and infalling satellites.

In the present paper, we expand upon this initiative by performing a
{\it systematic} numerical study aiming to elucidate the effects of
growing thin disks on both the morphology and kinematics of realistic
thick disks. For the latter, we adopt a subset of the models presented
in \citetalias{villalobos-helmi2008}.  Given the complex interplay of
effects (e.g., gas cooling, star formation) relevant to the formation
and evolution of spiral galaxies and the outstanding issues regarding
disk galaxy formation in CDM cosmogonies \citep[e.g.,][]{mayer2008},
we restrict ourselves to modeling the gravitational potential of a
thin disk that slowly grows in mass over time using an ensemble of
collisionless $N$-body simulations.

Our simulation set is carefully designed to permit an investigation of
a large parameter space and we explore several aspects of the growth
of a thin disk that are likely to have an effect on the morphological
and kinematical evolution of thick disks. In particular, we vary the
growth timescale, final mass, scale-length, scale-height, and
orientation of the growing disk total angular momentum vector. In
addition, we perform experiments where the same thin disk grows inside
thick disks that rotate in either a prograde or retrograde sense 
with respect to their halos.  
The present work establishes that the reformation of a
thin disk may play a significant role in setting the structural and
kinematic properties of thick-disk stars, and thus it is an important
ingredient in models of thick-disk formation.

The outline of this paper is as follows. In Section~\ref{methods} we
briefly describe the models adopted and the setup of the numerical
experiments performed in the present study. Section~\ref{results}
contains the results regarding the thick-disk morphological and
kinematical evolution induced by the growth of a thin-disk component.
A detailed comparison between the properties of the initial and
post-growth thick disks is presented in Section~\ref{comparison}.
Lastly, Section~\ref{summary-conclusions} summarizes our main
conclusions. Throughout this work we use the terms ``thick disk'' and
``heated disk'' interchangeably to indicate the perturbed stellar
component that resulted from the original interaction between a thin
primary disk galaxy and an infalling satellite, and consists of stars
from both the stellar disk of the primary galaxy and the disrupted
system.

\begin{table*}
 \caption{Summary of Numerical Experiments}
 \label{table-exper}
  \begin{center}
  \begin{tabular}{@{}cccccccccc}
\tableline\\
    & \multicolumn{5}{c}{Growing thin disk} &&&  \multicolumn{2}{c}{Thick disk} \\
&  & $R_d$ & $z_d$ & $\tau$&  \\
Run    & $M_d/M_{\rm thick}$ & (kpc) & (pc) & (Gyr) & alignment &&& rotation & inclination\\
(1) & (2) & (3) & (4) & (5) & (6) &&& (7) & (8)  \\
\tableline
\tableline
\\
Reference & 5                     & 3     & 125  & 1      & halo     &&& prograde    & 30$\degr$   \\
model &&&&&&&&& \\
\tableline
\\
A1     & 5          & 3            & 125         & 1      & halo     &&& prograde    &  0$\degr$   \\
A2     & 5          & 3            & 125         & 1      & halo     &&& prograde    & 60$\degr$   \\
\tableline
\\
B1   & 5          & 3            & 125          & 0      & halo     &&& prograde    & 30$\degr$   \\
B2   & 5          & 3            & 125          & 5      & halo     &&& prograde    & 30$\degr$    \\
\tableline
\\
C1    & 2          & 3            & 125         & 1      & halo     &&& prograde    & 30$\degr$   \\
C2    & 5          & 3            & 25          & 1      & halo     &&& prograde    & 30$\degr$   \\
C3    & 5          & 1            & 125         & 1      & halo     &&& prograde    & 30$\degr$   \\
\tableline
\\
D     & 5          & 3            & 125         & 1      & thick disk      &&& prograde    & 30$\degr$   \\
\tableline
\\
E1     & 5          & 3            & 125         & 1      & halo     &&& prograde    & 0$\degr$   \\
E2     & 5          & 3            & 125         & 1      & halo     &&& retrograde  & 0$\degr$   \\
  \tableline
  \tableline
 \end{tabular}
 \end{center}
{\sc Notes.}--- Column (1): Labels for the numerical simulations.
Column (2): Mass of the growing thin disk in units of the initial thick disk.
Column (3): Radial scale-length of the growing thin disk in kpc.
Column (4): Vertical scale-height of the growing thin disk in pc.
Column (5): Growth timescale of the thin disk in Gyr.
Column (6): Alignment of the rotation axis of the growing thin disk. 
Column (7): Sense of rotation of the thick disk with respect to its halo.
Column (8): Initial orbital inclination of the infalling satellite that produced 
the initial thick disk.
\end{table*}

\section{Methods}
\label{methods}

We begin this section by briefly describing the primary disk models
used in this study. In Section~\ref{thin-disk_models} we introduce the
thin-disk models and discuss the method we employ to grow them inside
the thick disks. Lastly, in Section~\ref{simulations} we
present the numerical experiments with growing disks of the adopted
simulation campaign.

\subsection{Models of Thick Disks}
\label{thick-disk_models}

In \citetalias{villalobos-helmi2008}, $25$ dissipationless $N$-body
simulations were carried out to study the general properties of thick
disks formed by single encounters between a primary thin disk galaxy
and a massive satellite system. These simulations explored: (1) two
models for the primary disk galaxy, whose structure and kinematics
resembled those of the MW at present (``$z$=0'' experiments) and at
z=1 (``$z$=1'' experiments); (2) two morphologies for the stellar
component of the satellite, spherical and disky; (3) two mass ratios
between the satellite and the primary disk galaxy ($10\%$ and $20\%$);
and (4) three initial satellite orbital inclinations ($i=0\degr,
30\degr$, and $60\degr$), in both prograde and retrograde directions
with respect to the rotation of the primary disk.

Both the primary disk galaxy and the satellite were initialized as
``live'' $N$-body models consisting of a dark matter and a stellar
component. The dark halos of both systems followed the
\citet{navarro1997} density profile and have been adiabatically
contracted to account for the presence of the stellar component
\citep[see e.g.,][]{blumenthal1986,mo1998}.  The stellar component of
the spherical satellite was set up with structure and kinematics
consistent with the observed fundamental plane of dE+dSphs galaxies
\citep{derijcke2005}. On the other hand, the disky satellites have
been constructed in the same way as the primary disk galaxies. Their
scale-lengths and scale-heights are smaller than those of the primary
disks, following the proportion between the virial radii of the
corresponding dark matter halos (see \citetalias{villalobos-helmi2008}
for details).  The orbital parameters of the encounters were drawn
from studies of infalling substructures in a cosmological context
\citetext{\citealt{benson2005}, see also
  \citealt{tormen1997,khochfar2006}}, and the satellites were released
far away from the centers of the disk galaxies, $\sim 35$ ($\sim 50$)
times the disk scale-length for ``$z$=0'' (``$z$=1'') experiments.

The simulations of thick-disk formation were evolved for $5$~Gyr
($4$~Gyr) in the ``$z=0$'' (``$z=1$'') experiments and the satellites
have typically merged by $t=3$~Gyr ($t=2$~Gyr). By the final time, the
systems are relaxed. The simulations were carried out with $5 \times
10^5$ particles in the host dark matter halo, and $10^5$ particles in
each of all the other components. The softenings of thick-disk stars
range from 12 pc to 70 pc (we refer the reader to \citetalias{villalobos-helmi2008} 
for a complete description of the experiments), and the softening
of growing disk particles correspond to 1/5 of the growing disk scale-height
(see Section~\ref{simulations}).

The primary aim of the present work is to study the most generic
features of the evolution of thick disks in response to the growth of
an embedded thin disk.  For this reason, we select a subset of the
thick-disk models presented in \citetalias{villalobos-helmi2008} (see
Section~\ref{simulations} below). In particular, we focus on models
that were produced via an encounter with a satellite on a
low/intermediate orbital inclination. Assuming a dynamical heating
model for the origin of thick disks \citep[e.g.][]{kazantzidis2008},
such accretion events are relevant to the formation of the thick disk
of the MW \citepalias{villalobos-helmi2008}. This is based on the
possible existence of a vertical gradient in the rotational velocity
of the Galactic thick disk and on the observed value of
$\sigma_z/\sigma_R$ in the solar neighborhood
\citep[e.g.][]{girard2006,vallenari2006}. In addition, due to the fact
that satellite structure is not essential in establishing the
structural properties of thick disks
\citepalias{villalobos-helmi2008}, we only consider thick-disk models
that were produced by interactions with spherical satellites.

\subsection{Models of Growing Disks}
\label{thin-disk_models}

%
\begin{figure*}
  \begin{center}
  \includegraphics[width=88mm]{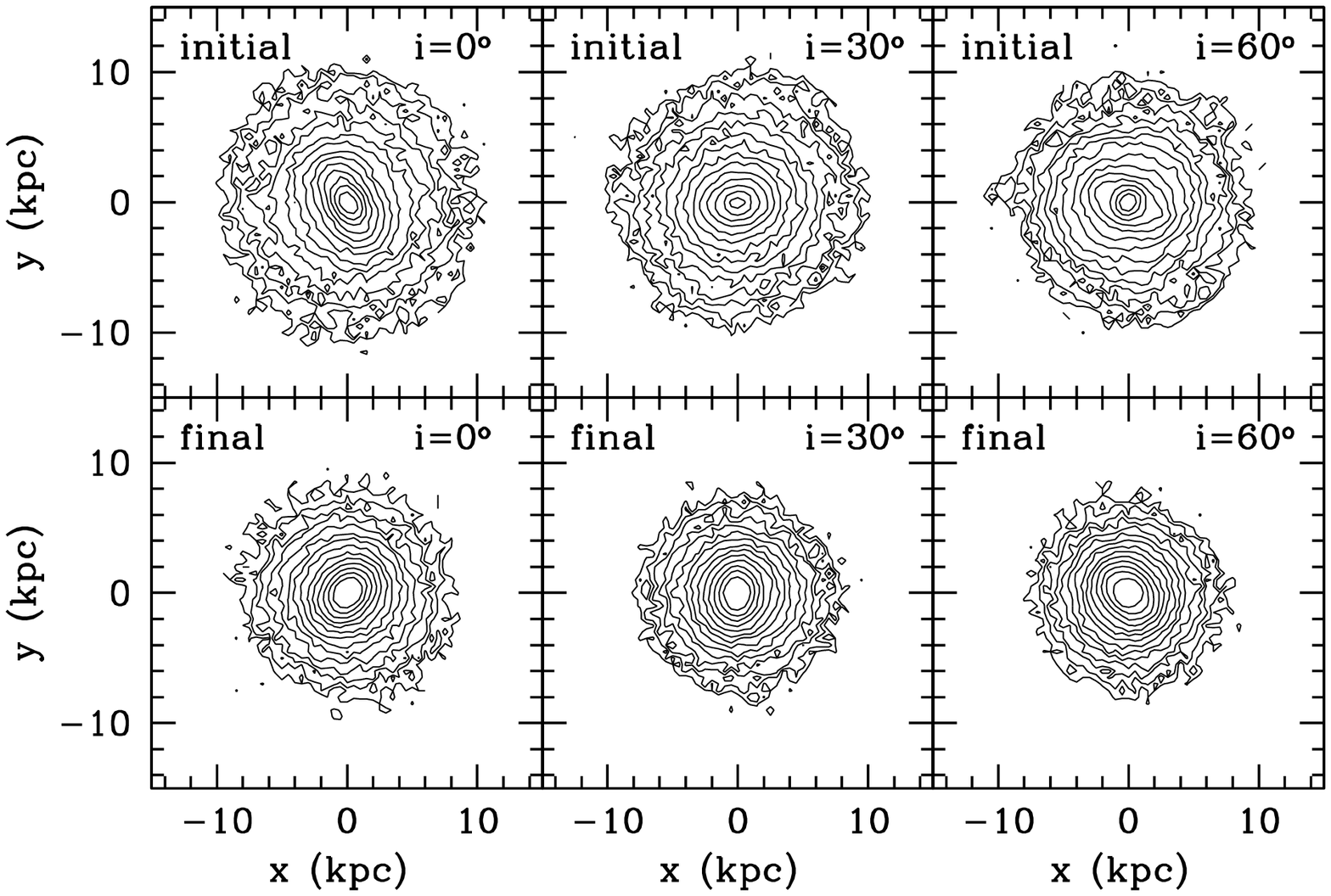}
  \hspace*{0.01cm}
  \includegraphics[width=88mm]{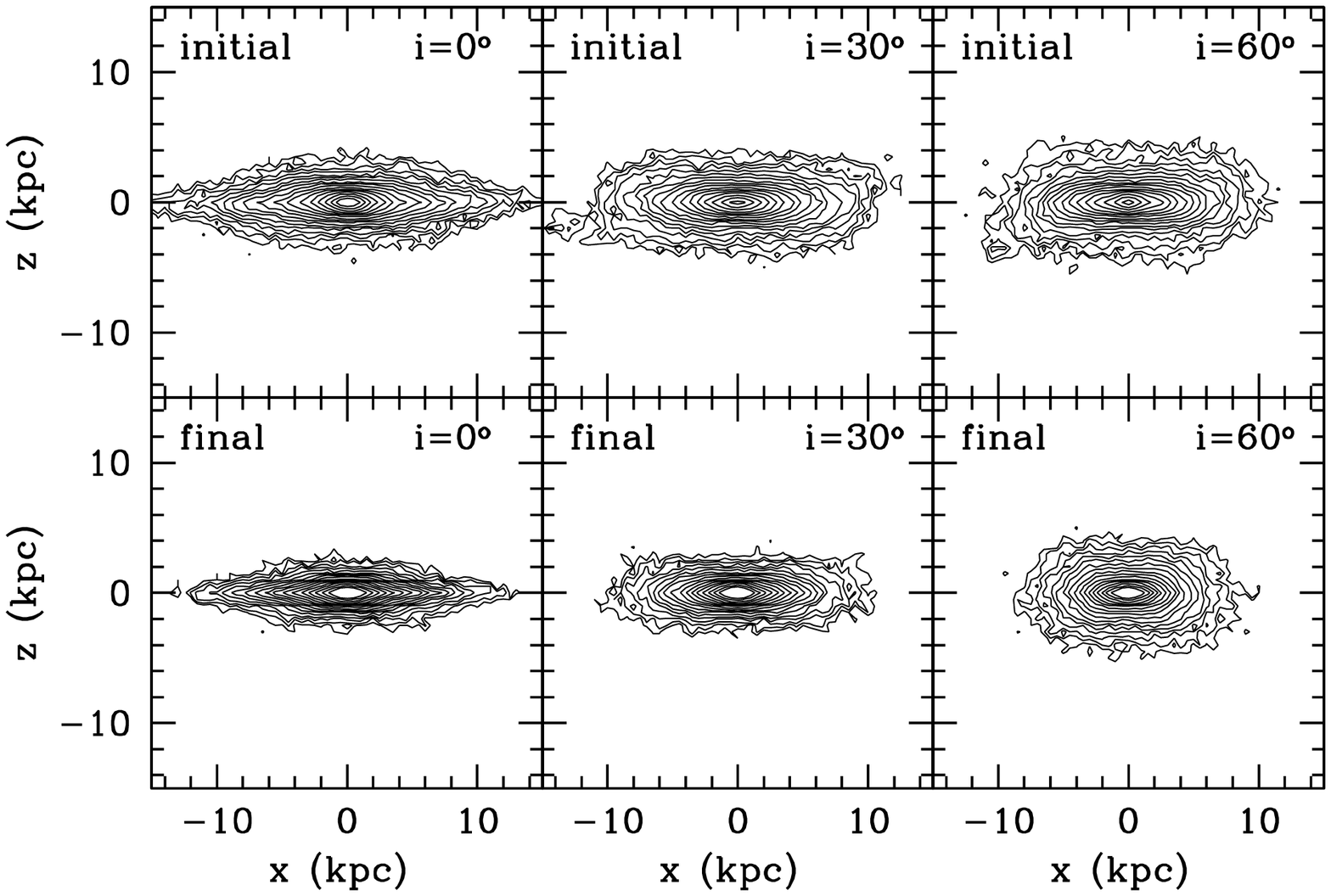}
  \caption{Initial (upper panels) and final (bottom panels) surface
    brightness contours of thick-disk stars viewed face-on (left
    panels) and edge-on (right panels) in the simulations with growing
    thin disks.  Results are presented for experiment A with a thin-disk
    growth timescale of $\tau=1$~Gyr and the orbital inclination of
    the accretion event that formed the initial heated disk is
    indicated in each panel. Contours correspond to equal surface
    brightness levels (0.39 \massq) in the V-band for face-on
    (20.25-25.35 \massq) and edge-on (19.09-25.35 \massq) views. Thick
    disks experience strong radial and vertical contraction in
    response to the growth of a new thin disk component.
    \label{contour-contraction-expB}}
    \end{center}
\end{figure*}

The growing disks follow an exponential distribution in cylindrical 
radius $R$, and their vertical structure is modeled in the standard 
way as a collection of isothermal sheets \citep{spitzer1942}:
\be 
  \rho_{\rm thin}(R,z) = K \exp\left(-\frac{R}{R_d}\right) {\rm
  sech}^2\left(\frac{z}{2 z_d}\right) \ ,
   \label{disc_density}
\ee
where $M_d$, $R_d$, $z_d$ denote the mass, radial scale-length, and
(exponential) vertical scale-height of the disk, respectively, and
$K=M_d/8\pi R_d z_d$.

Each growing-disk simulation was performed using the following
procedure: (1) insert a {\it massless} Monte Carlo particle
realization of the desired disk model inside the primary disk
galaxies. (2) increase the mass of this distribution to its final
value linearly over a timescale $\tau$, according to the following
law:
\be
   M(t) = M_d\,(t/\tau) \qquad\hbox{$0\leq t\leq\tau$}. 
\ee

During the growth period, the growing disk remains rigid and their
particles are fixed in place (i.e., behaving like a fixed disk potential), 
while the ``live'' particles composing the thick disk and its dark matter 
halo are allowed to achieve equilibrium as the mass of the system grows. 
Throughout the experiments, all other properties of the growing disk 
(e.g., scale-length, scale-height) are kept constant. All numerical 
simulations of growing disks were carried out with the multi-stepping, 
parallel, tree $N$-body code PKDGRAV \citep{stadel2001}.

\subsection{Numerical Experiments with Growing Disks}
\label{simulations}

Overall, we have performed $11$ simulations of the growth of a new
thin disk within a subset of the thick disks presented in
\citetalias{villalobos-helmi2008}. We have explored a variety of
parameters related to the growing disks that may affect the thick-disk
evolution, including: (1) the growth timescale: $\tau=0$~Gyr,
$1$~Gyr, and $5$~Gyr\footnote{Note that these values bracket the range
  of potential timescales for the disk growth including strongly
  non-adiabatic and adiabatic ones. While adopting an instantaneous
  growth for the disk is obviously unphysical, it still represents an
  interesting limiting case for the disk growth timescale.}; (2) the
final mass: $M_d=2M_{\rm thick}$ and $5M_{\rm thick}$, where $M_{\rm
  thick}$ denotes the mass of the initial thick disk\footnote{Note
  that the value of $M_d=2 M_{\rm thick}=2.8 \times 10^{10}\Msun$ is
  within the range of the total gas mass accreted by the MW disk since
  $z\sim 1$ ($1.5-3 \times 10^{10} \Msun$ assuming a constant infall
  rate of $2-4$ $\Msun/yr$) according to analytical models for the
  evolution of the MW disk in a cosmological context
  \citep{naab_ostriker2006}. Although there are still large
  uncertainties in this estimate, the larger value of $M_d=5 M_{\rm
    thick}$ is relevant to a system whose thin-thick mass ratio is
  similar to that of the Galaxy at present \citep[e.g.,][]{juric2008}};
  (3) the vertical scale-height: $z_d=25$~pc and
  $125$~pc\footnote{Note that these values are consistent with the
    scale-heights of known, young, star-forming disks observed both in
    external galaxies \citep[e.g.,][]{wainscoat1989,matthews2000} and
    in the MW \citep[e.g.,][]{Bahcall1980,reid1993}.}; (4) the radial
  scale-length: $R_d=1$~kpc and $3$~kpc; (5) the orientation of the
  angular momentum vector: aligned either to the angular momentum of
  the dark matter halo or to the angular momentum vector of the thick
  disk; and (6) the sense of rotation of the initial thick disk with 
  respect to its halo: prograde or retrograde.  Lastly, we also 
  compare the evolution of initial thick disks produced by infalling 
  satellites on different orbital inclinations, 
  $i=0\degr$, $30\degr$, and $60\degr$.

The runs are labeled as A, B, C, D, and E and summarized in
Table~\ref{table-exper}. Our reference model is characterized by the
following set of parameters for the growing thin disk: $\tau=1$~Gyr,
$M_d = 5 M_{\rm thick}$, $z_d= 125$~pc, $R_d =3$~kpc, and a thick disk
that is the result of a (prograde) satellite accretion event with an initial
orbital inclination of $i=30\degr$. In what follows, we
refer to experiment ``X'' to denote comparisons between Run ``X'' and
the reference model, except for run E.

\begin{figure*}
  \begin{center}
    \includegraphics[height=50mm]{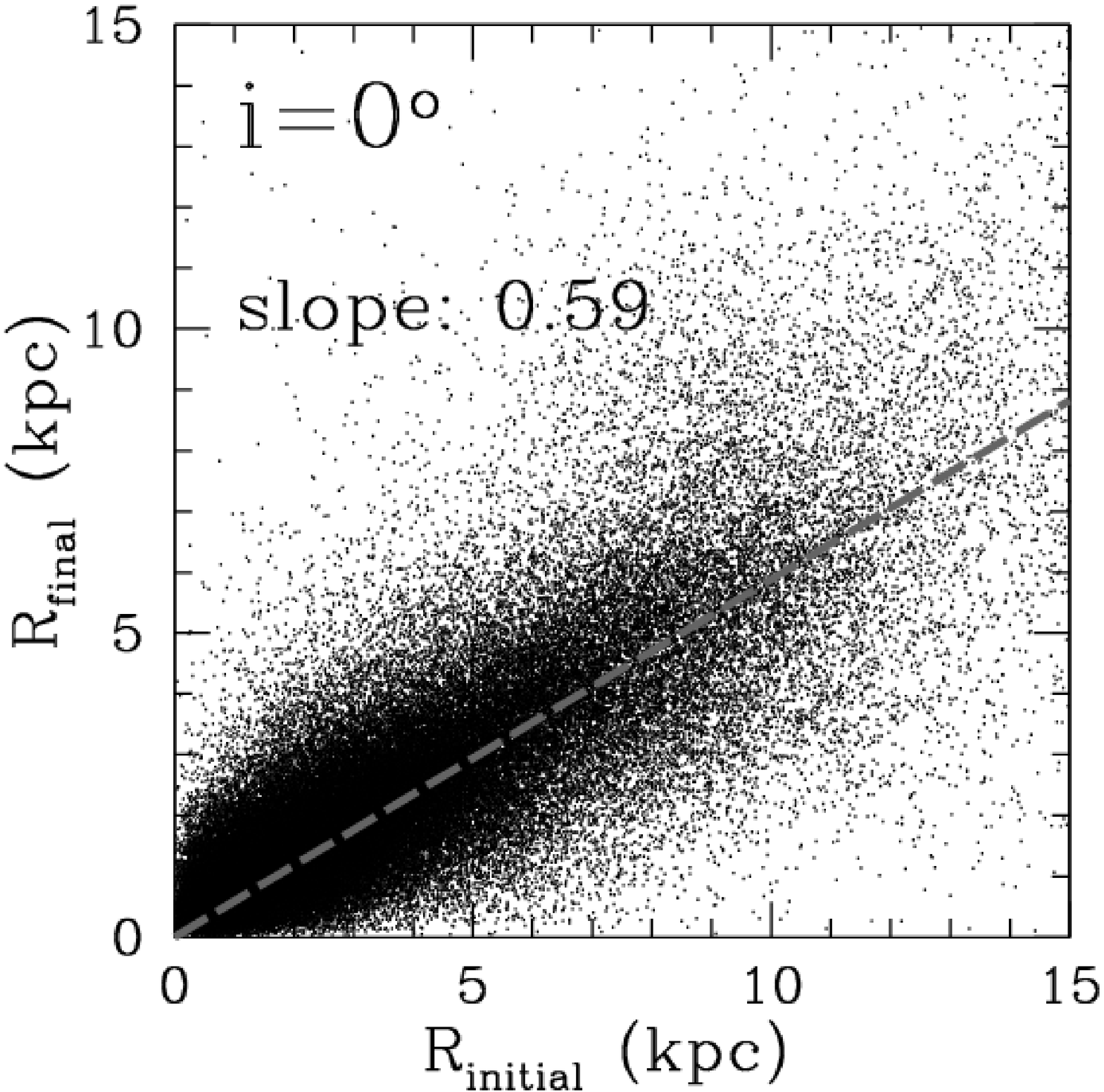}
    \hspace*{-0.1cm}
    \includegraphics[height=50mm]{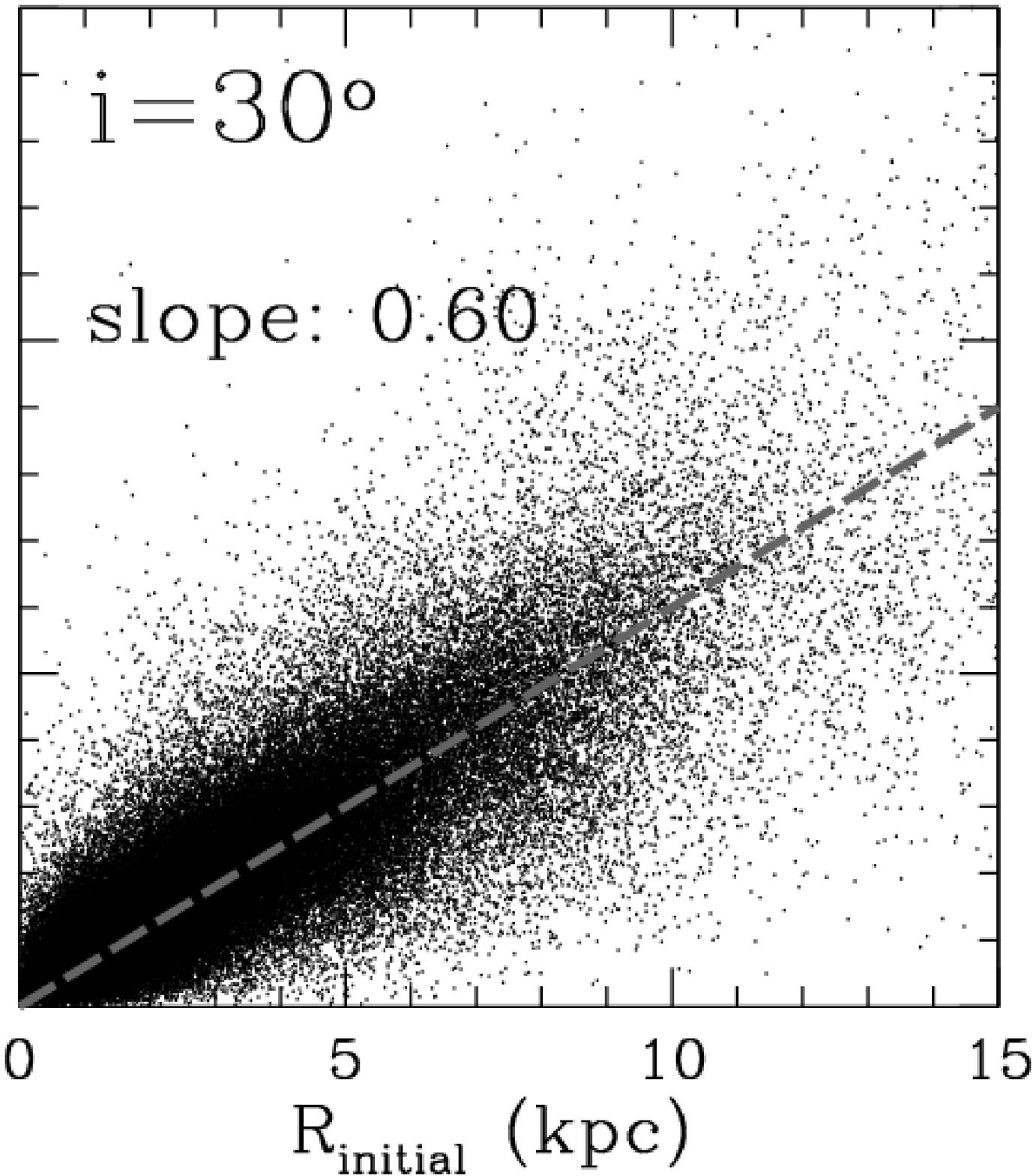}
    \hspace*{-0.1cm}
    \includegraphics[height=50mm]{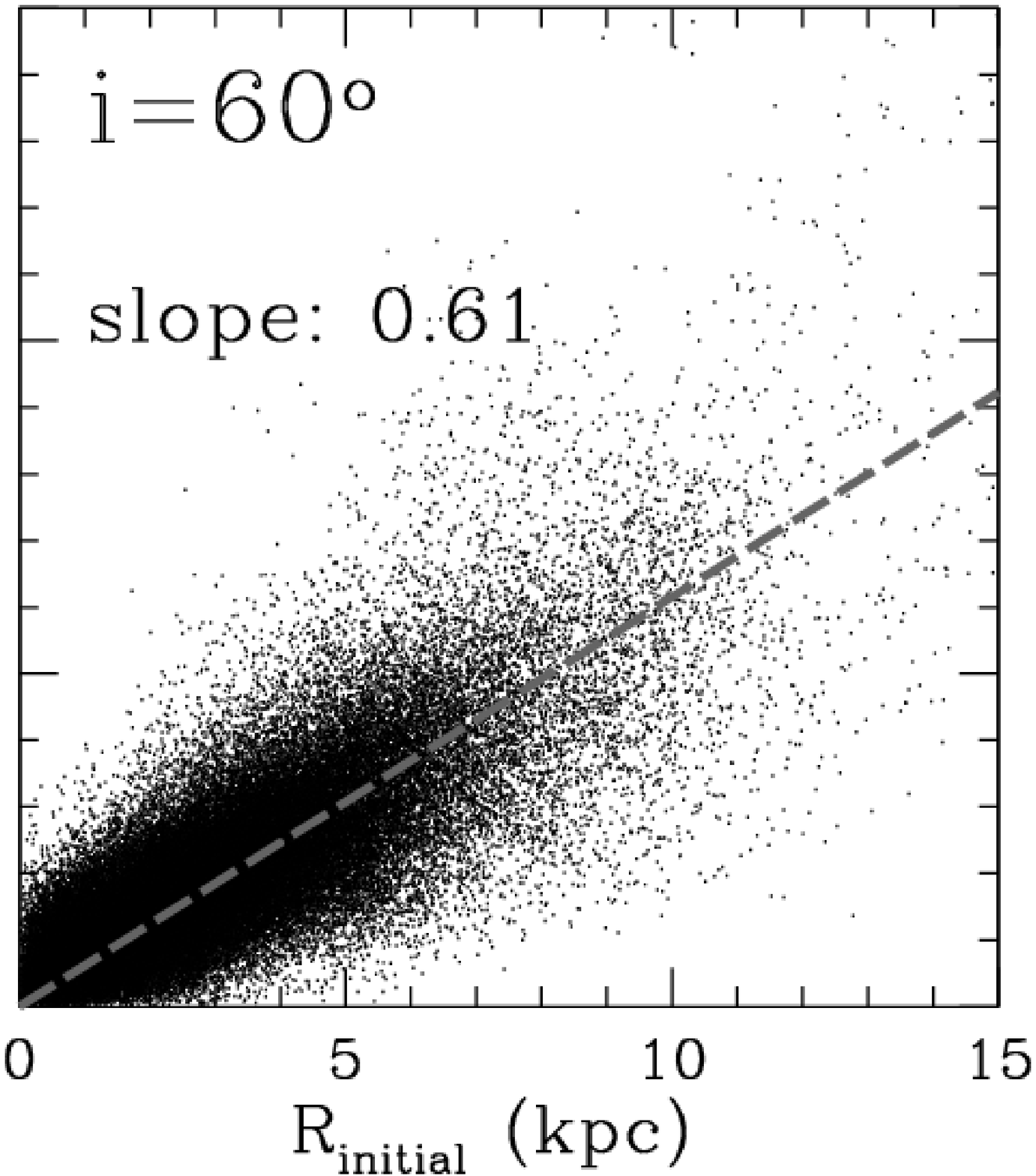}
    \caption{Initial and final cylindrical
      radial distance of heated disk stars before and after the
      structural contraction. Results are presented for experiment A.
      The dashed line shows the best linear fit.  For clarity, only
      heated disk stars are shown.}
    \label{radial-pos-evol-expB}
  \end{center}
\end{figure*}

All numerical experiments of growing disks correspond to a total time
of $t=\tau+0.5$~Gyr (except for the case of instantaneous growth which
is evolved for $1.5$~Gyr to match the simulations with a growth
timescale of $\tau=1$~Gyr).  Apart from the production simulations
described above, we performed two additional sets of experiments.
First, we evolved a subset of the initial thick disks in isolation for
$1.5$~Gyr and confirmed that their properties remained fairly
unmodified. Second, after the full growth of the new thin disks, we
evolved the resulting composite disk galaxies for another $4$~Gyr in
isolation, confirming that their properties do not evolve appreciably
with time. Therefore, we conclude that the evolution of the thick-disks
structural and kinematic properties presented next is due solely to
the adiabatic compression caused by the growth of the thin disks.

\section{Results}
\label{results}

%
\begin{figure}
  \begin{center}
    \includegraphics[width=87mm]{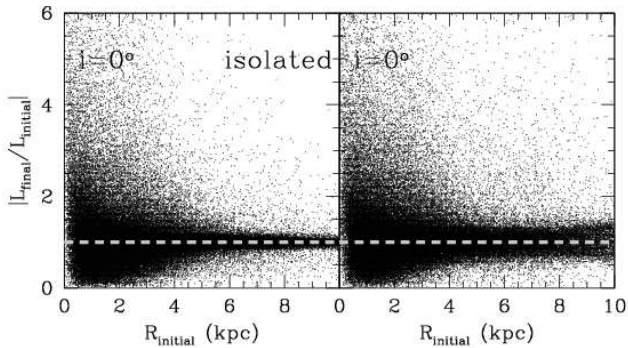}
    \caption{
      Ratio of final-to-initial angular momenta of particles belonging 
      to the isolated thick disk (i.e., evolved without the growing new 
      thin disk) (left panel) and the corresponding contracted thick 
      disk in experiment A1 (right panel). In both cases the thick disks 
      have been evolved for the same amount of time. The dashed line 
      indicates where $|L_{\rm final}/L_{\rm initial}|=1$.  
}
    \label{ang-momenta-a1}
  \end{center}
\end{figure}
\begin{figure}
  \begin{center}
    \includegraphics[width=46mm]{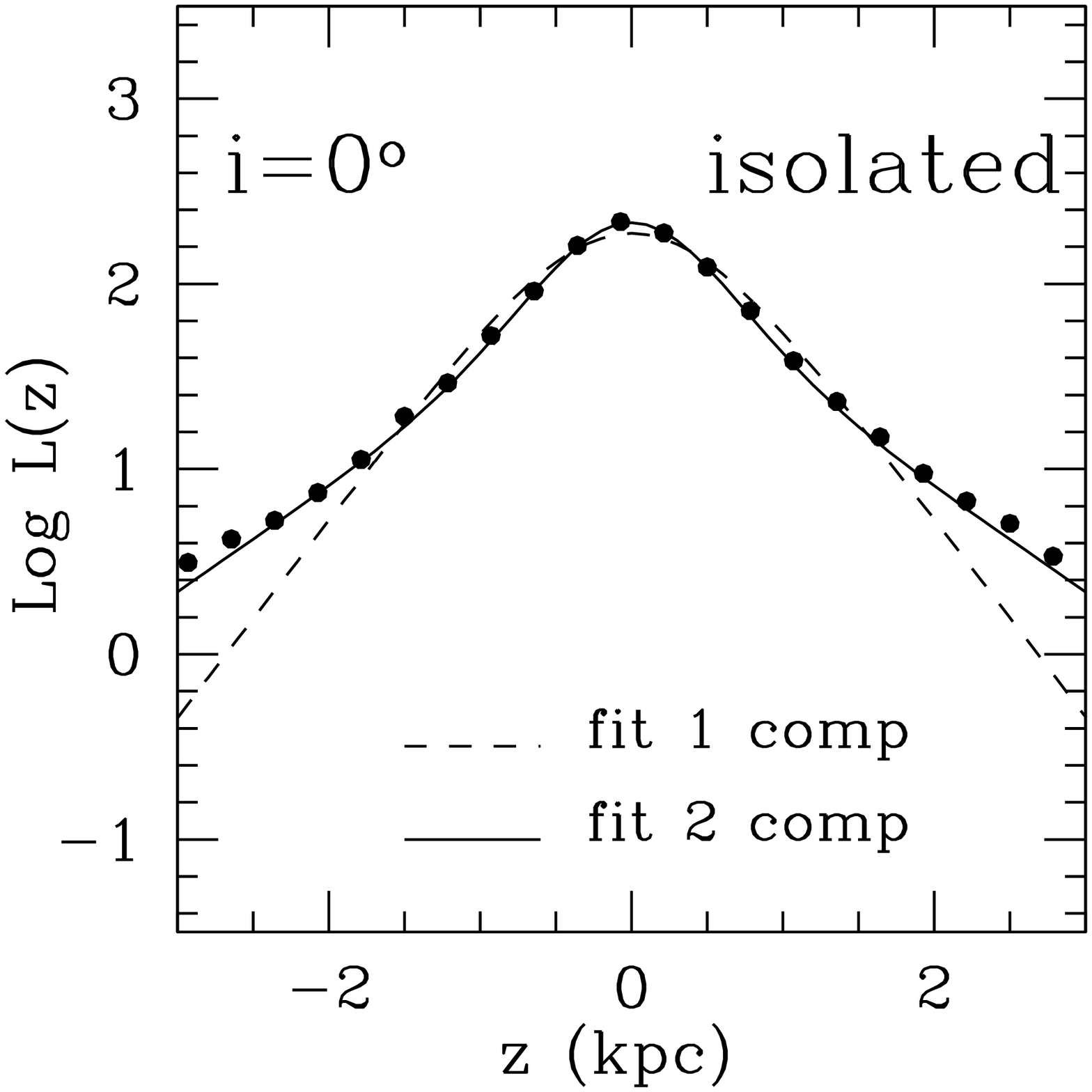}
    \hspace*{-1.092cm} 
    \includegraphics[width=46mm]{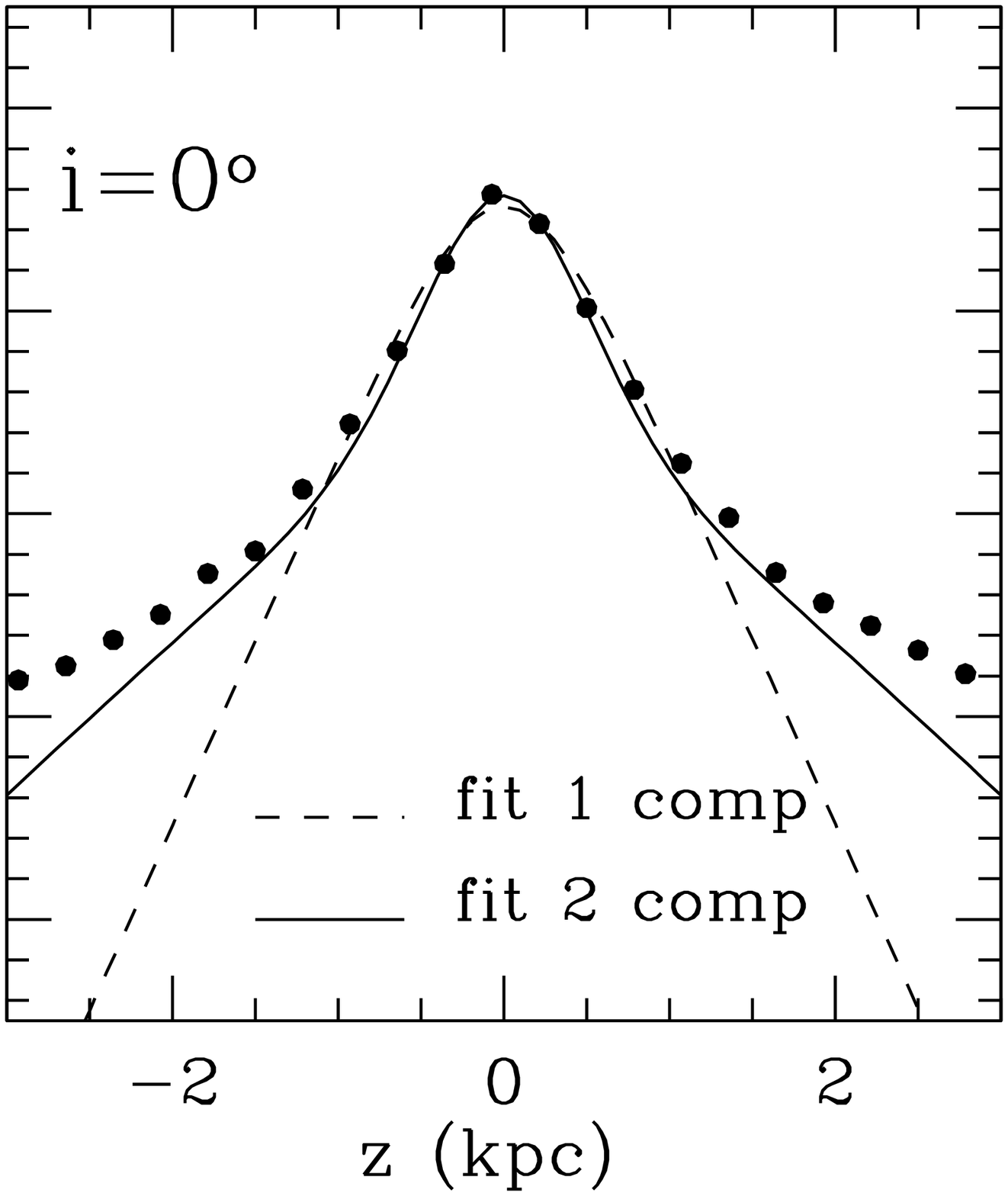}
    \caption{
      Final vertical surface brightness profiles of the
      isolated thick disk (i.e., evolved without the growing new thin disk) 
      (left panel) and the corresponding contracted thick
      disk in run A1 (right panel). The former has been evolved in
      isolation for the same amount of time as the latter and profiles
      are obtained via integrating within $R<10$~kpc. Dashed and solid
      lines show results for the best-fit with one- and two-component
      sech$^2$ decompositions, respectively. The growth of a new thin
      disk induces changes to the vertical structure of the initial 
      thick disks, highlighting the need for a more complicated functional 
      form to describe the final disk structure at large heights.
}
    \label{flare-experB}
  \end{center}
\end{figure}
%

\subsection{Structural Evolution of Thick Disks}

\subsubsection{General Features}
\label{struc-gral}

\begin{figure*}
  \begin{center}
    \includegraphics[width=52mm]{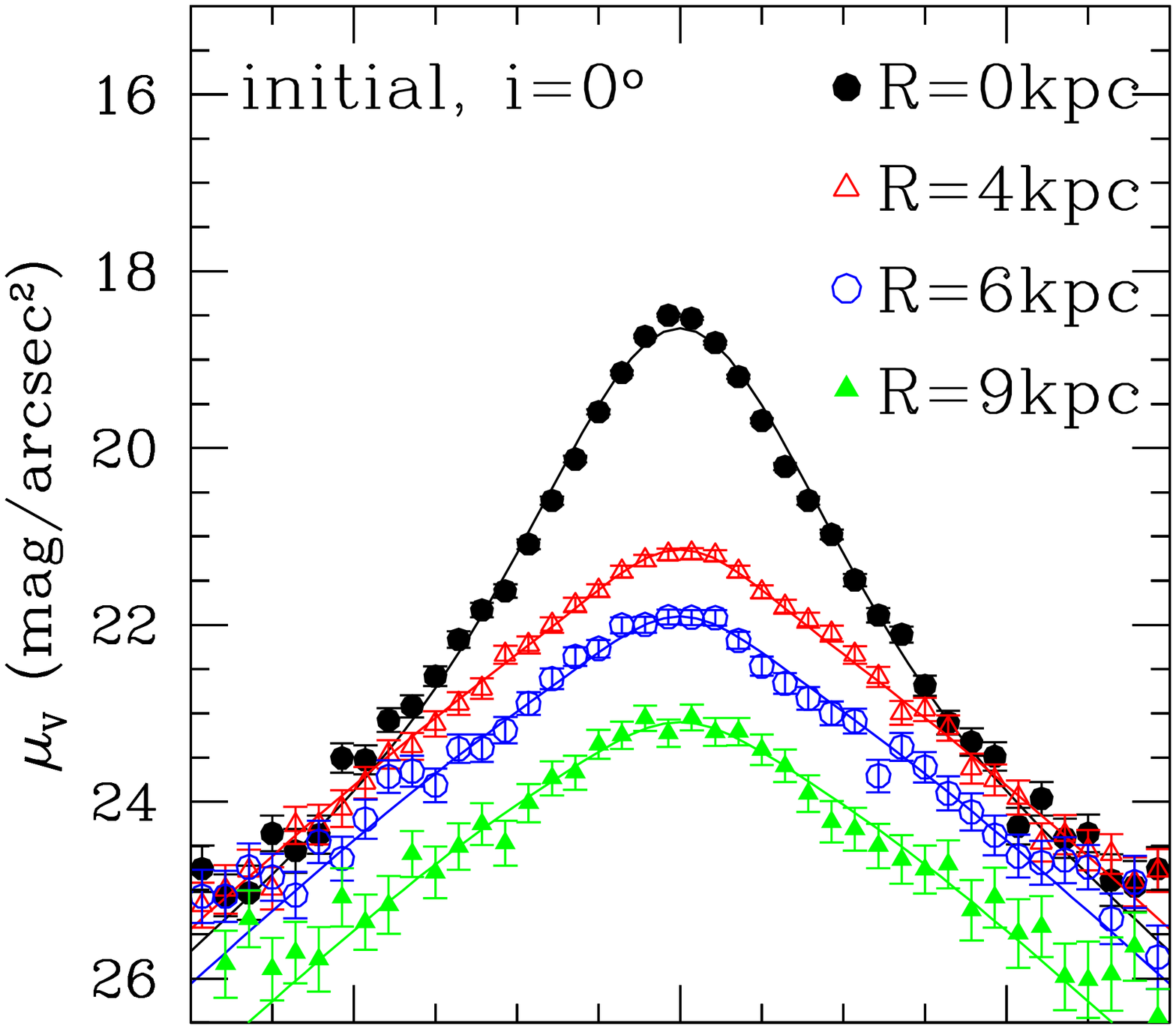}
    \hspace*{-1.2cm}
    \includegraphics[width=52mm]{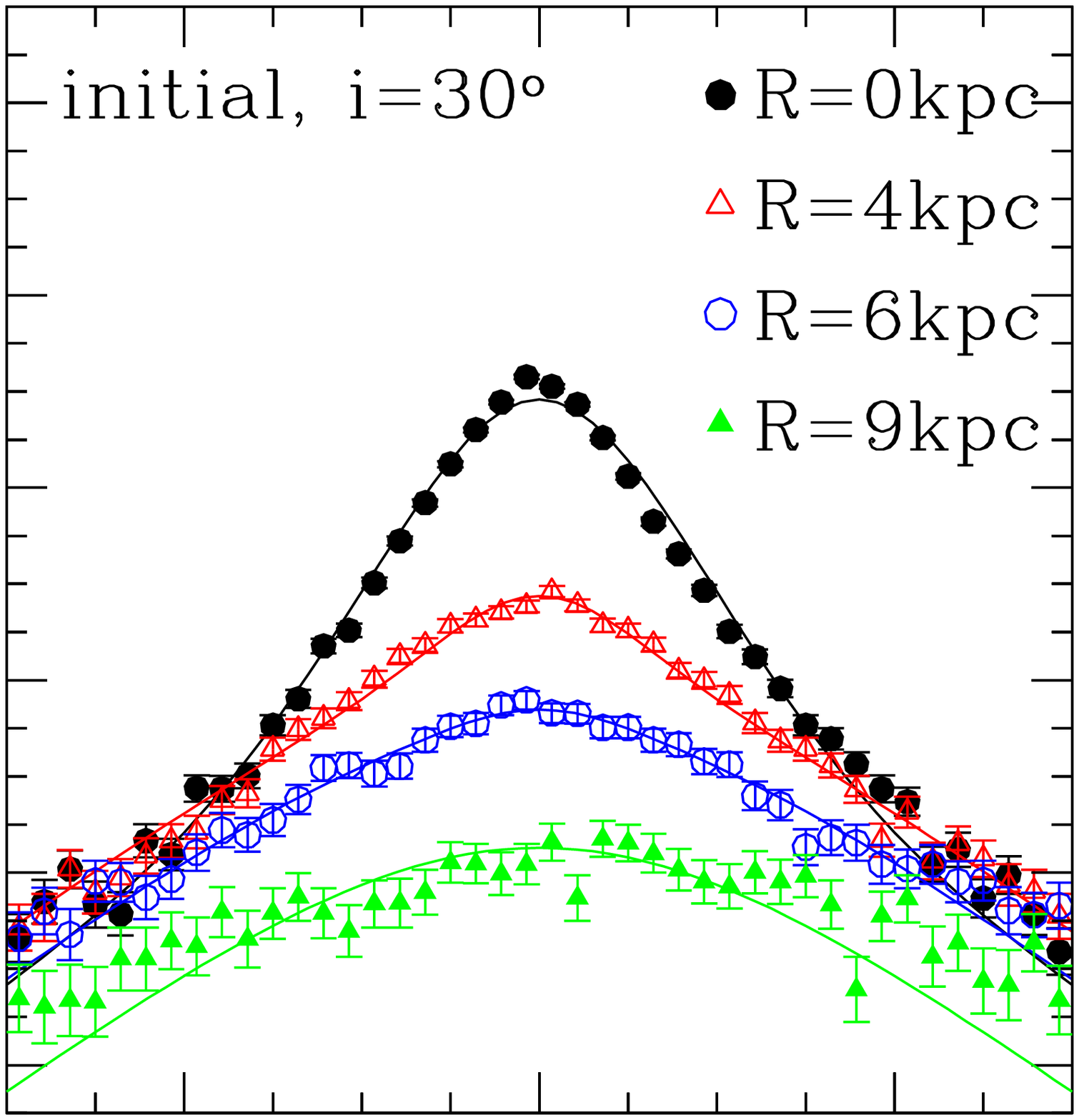}
    \hspace*{-1.2cm}
    \includegraphics[width=52mm]{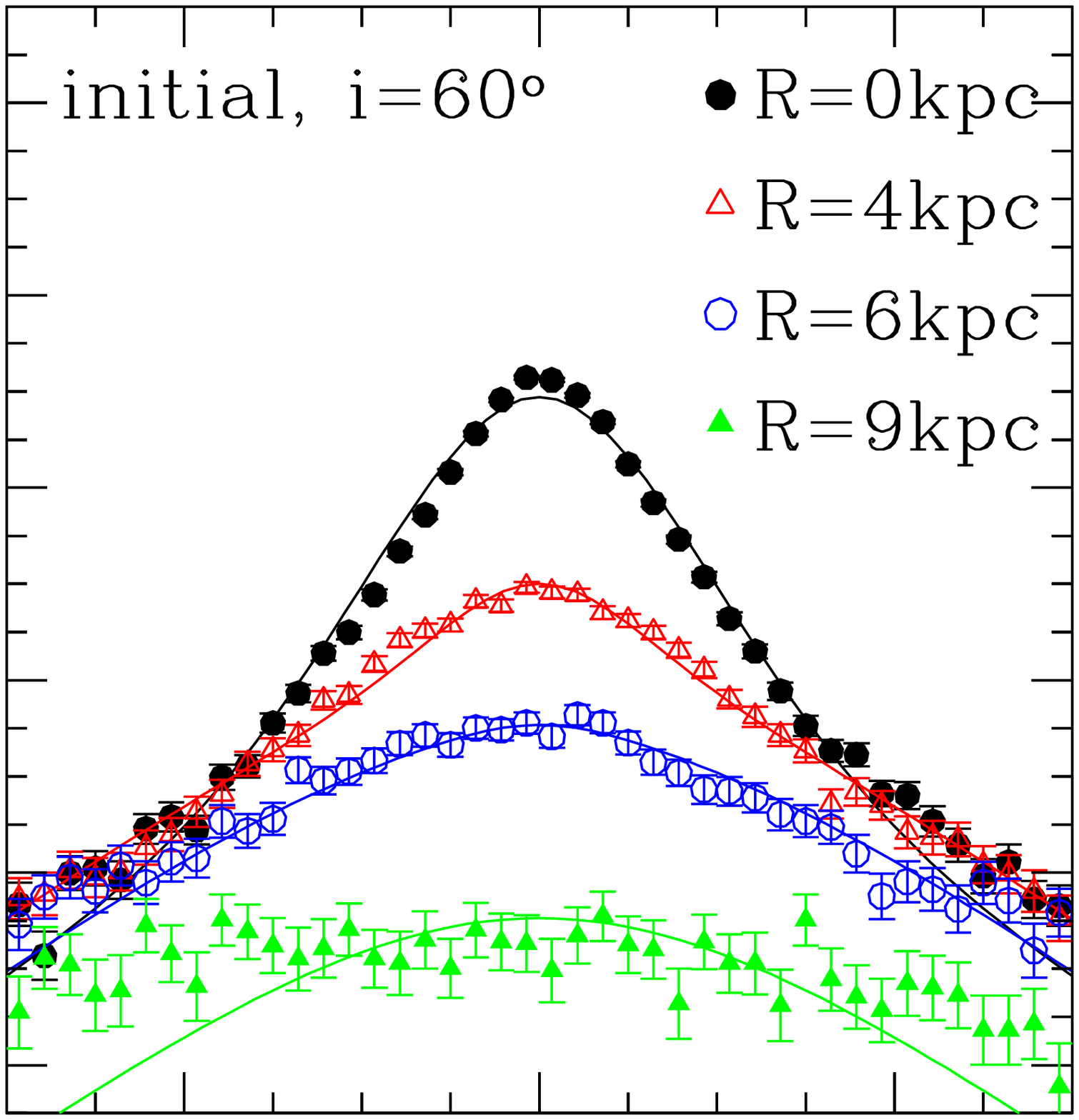}\\
    \vspace*{-0.94cm}
    \includegraphics[width=52mm]{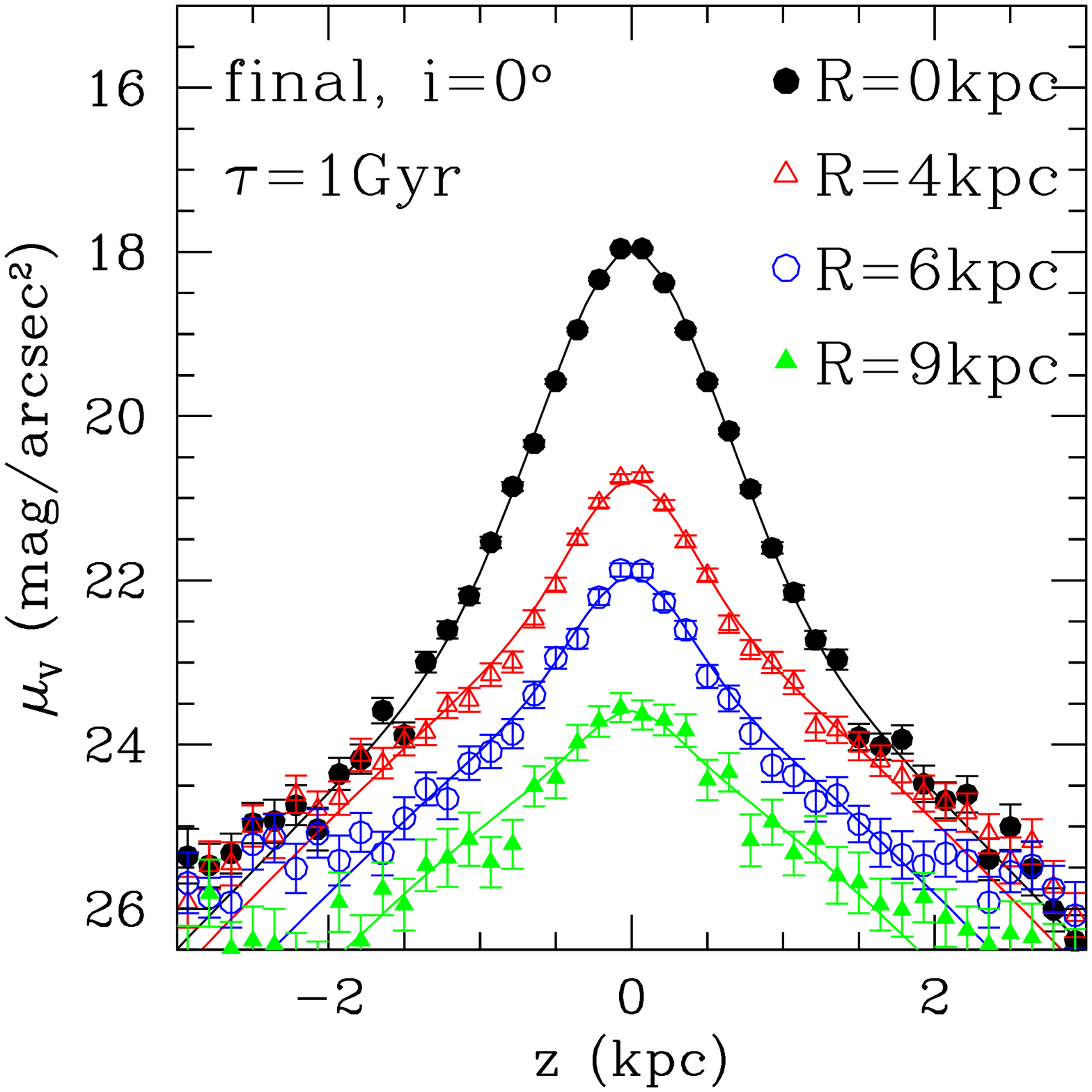}
    \hspace*{-1.2cm}
    \includegraphics[width=52mm]{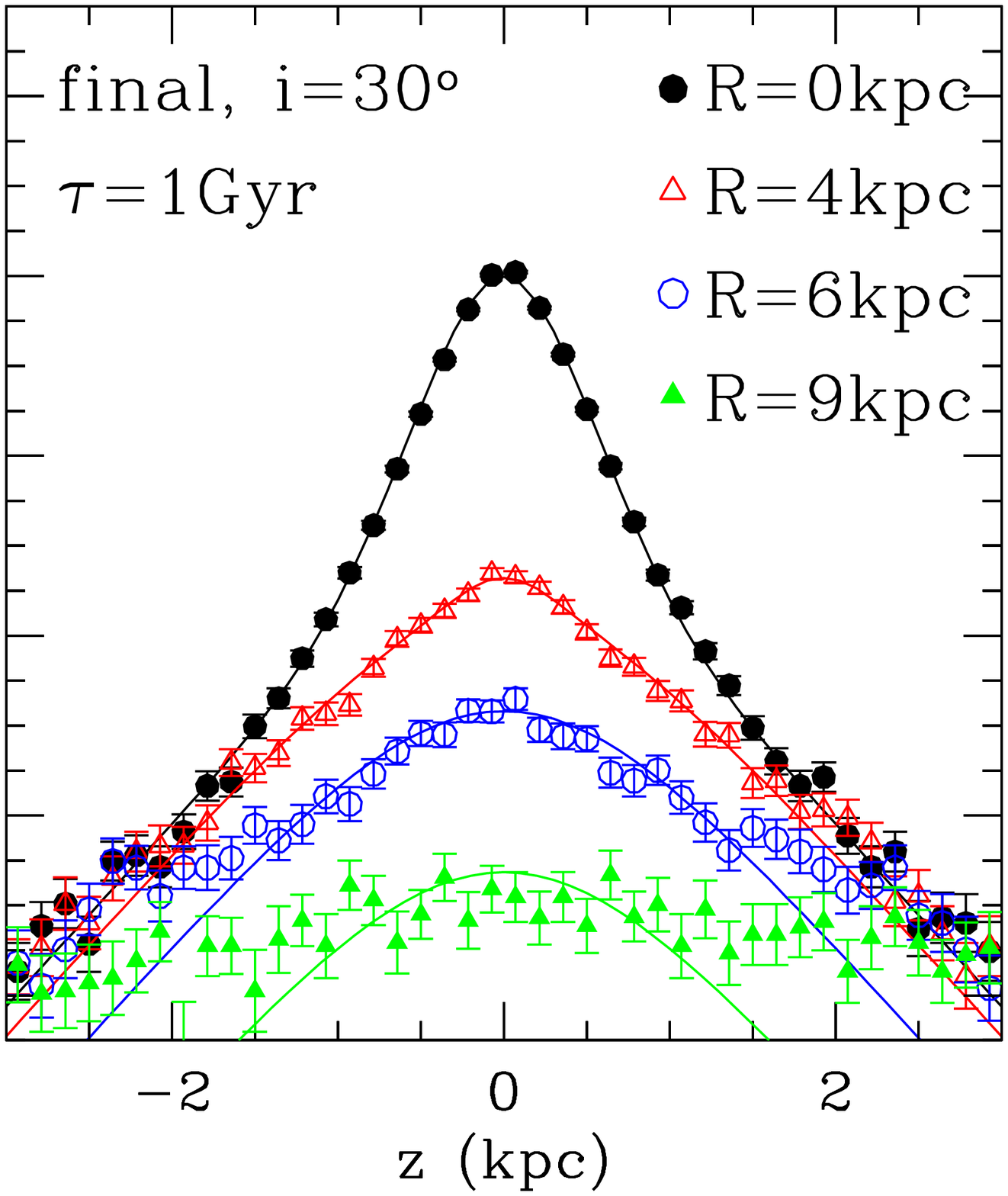}
    \hspace*{-1.2cm}
    \includegraphics[width=52mm]{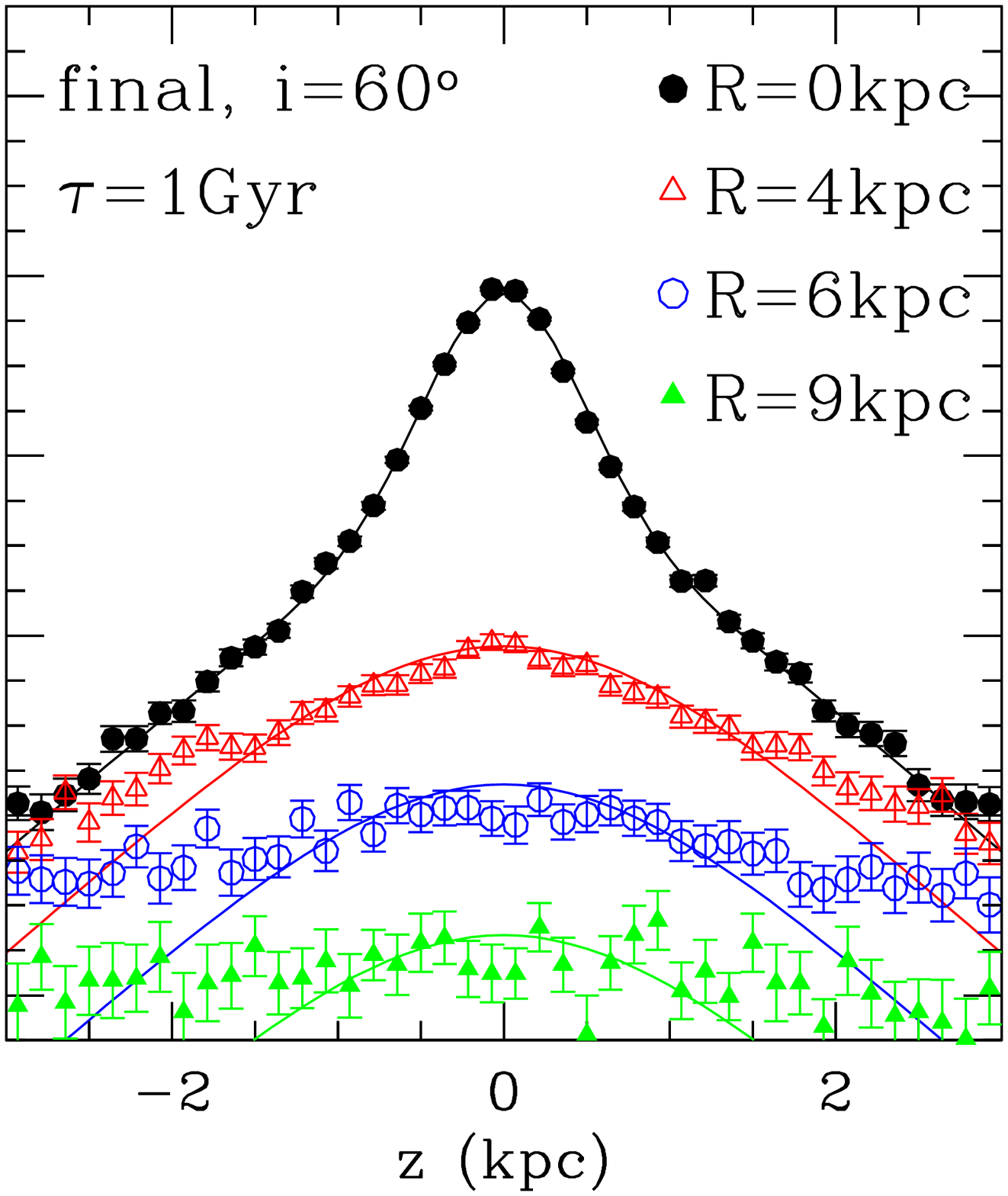}\\
    \includegraphics[width=52mm]{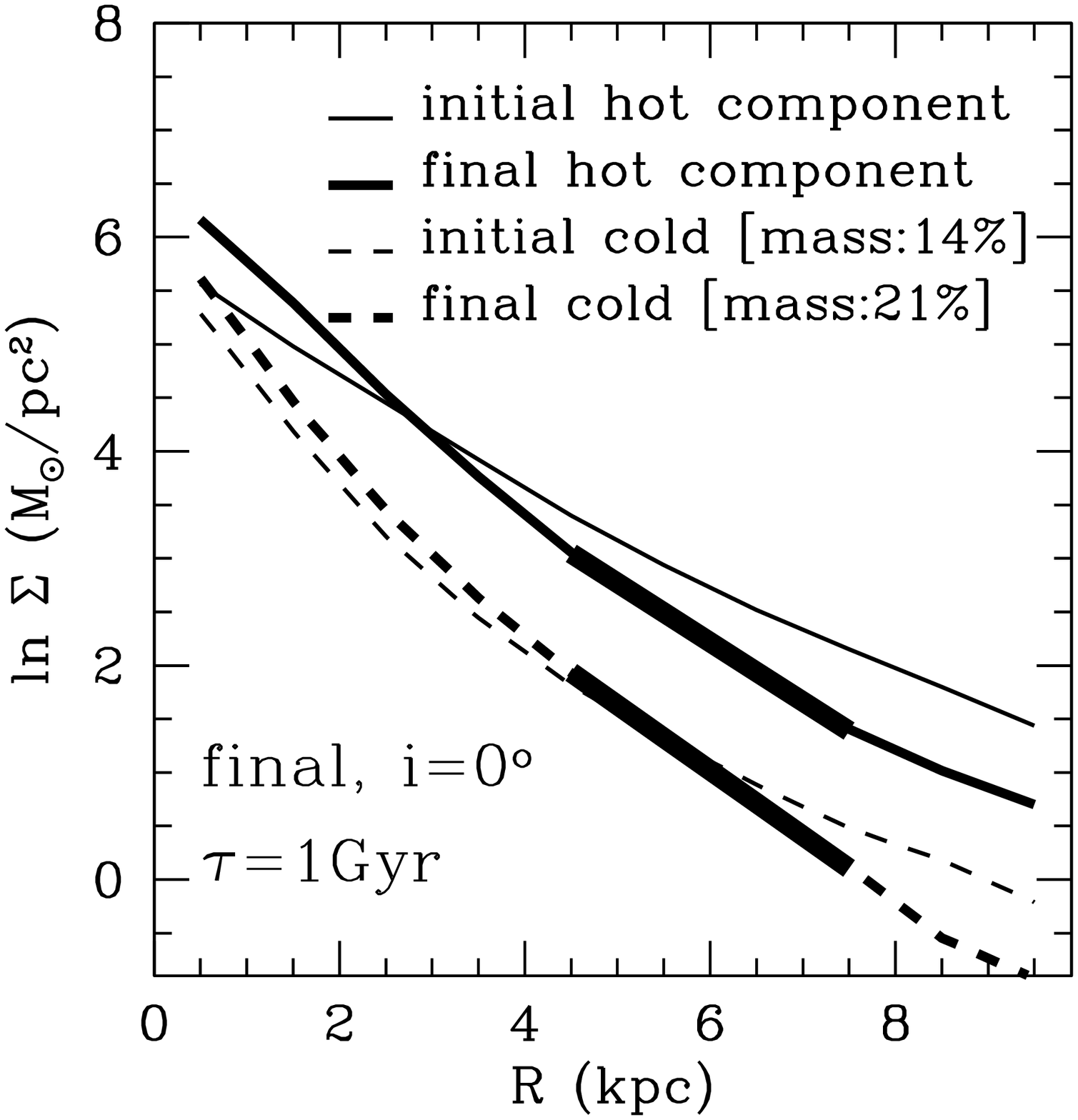}
    \hspace*{-1.2cm}
    \includegraphics[width=52mm]{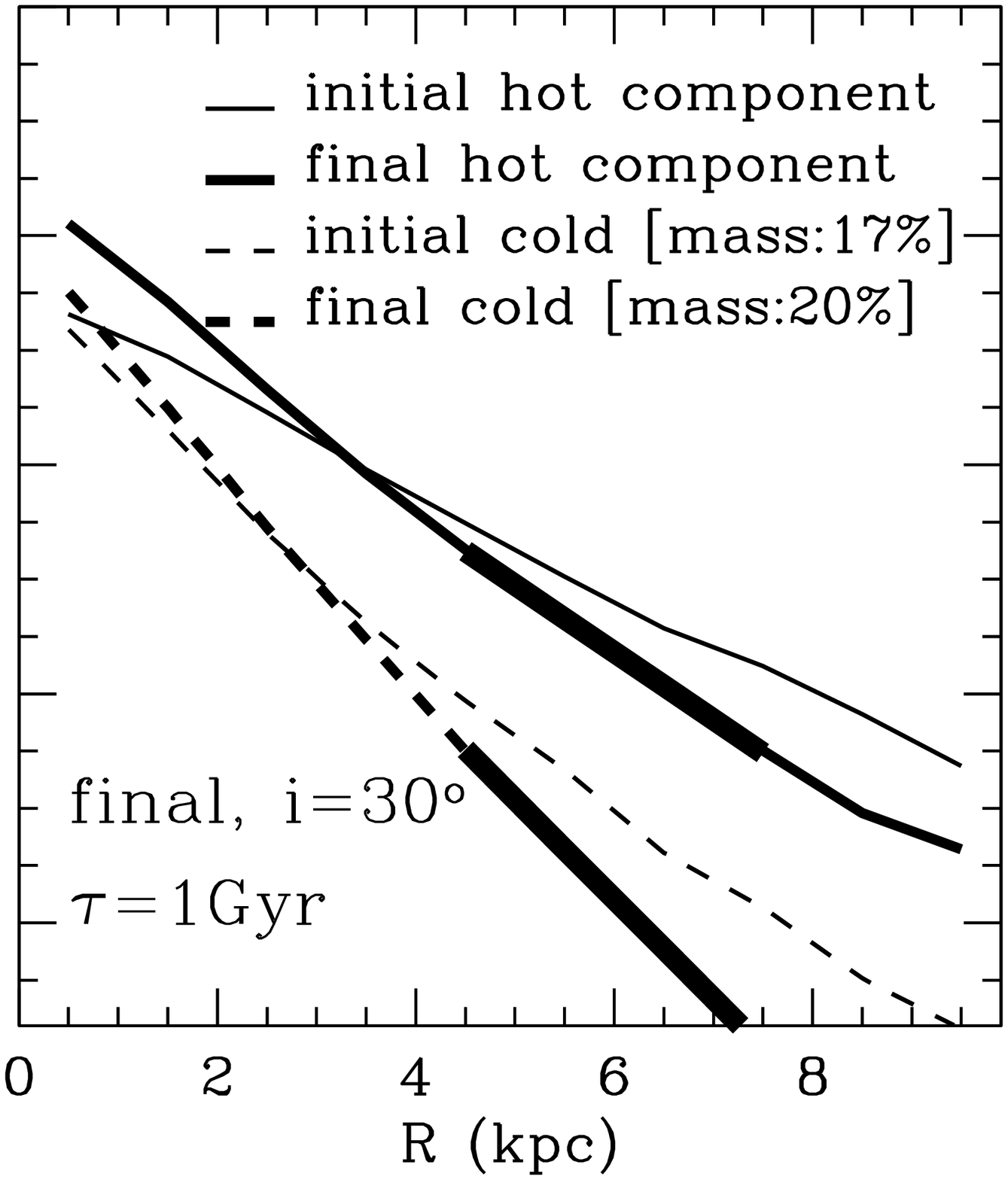}
    \hspace*{-1.2cm}
    \includegraphics[width=52mm]{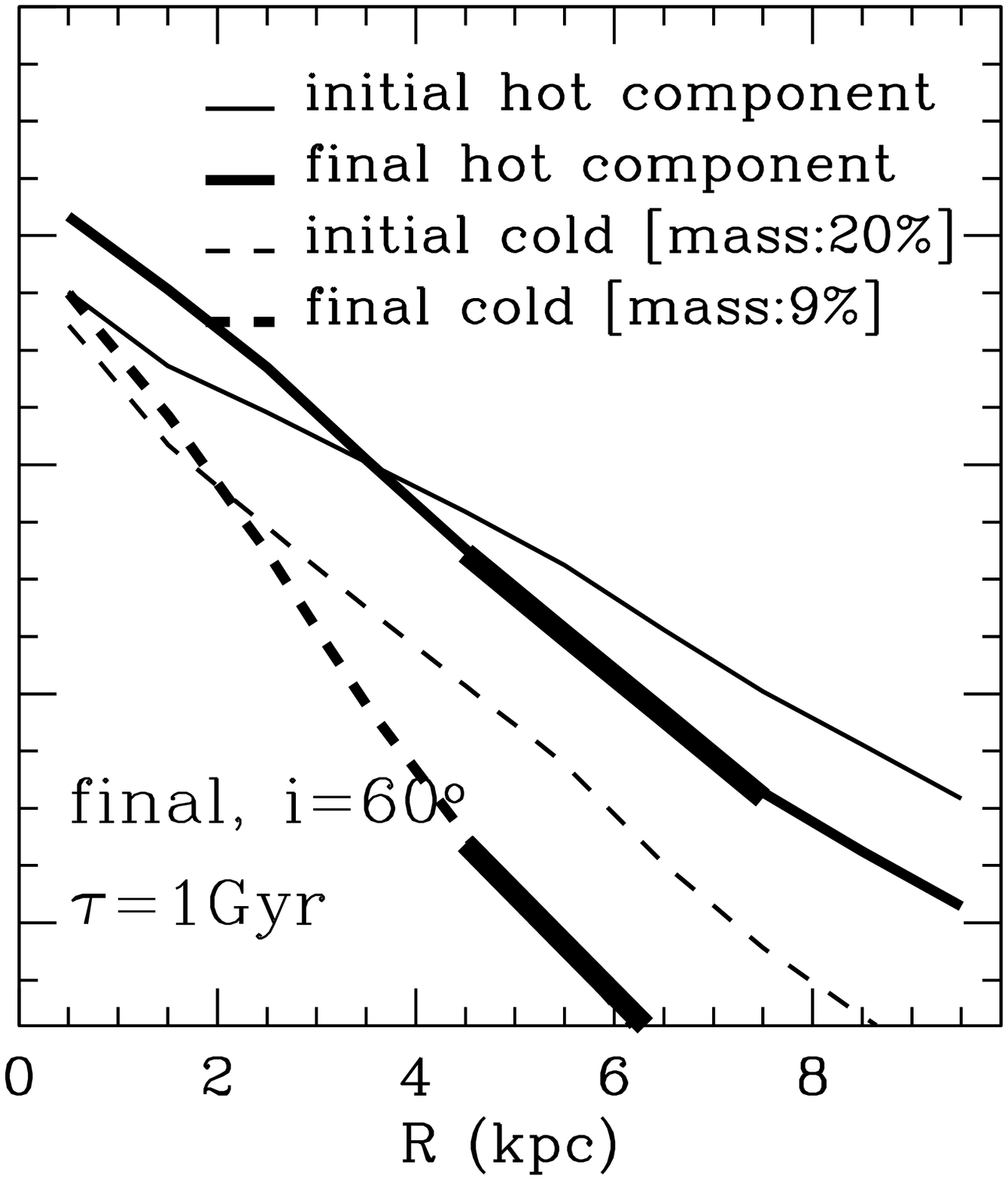}\\
    \includegraphics[width=72mm]{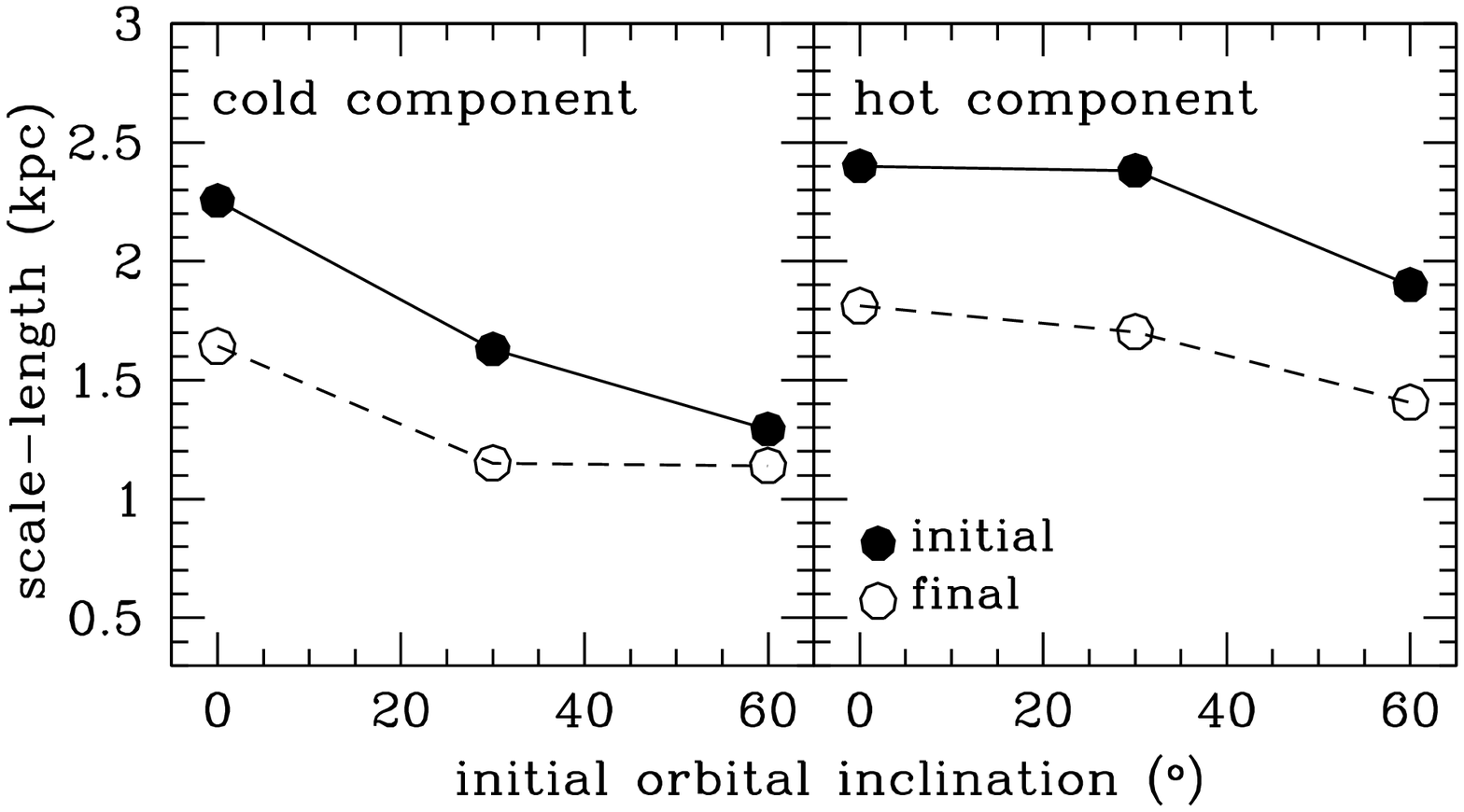}
    \includegraphics[width=72mm]{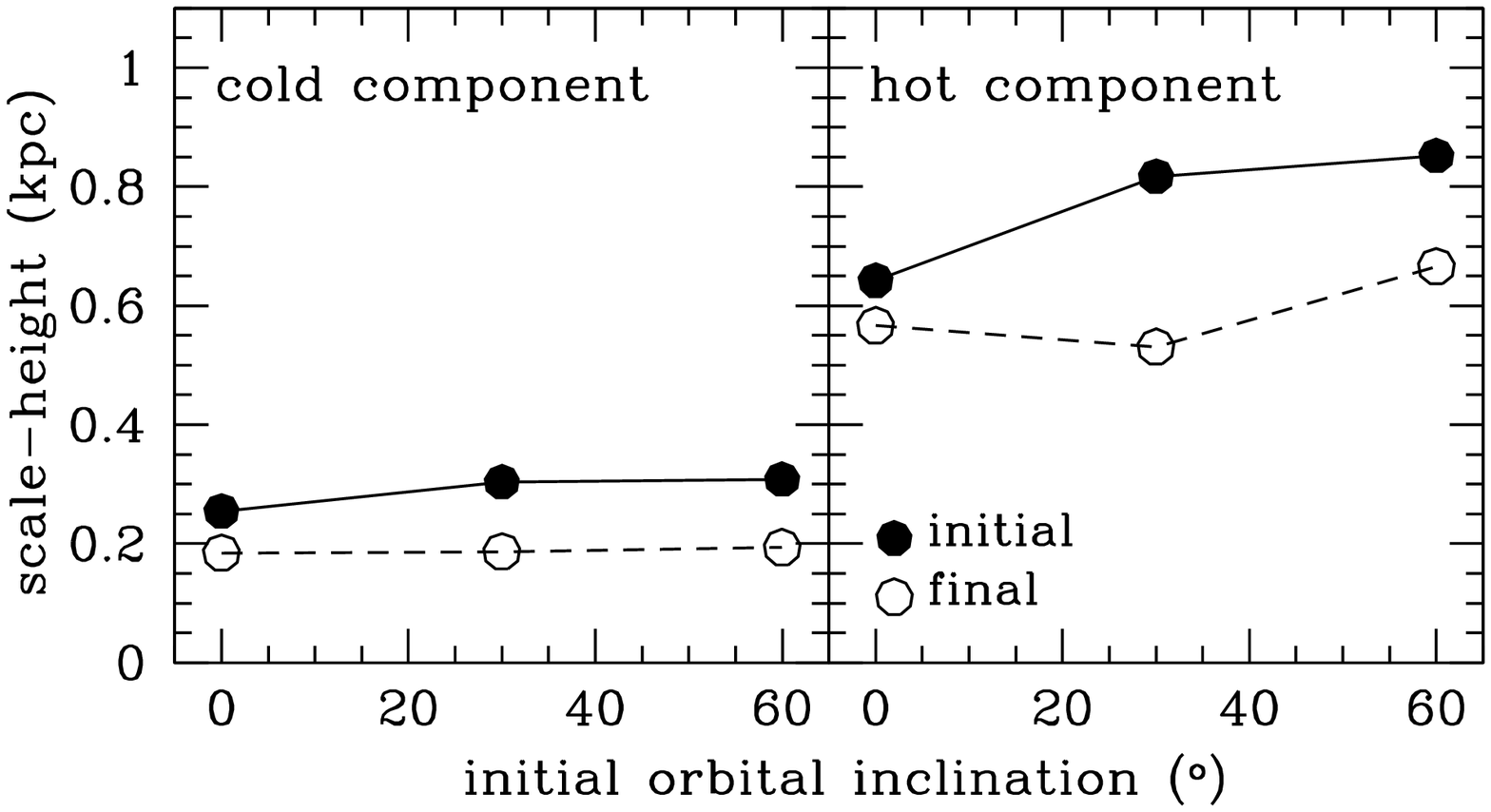}
    \caption{Experiment A. Final thick-disk properties induced by the
      growth of the fiducial thin disk on a timescale of $\tau=1$~Gyr.
      The initial thick disks were formed by satellite accretion
      events with initial orbital inclinations of $i=0\degr$ (left
      panels), $i=30\degr$ (middle panels), and $i=60\degr$ (right
      panels).  From top to bottom: Rows 1 and 2 show the initial and
      final vertical surface brightness profiles of the thick disks.
      Row 3 shows results for final thick-disk surface density
      profiles decomposed into a ``cold'' and a ``hot'' component. The
      mass fractions associated with these components are indicated in
      each panel. Row 4 presents initial (filled symbols) and final
      (open symbols) scale-lengths and scale-heights for the ``cold''
      and ``hot'' components as a function of the orbital inclination
      of the encounter that produced the initial thick disks.}
    \label{struc-a}
 \end{center}
\end{figure*}
\begin{figure*}
  \begin{center}
    \includegraphics[width=52mm]{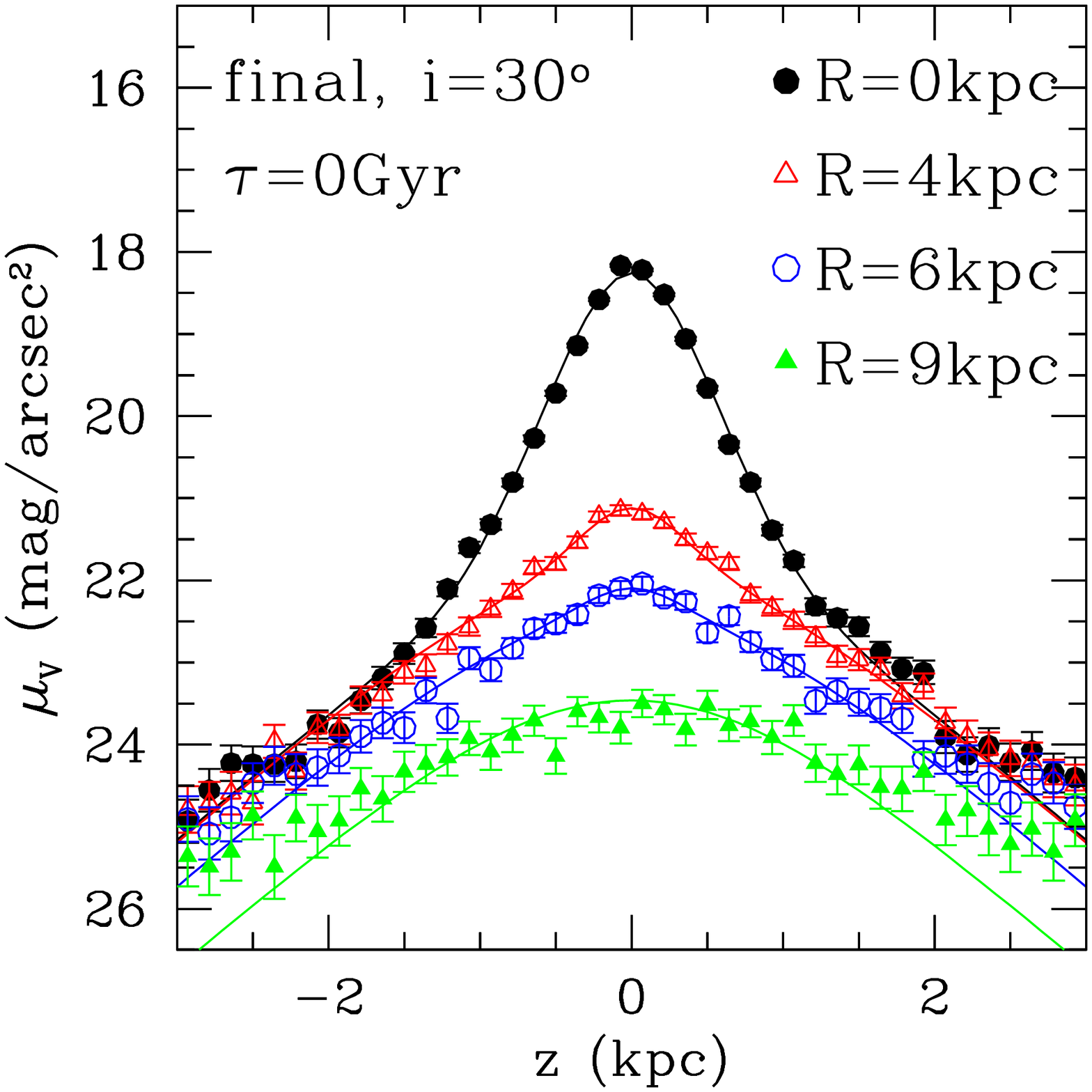}
    \hspace*{-1.2cm}
    \includegraphics[width=52mm]{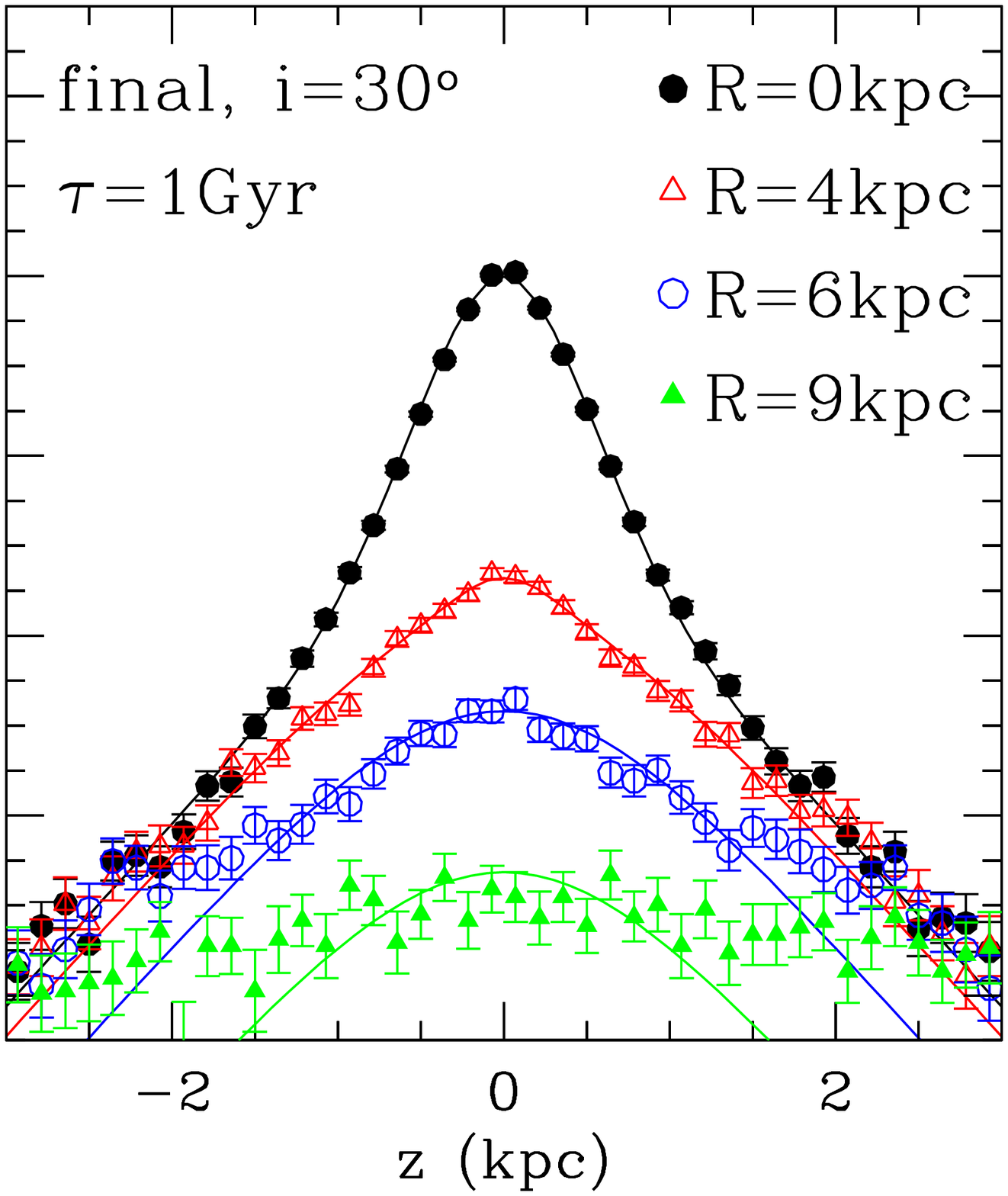}
    \hspace*{-1.2cm}
    \includegraphics[width=52mm]{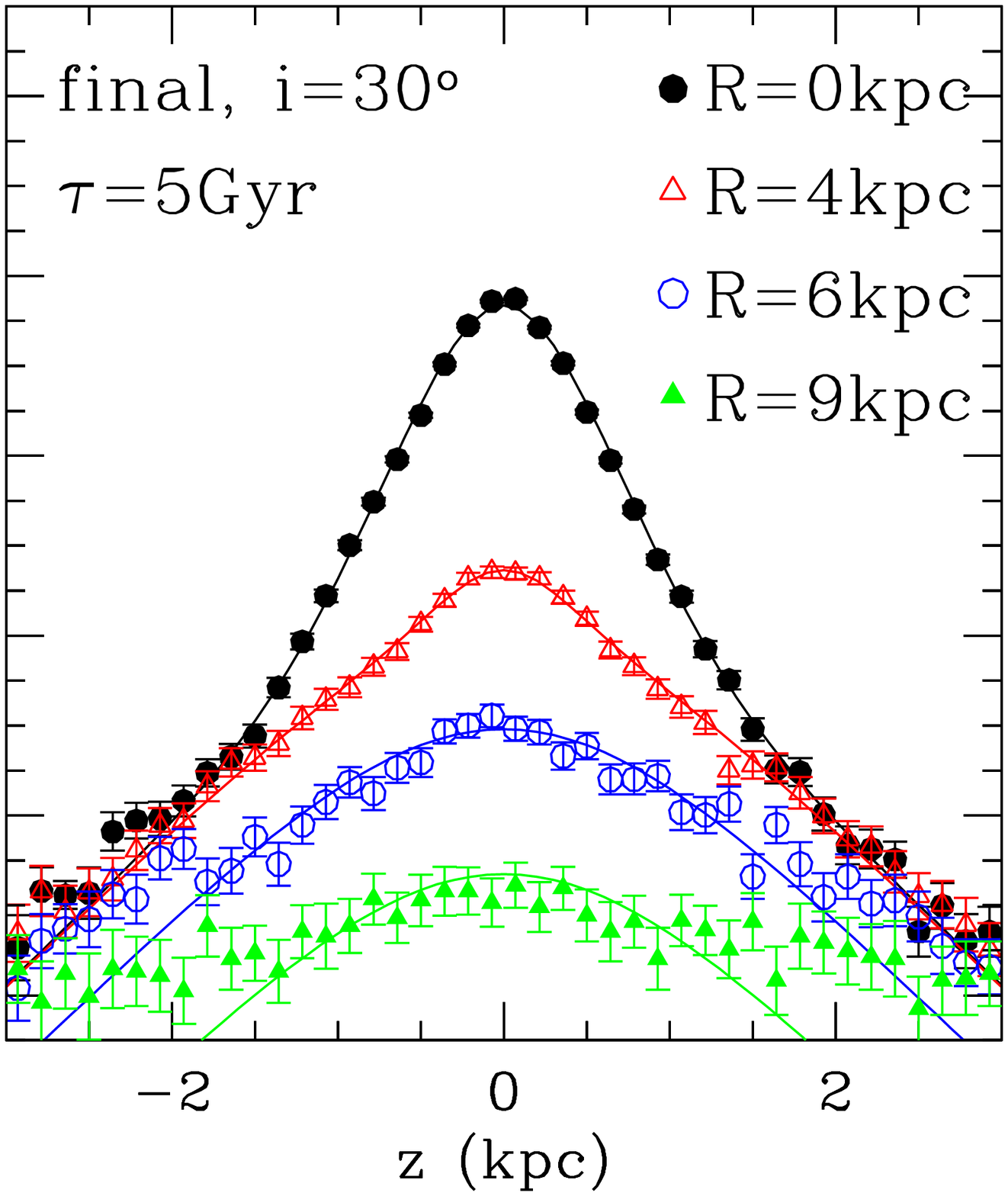}\\
    \includegraphics[width=52mm]{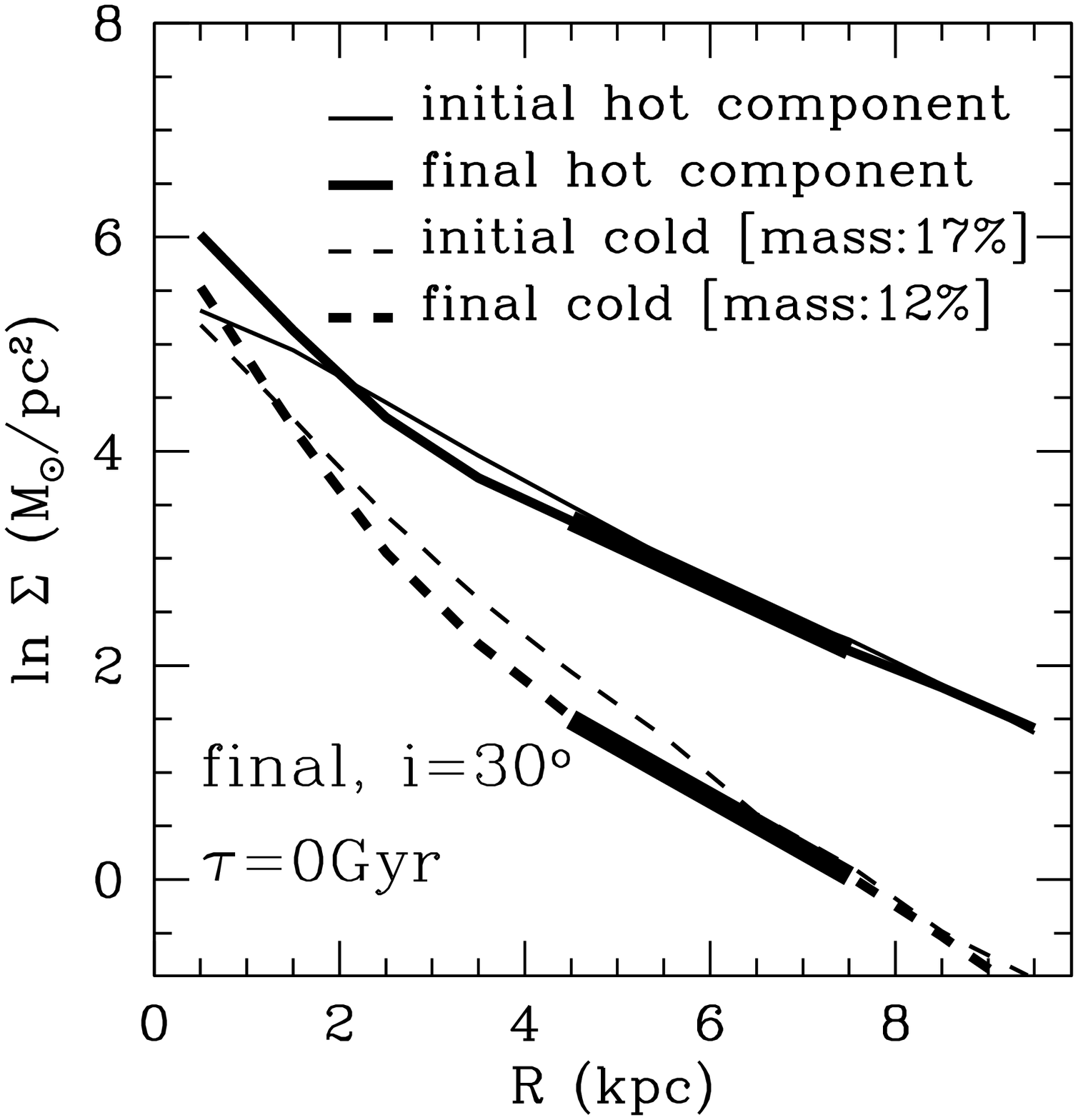}
    \hspace*{-1.2cm}
    \includegraphics[width=52mm]{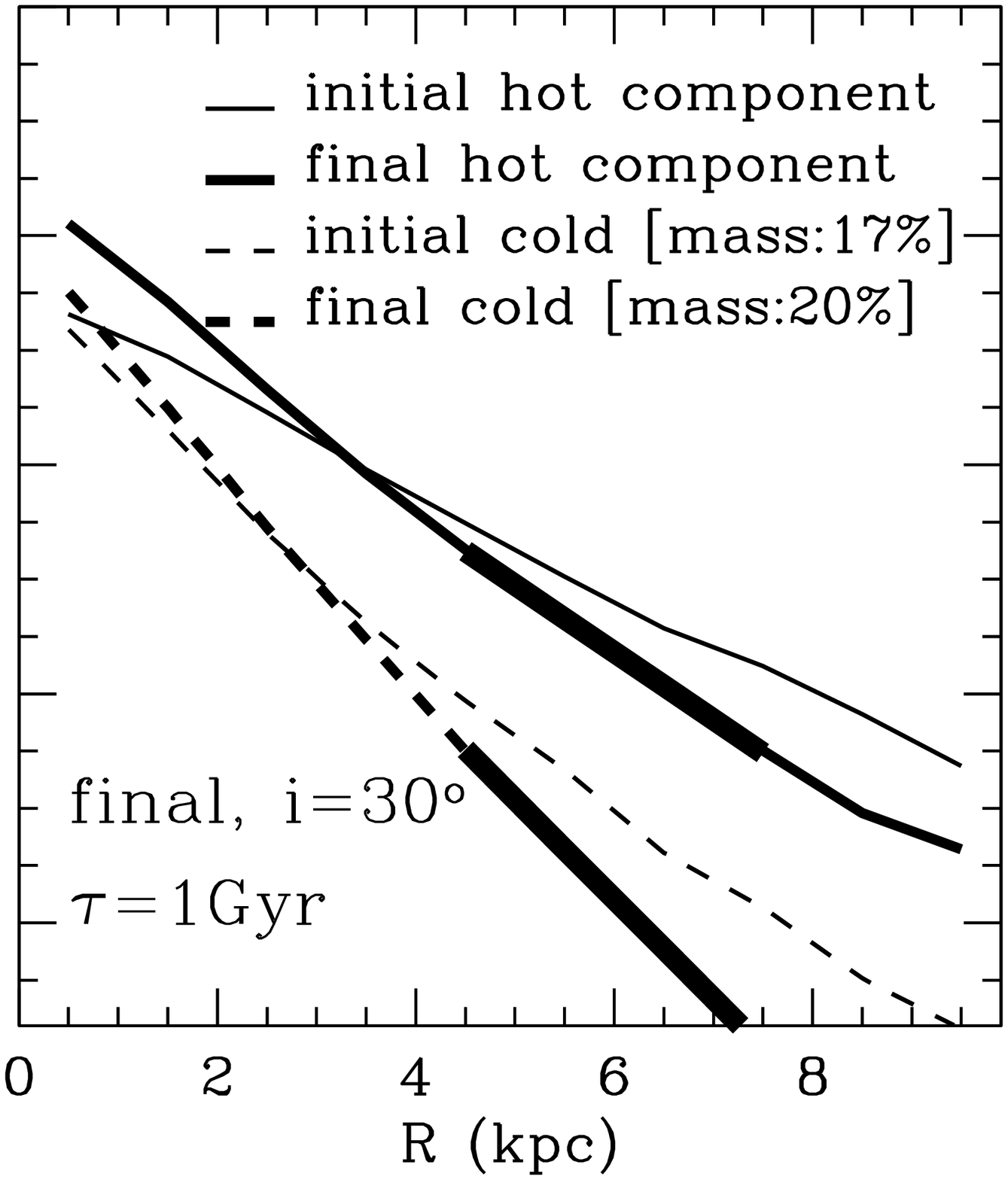}
    \hspace*{-1.2cm}
    \includegraphics[width=52mm]{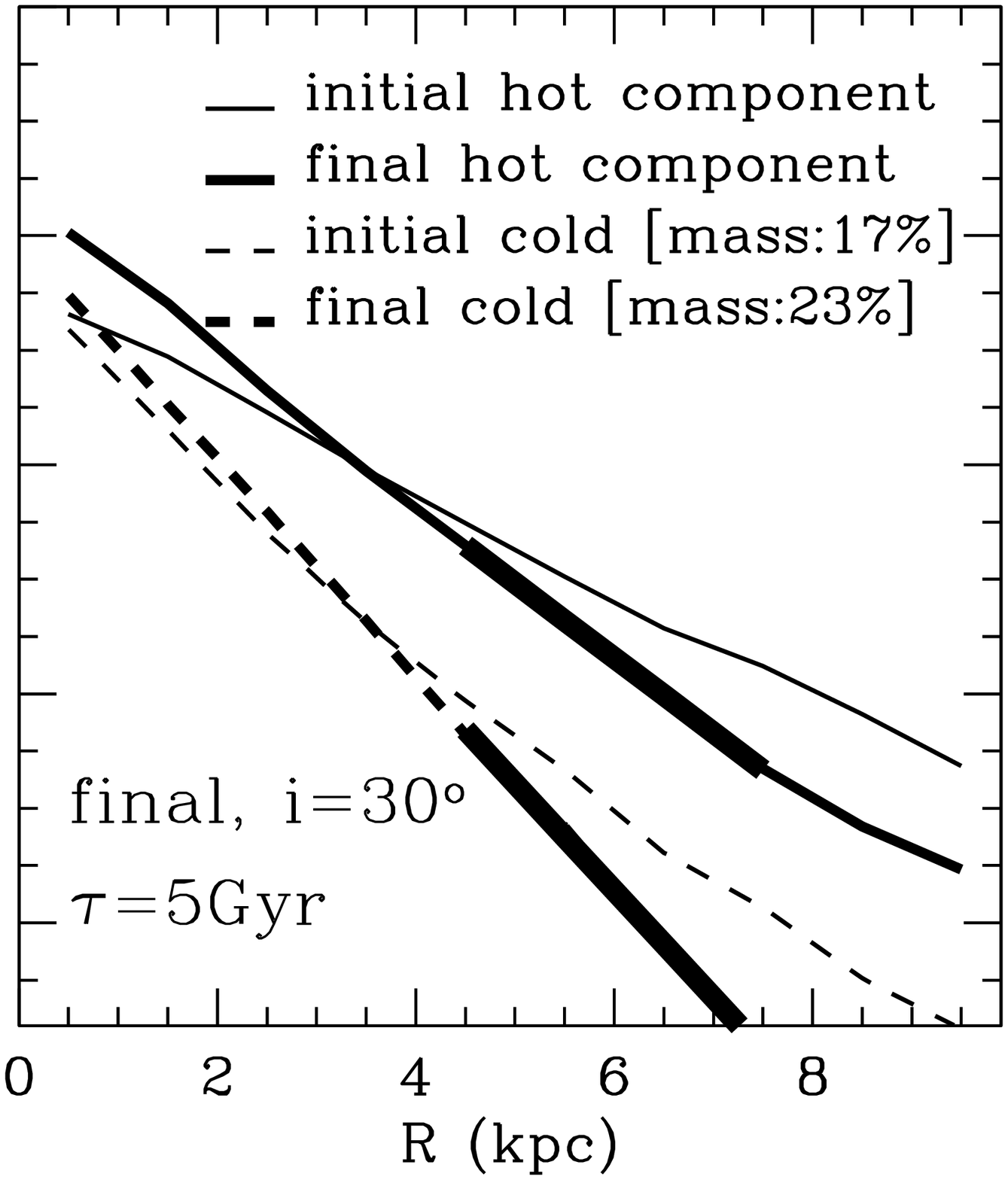}\\
    \includegraphics[width=72mm]{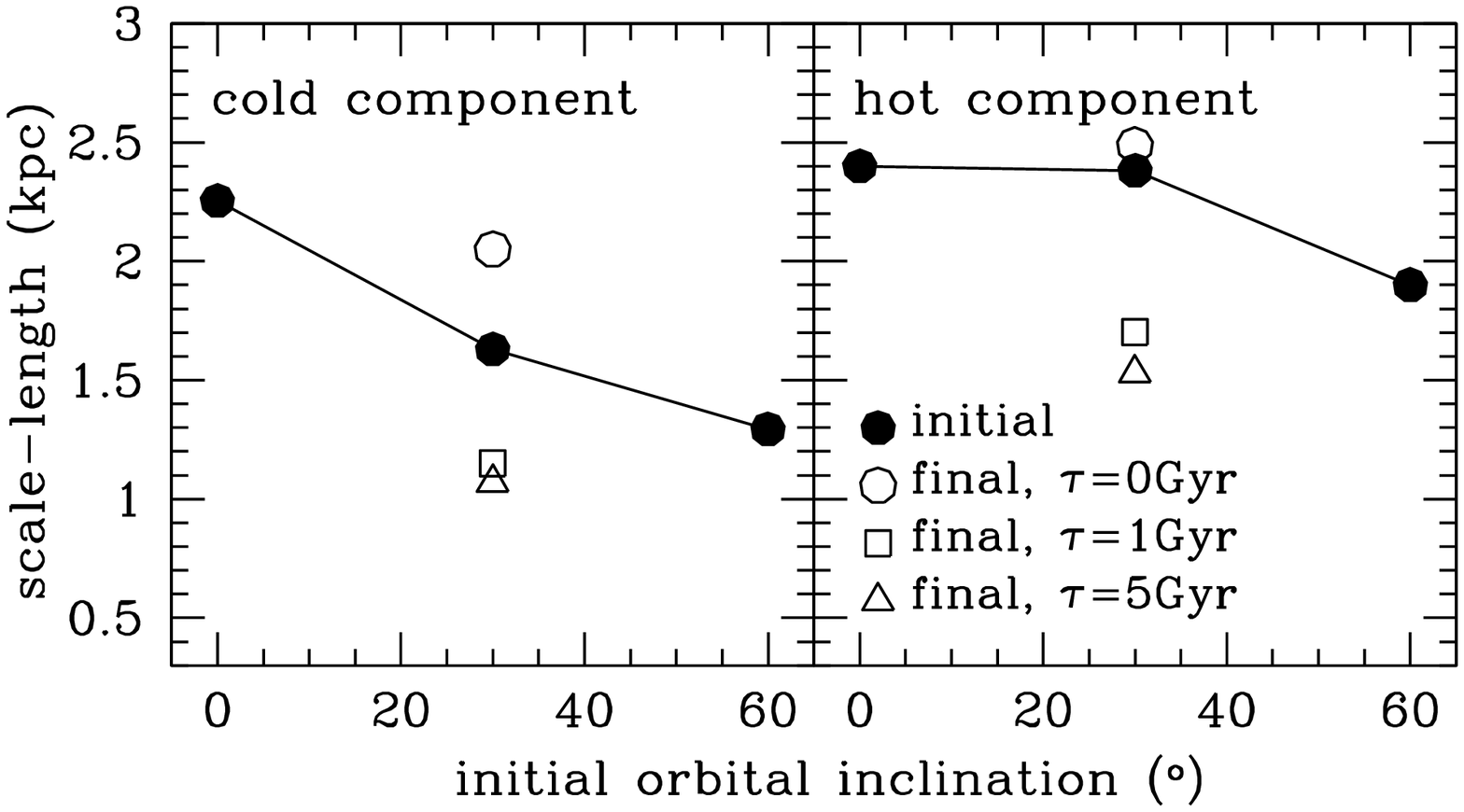}
    \includegraphics[width=72mm]{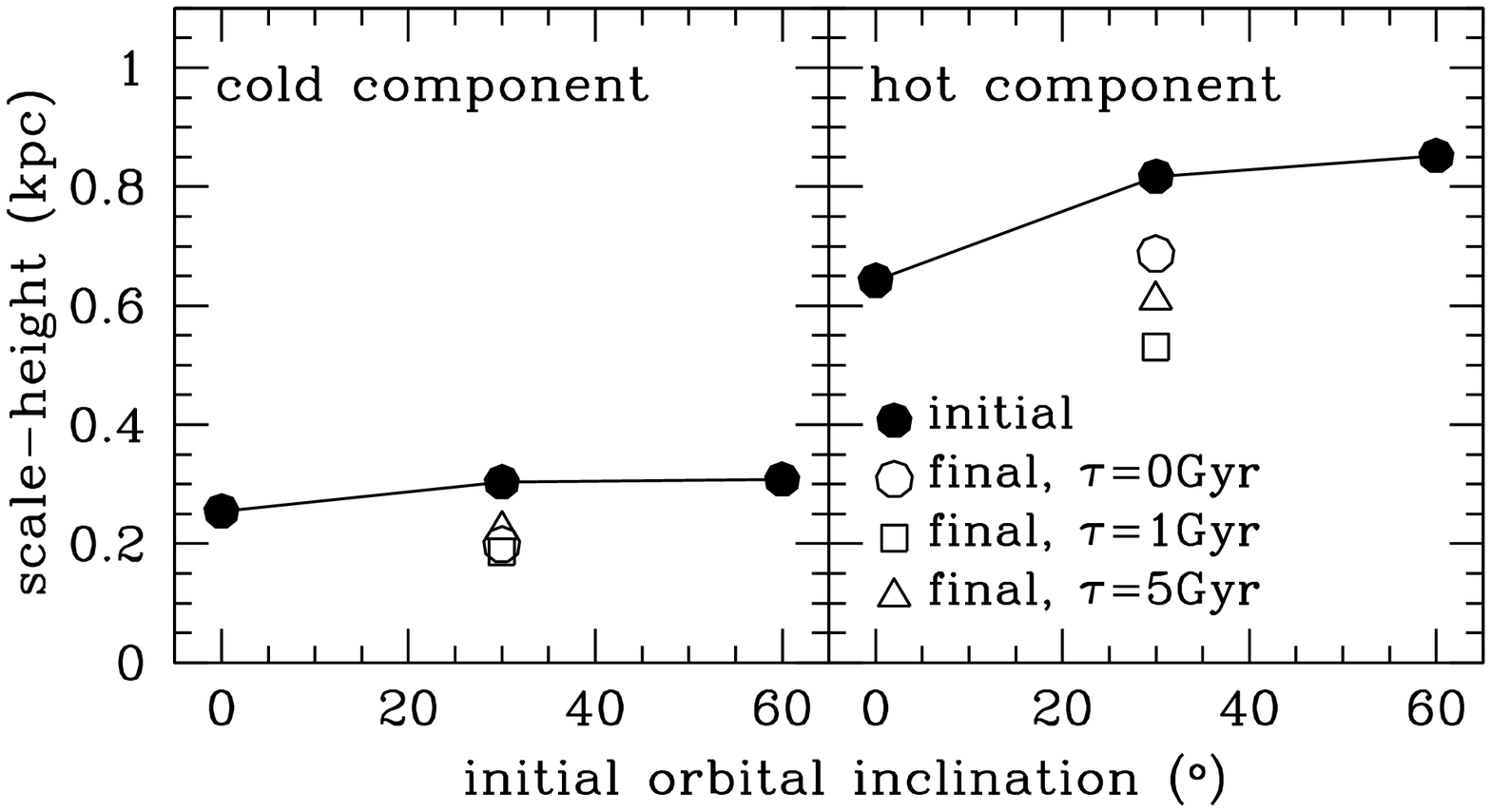}
    \caption{Experiment B. Final thick-disk structural properties for
      different growth timescales of the growing thin disk:
      $\tau=0$~Gyr (left panels), $\tau=1$~Gyr (middle panels), and
      $\tau=5$~Gyr (right panels). Rows 1 to 3 are the same as in
      Figure~\ref{struc-a}.}
    \label{struc-b}
  \end{center}
\end{figure*}
\begin{figure*}
  \begin{center}
    \includegraphics[width=52mm]{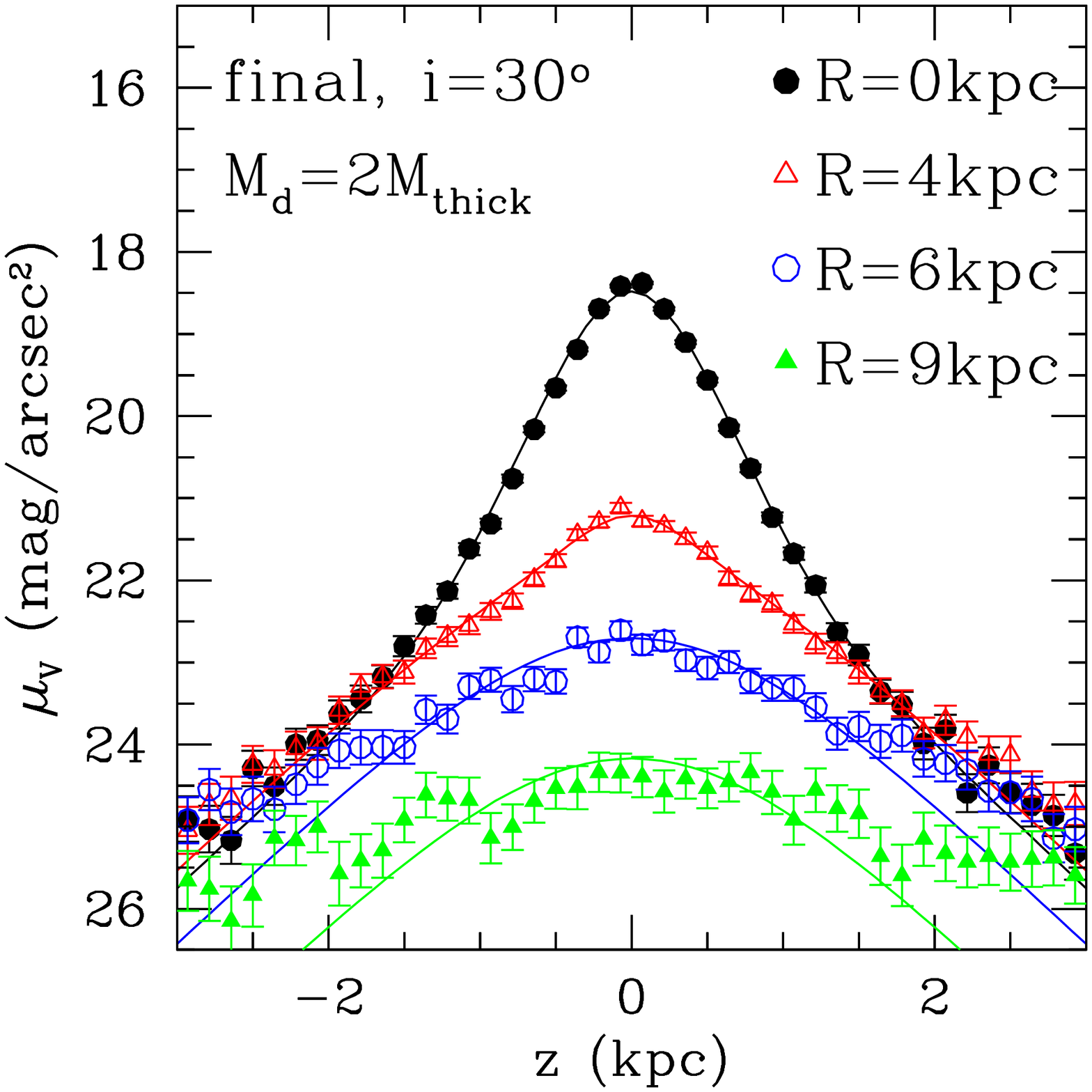}
    \hspace*{-1.2cm}
    \includegraphics[width=52mm]{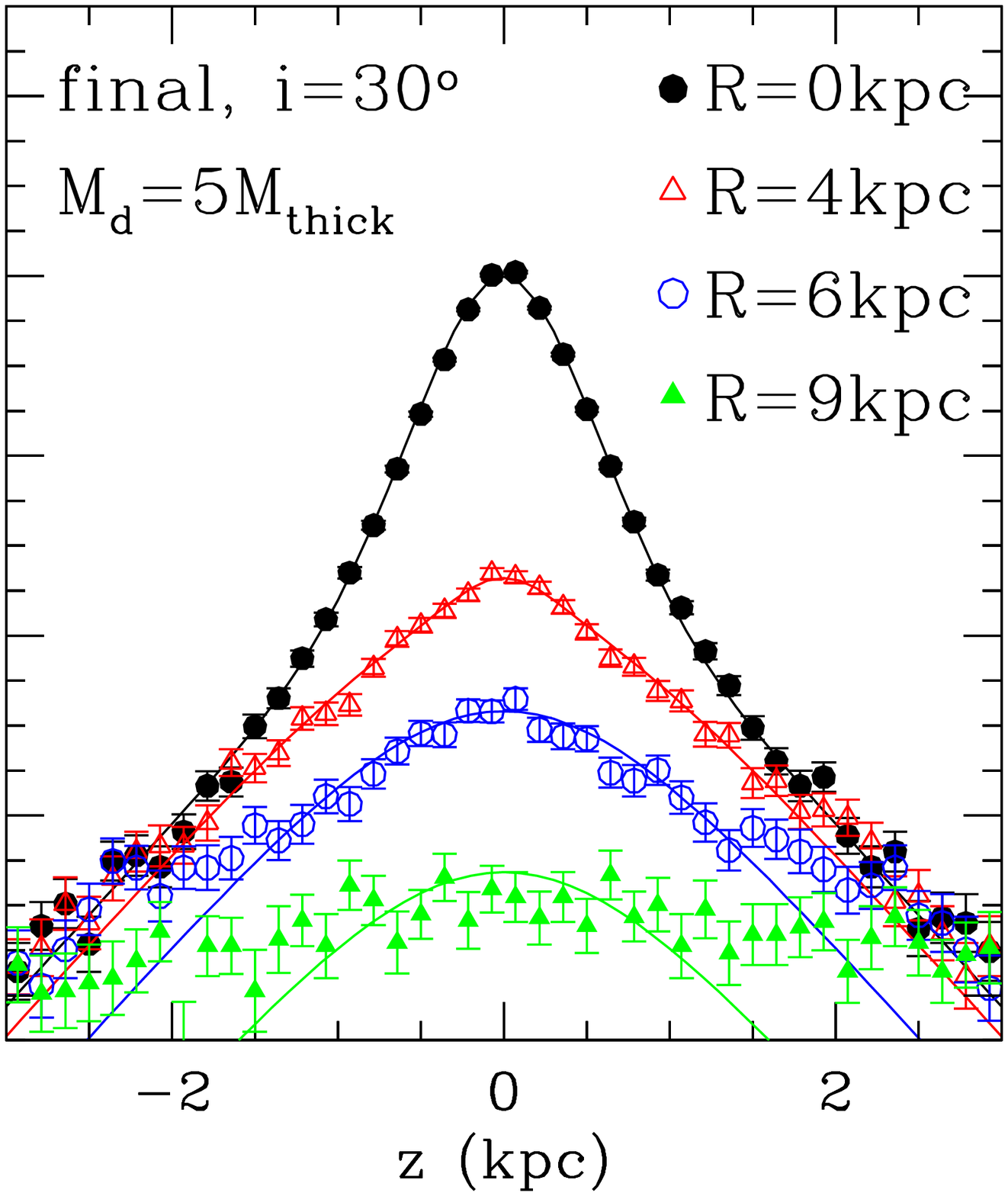}\\
    \includegraphics[width=52mm]{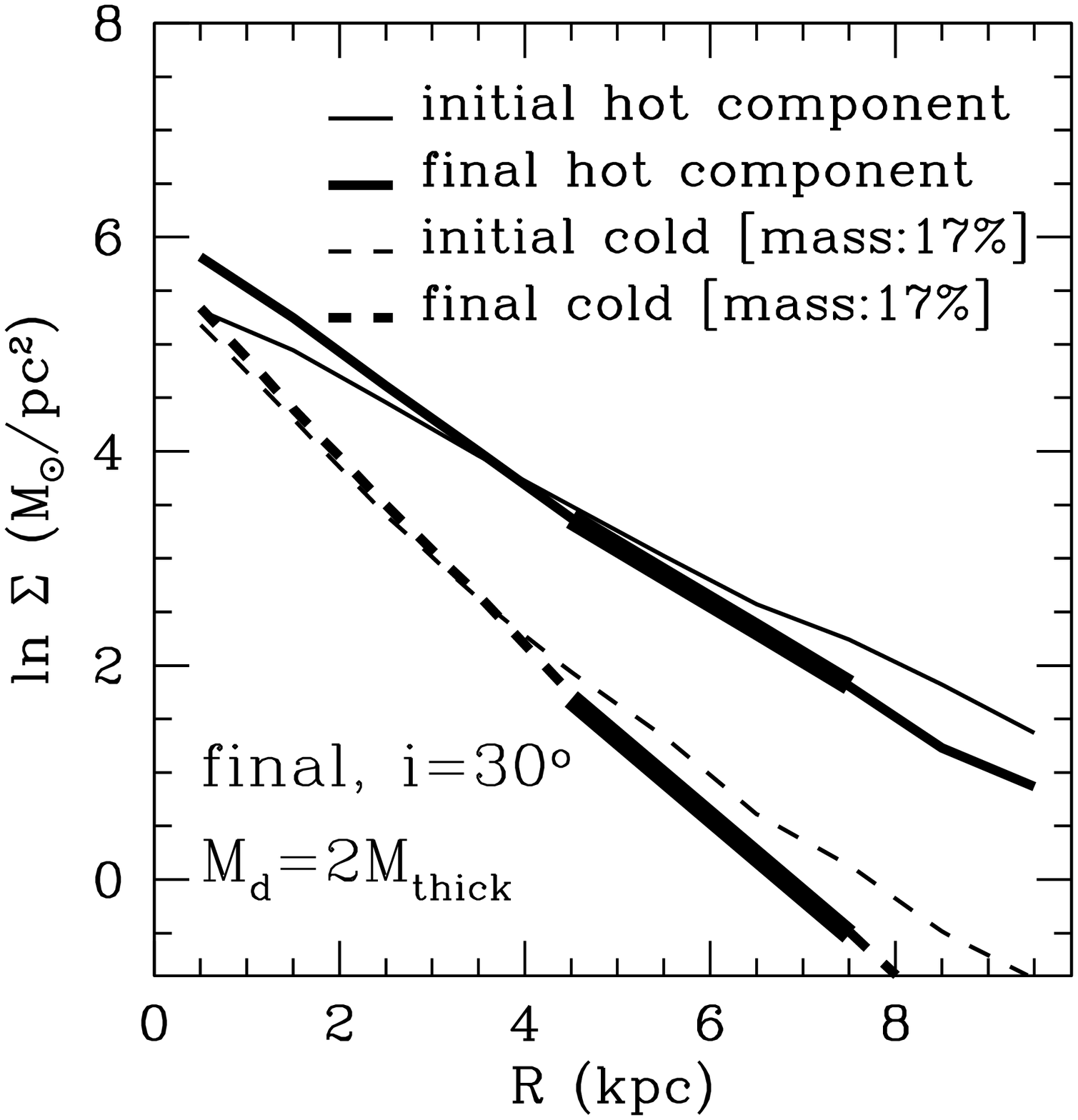}
    \hspace*{-1.2cm}
    \includegraphics[width=52mm]{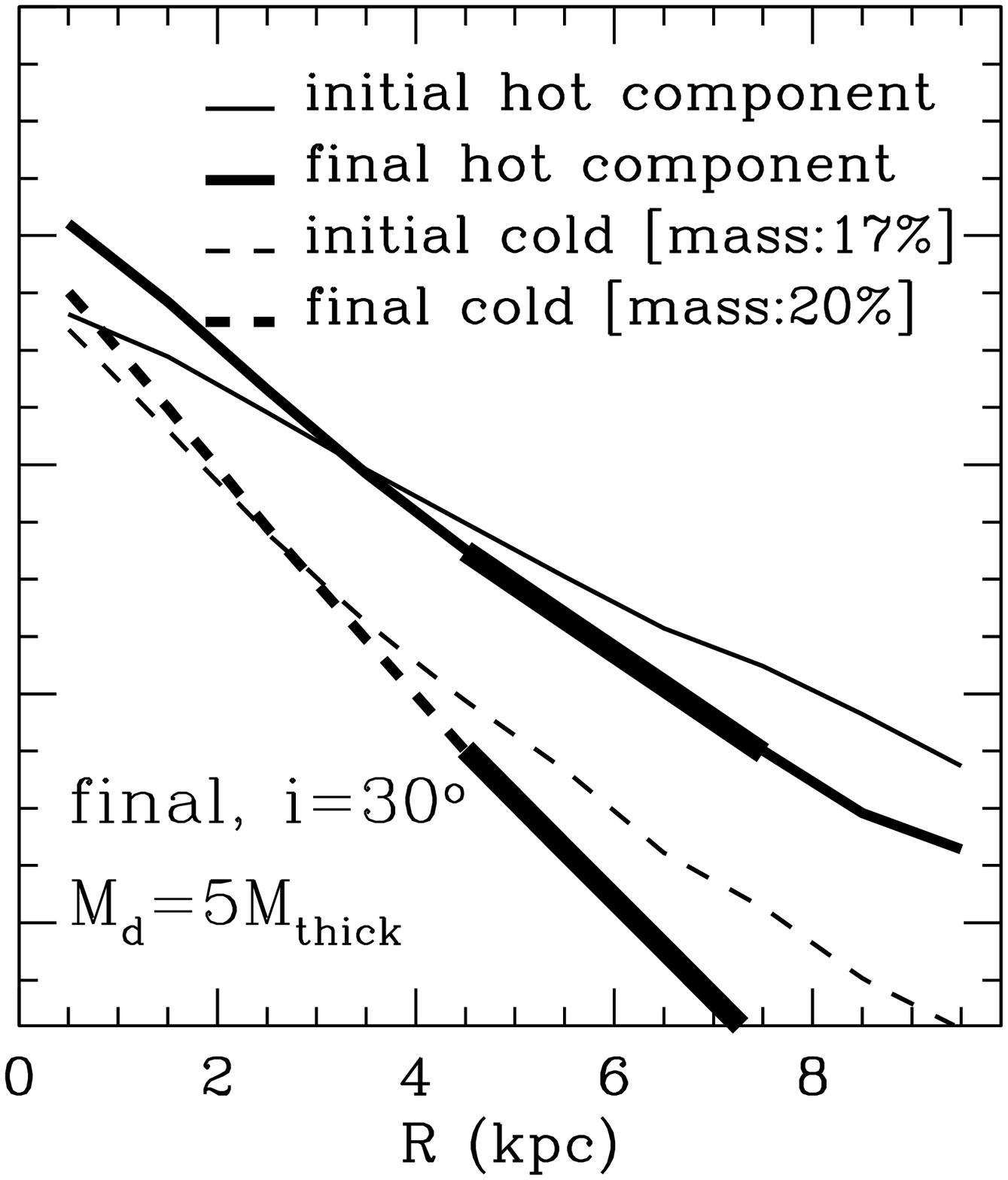}\\
    \includegraphics[width=72mm]{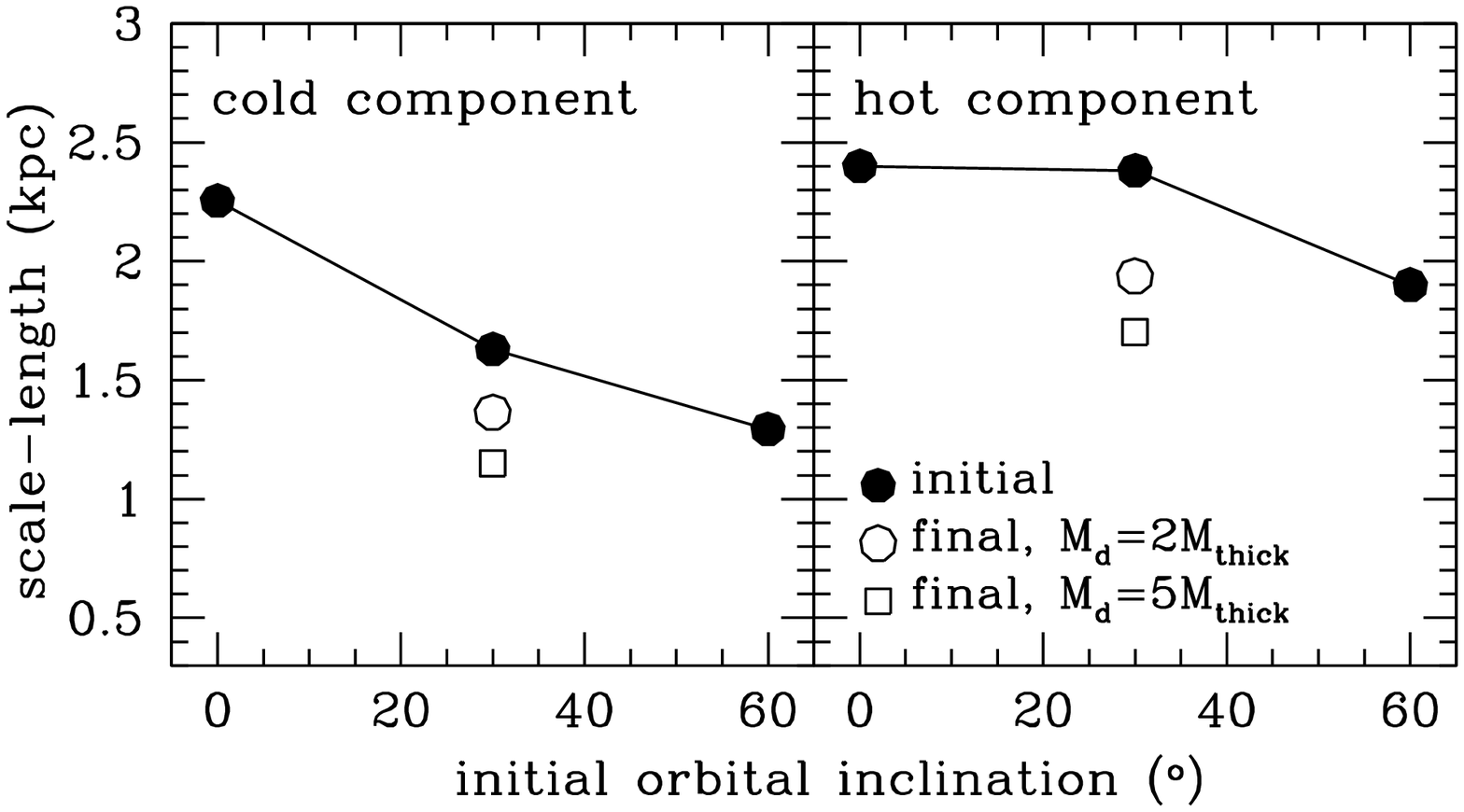}
    \includegraphics[width=72mm]{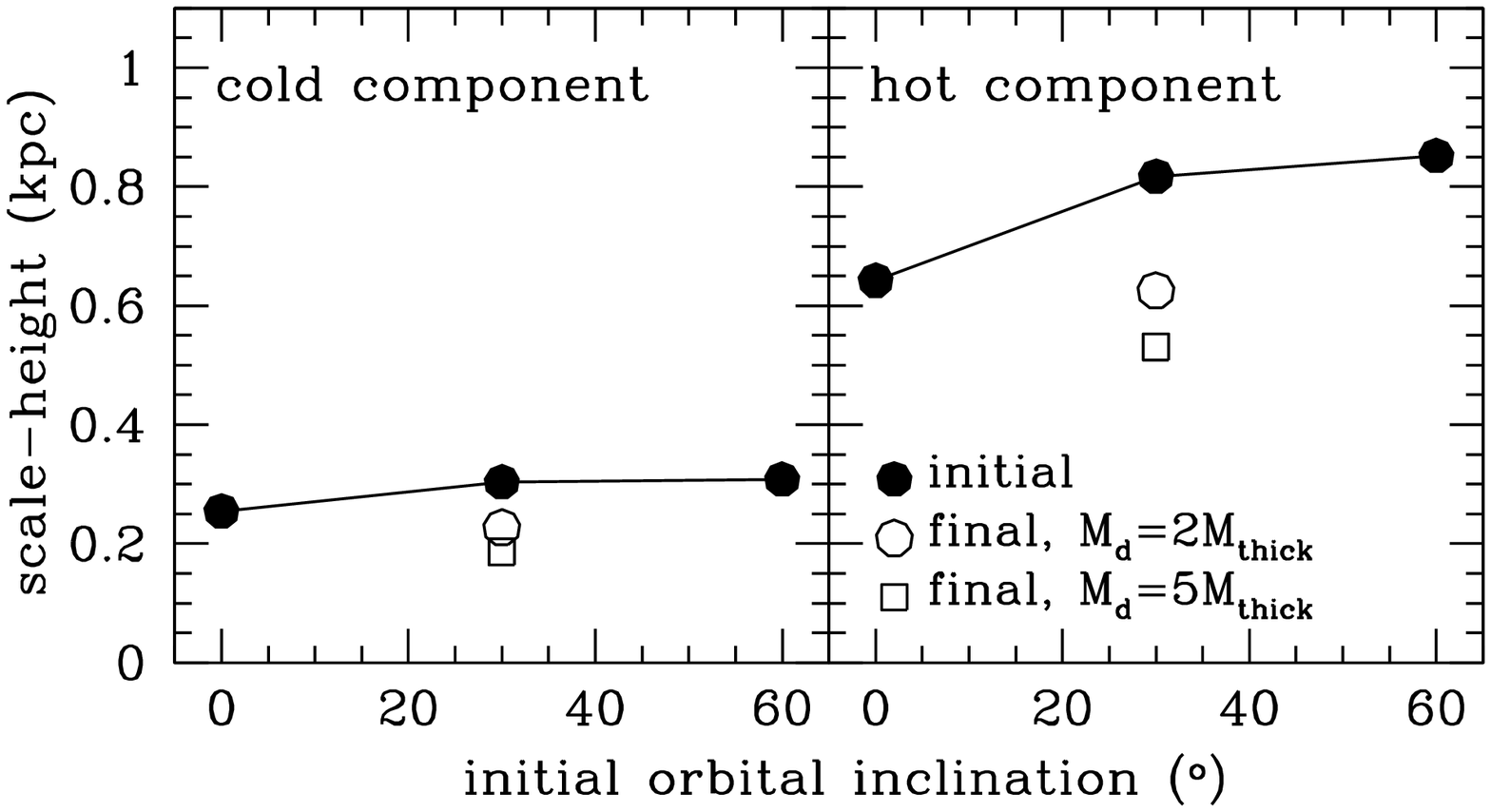}
    \caption{Experiment C1. Final thick-disk structural properties for
      different masses of the growing thin disk: $M_d=2 M_{\rm thick}$
      (left panels) and $M_d=5 M_{\rm thick}$ (right panels).  Rows 1
      to 3 are the same as in Figure~\ref{struc-a}.
    \label{struc-c1}}
  \end{center}
\end{figure*}
\begin{figure*}
  \begin{center}
    \includegraphics[width=52mm]{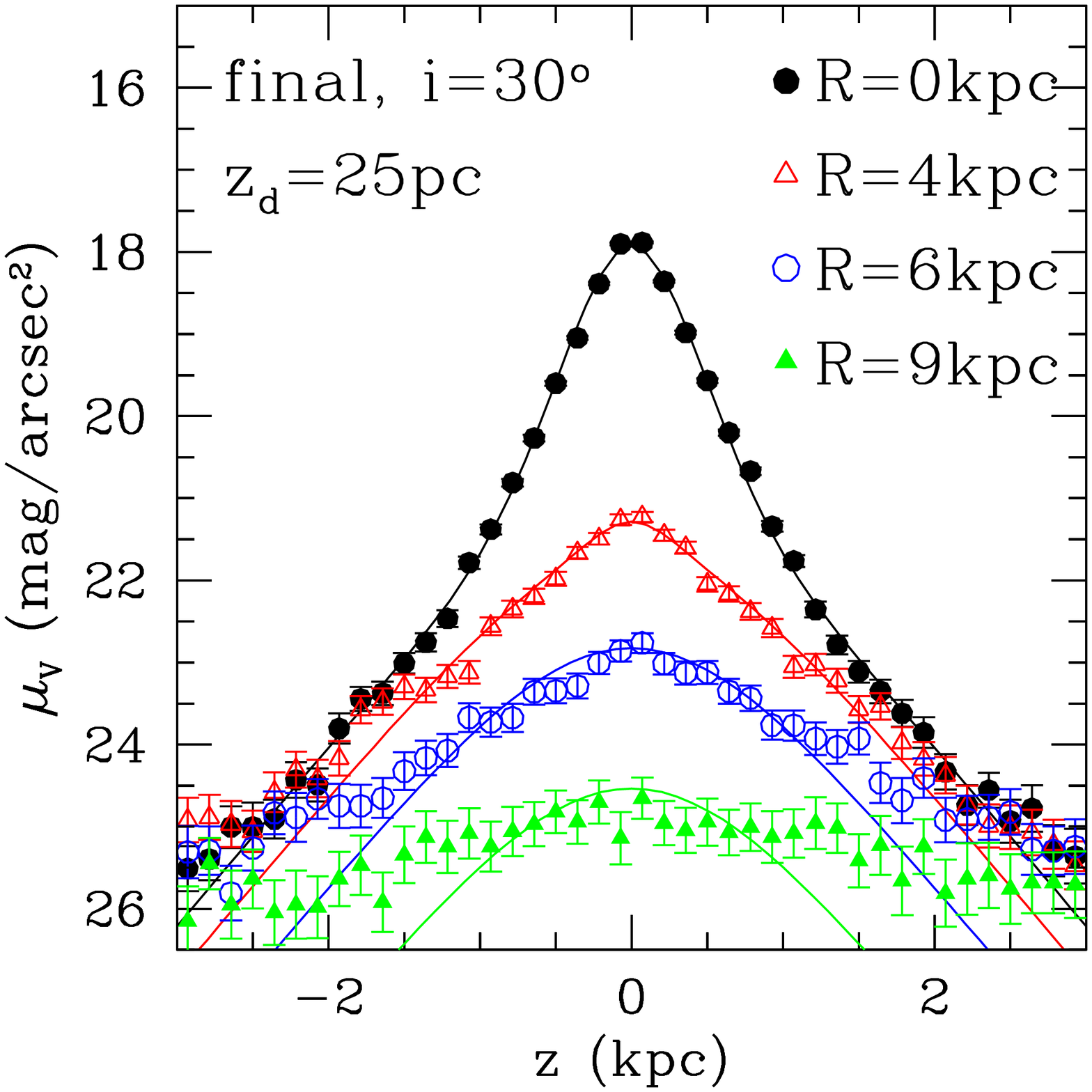}
    \hspace*{-1.2cm}
    \includegraphics[width=52mm]{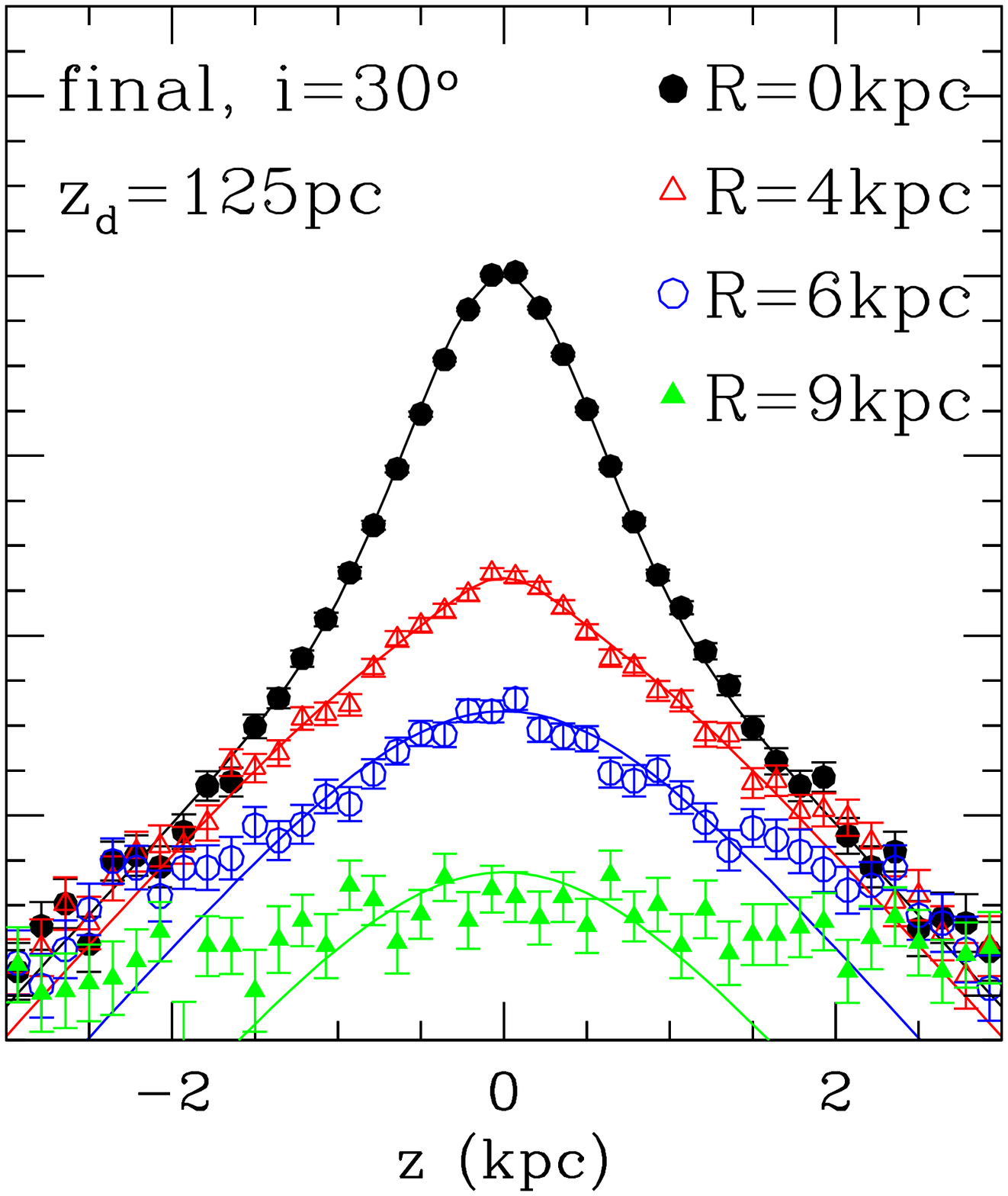}\\
    \includegraphics[width=52mm]{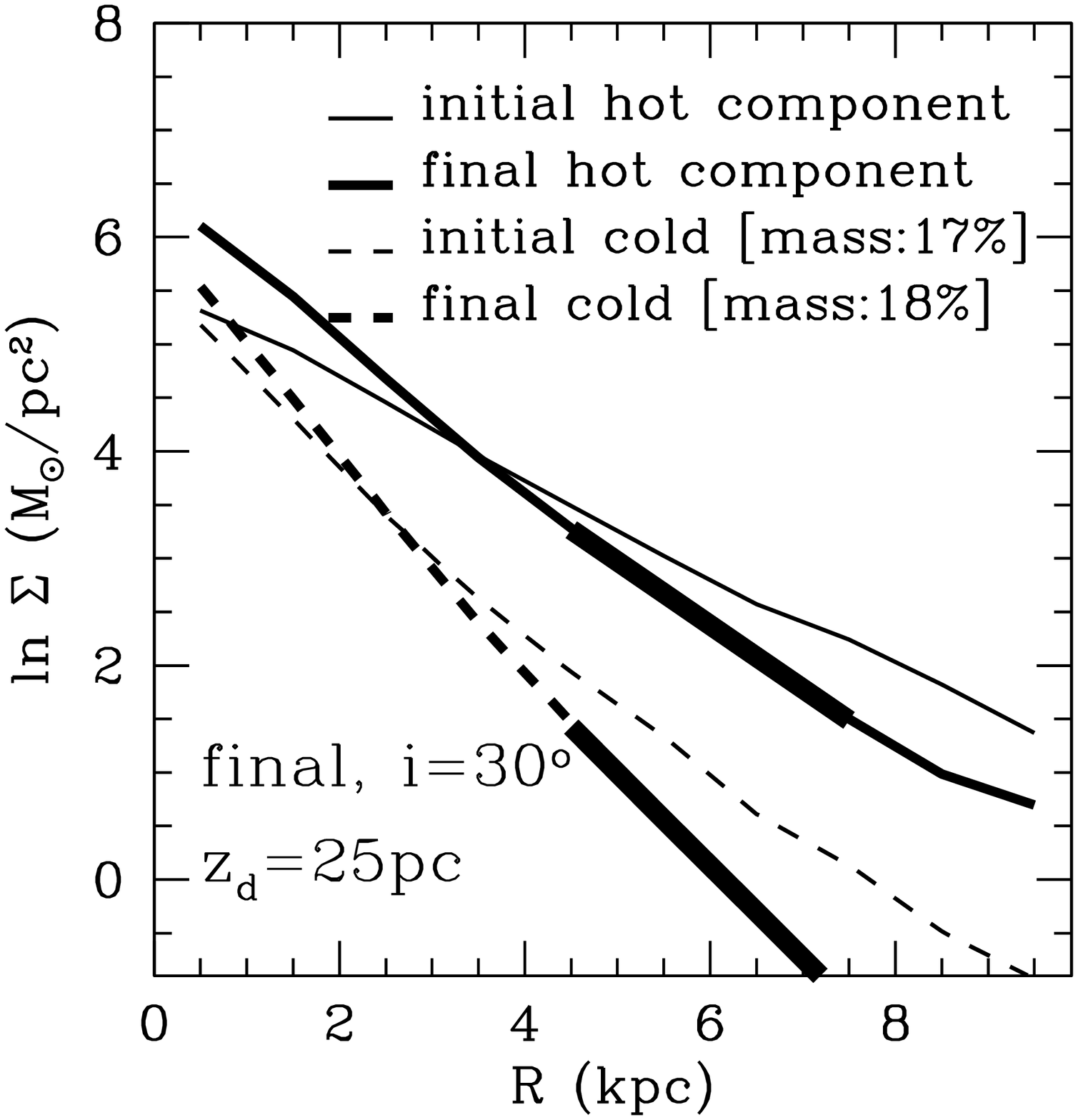}
    \hspace*{-1.2cm}
    \includegraphics[width=52mm]{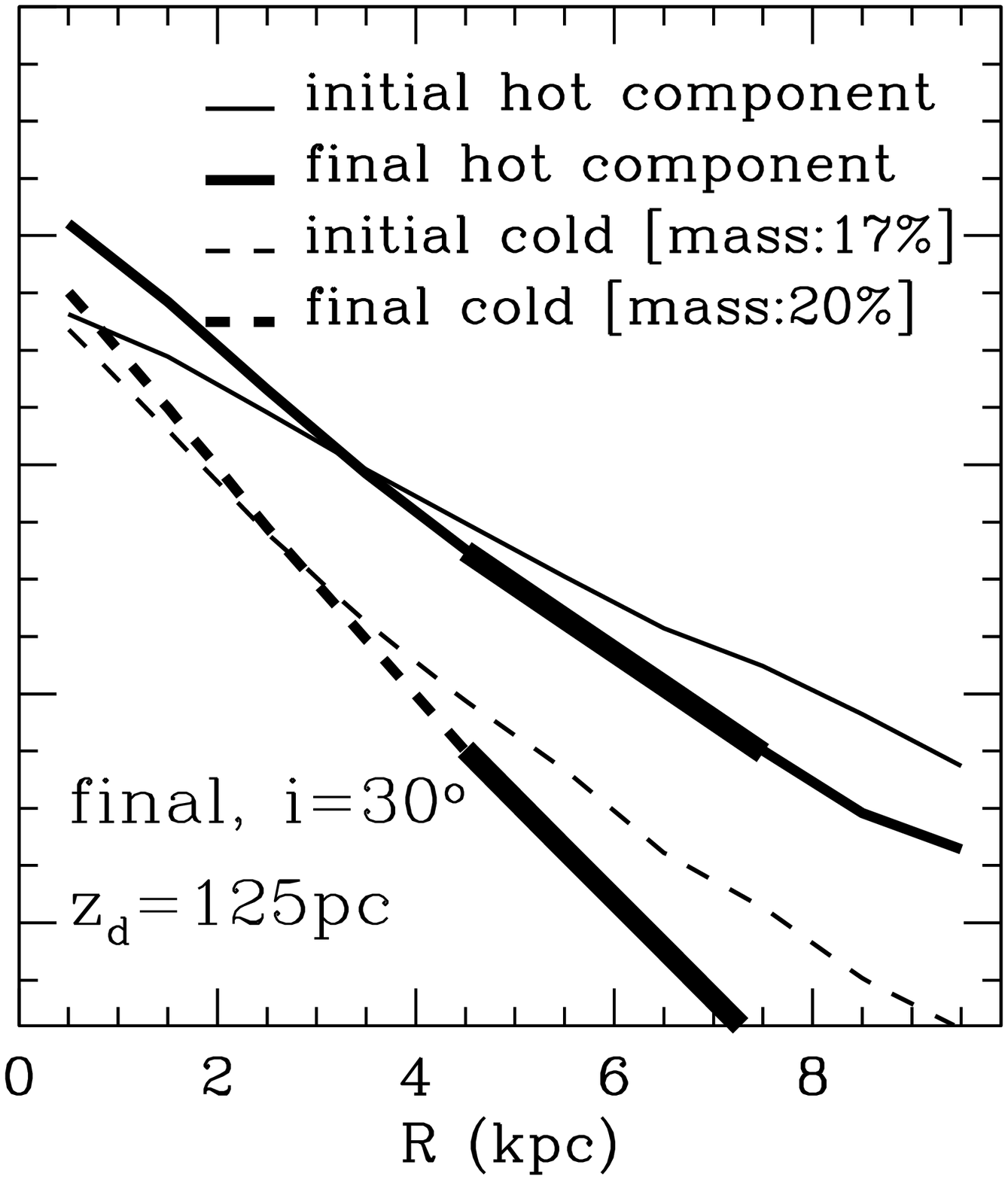}\\
    \includegraphics[width=72mm]{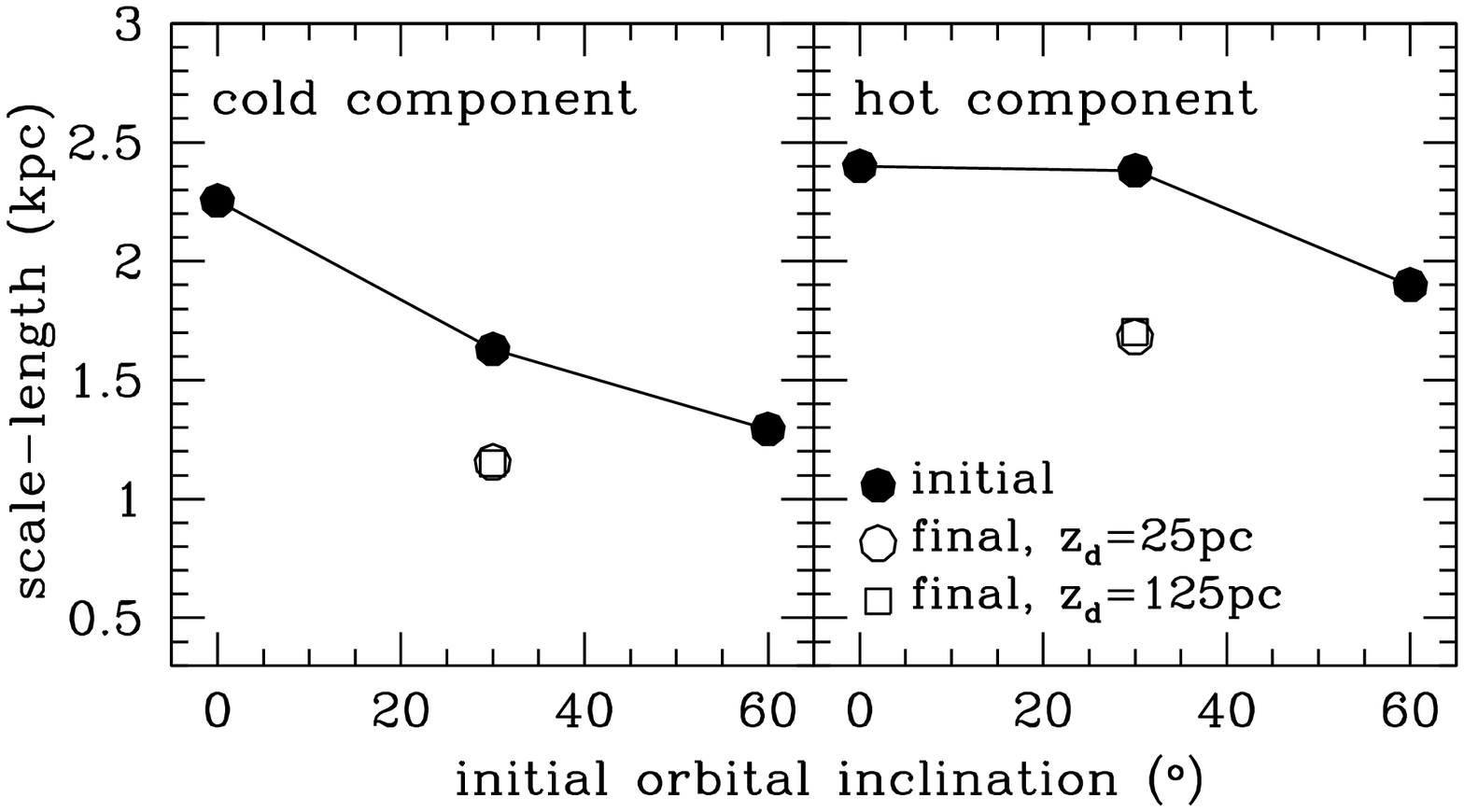}
    \includegraphics[width=72mm]{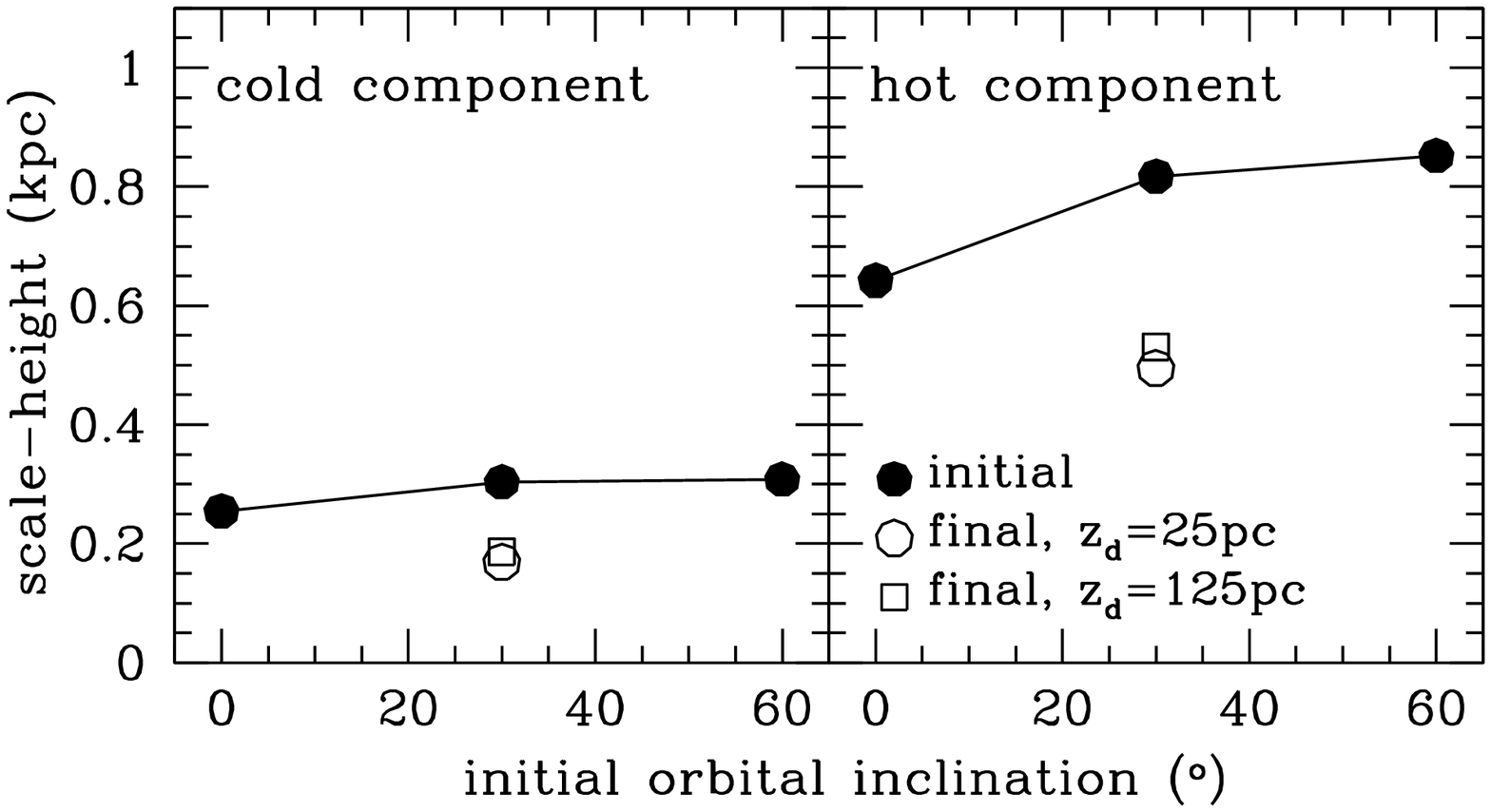}
    \caption{Experiment C2. Final thick-disk structural properties for
      different vertical scale-heights of the growing thin disk:
      $z_d=25$~pc (left panels) and $z_d=125$~pc (right panels). Rows
      1 to 3 as in Figure~\ref{struc-a}.}
    \label{struc-c2}
  \end{center}
\end{figure*}
\begin{figure*}
  \begin{center}
    \includegraphics[width=52mm]{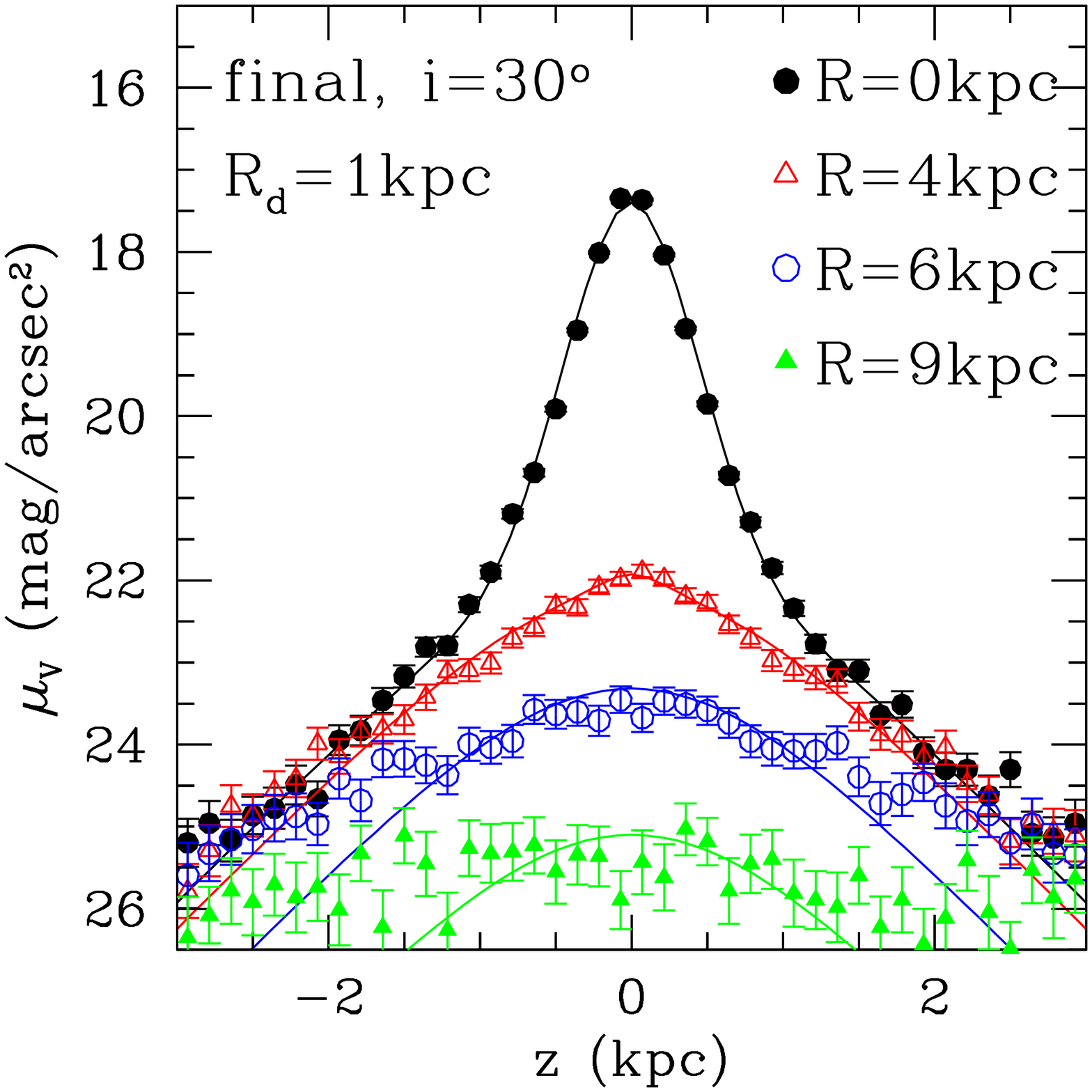}
    \hspace*{-1.2cm}
    \includegraphics[width=52mm]{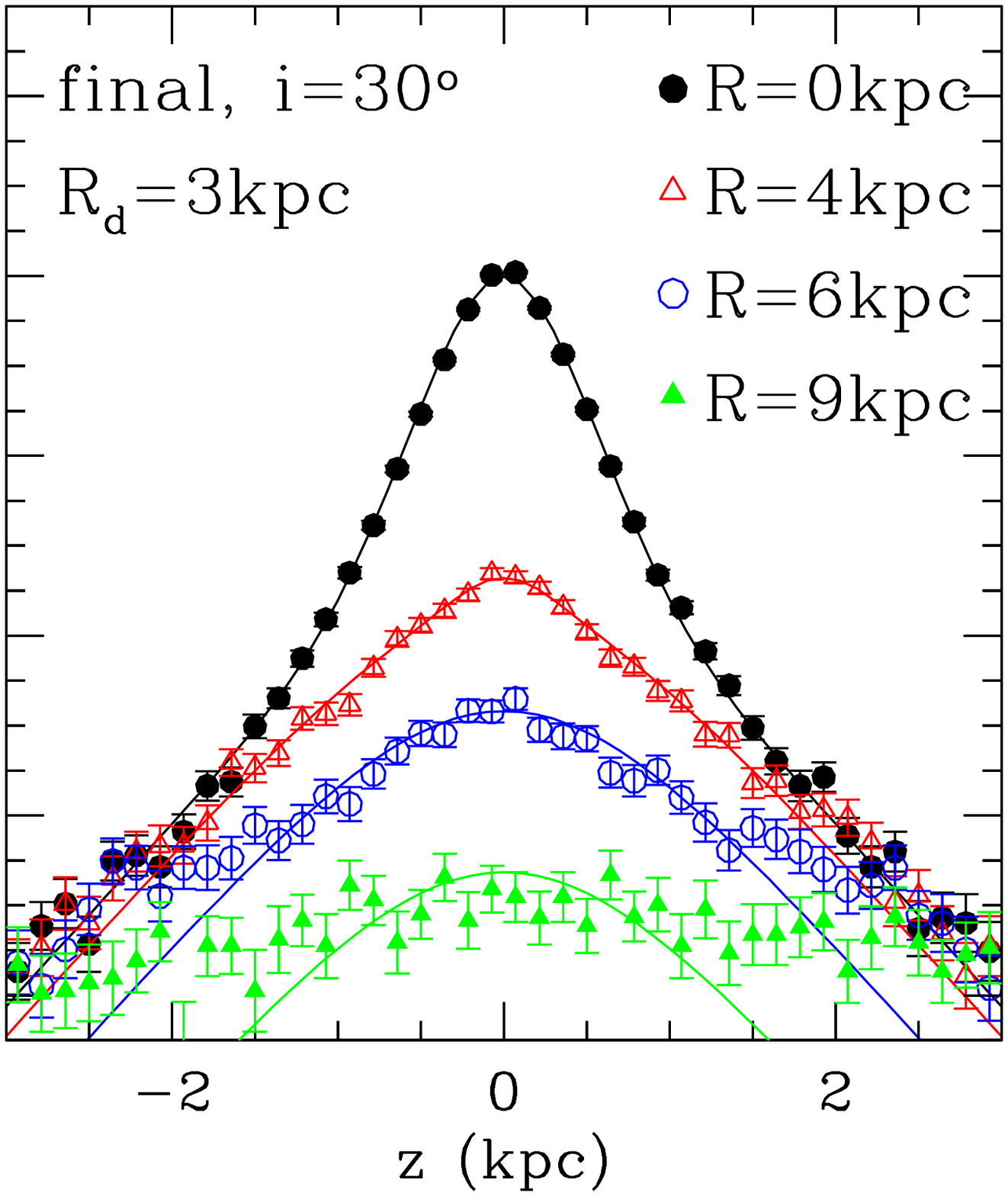}
    \hspace*{-0.8cm}
    \includegraphics[width=52mm]{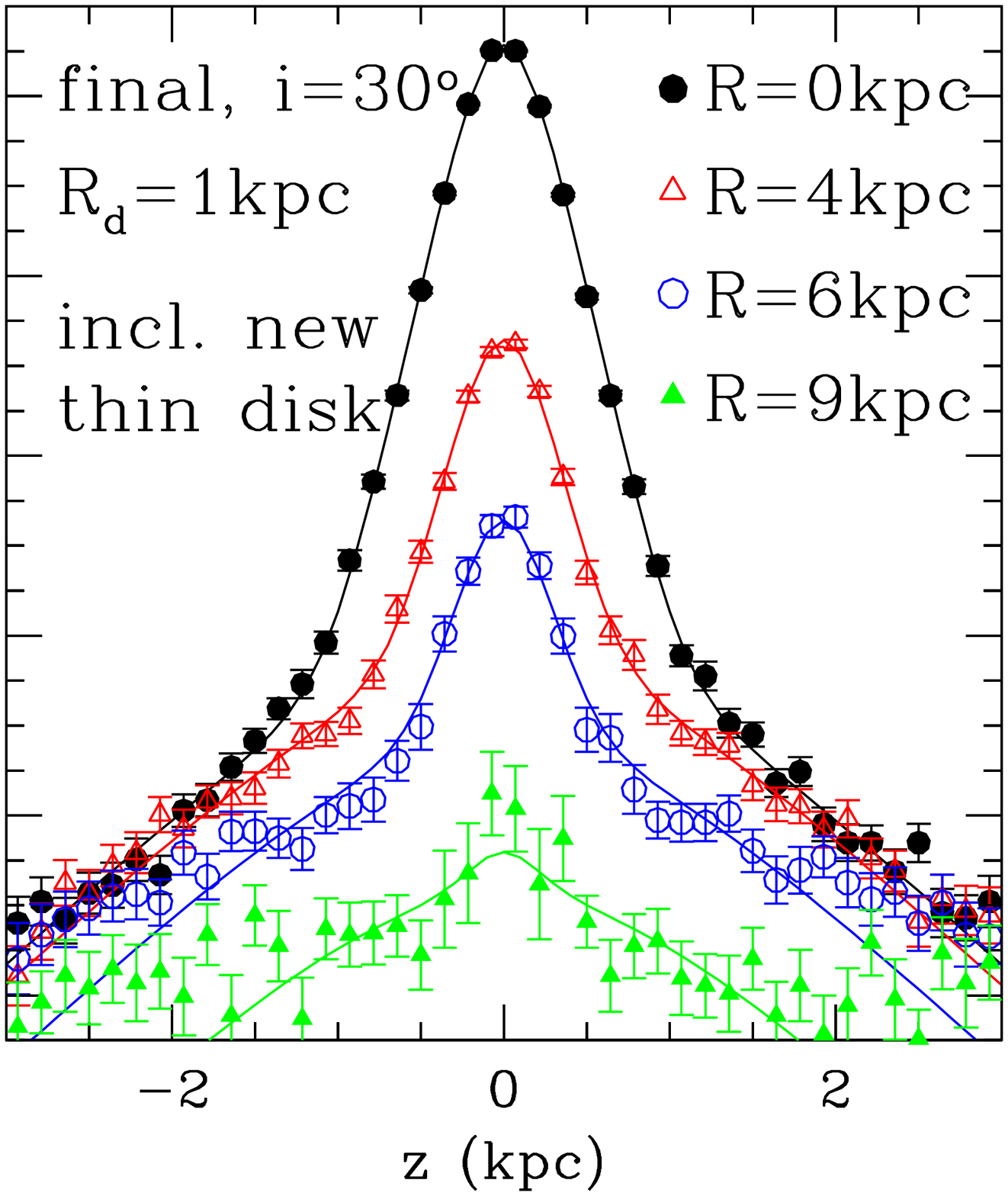}\\
    \includegraphics[width=52mm]{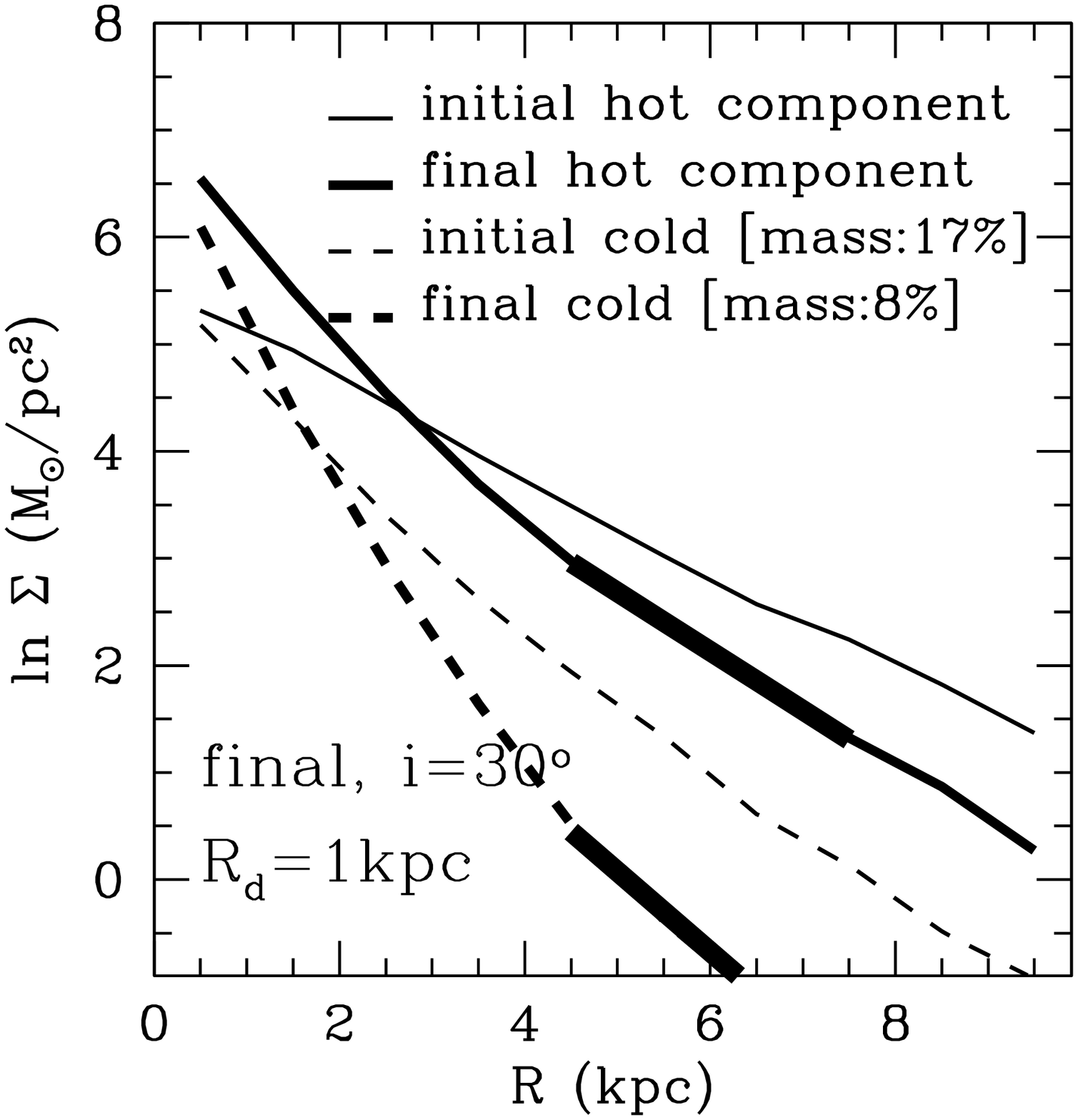}
    \hspace*{-1.2cm}
    \includegraphics[width=52mm]{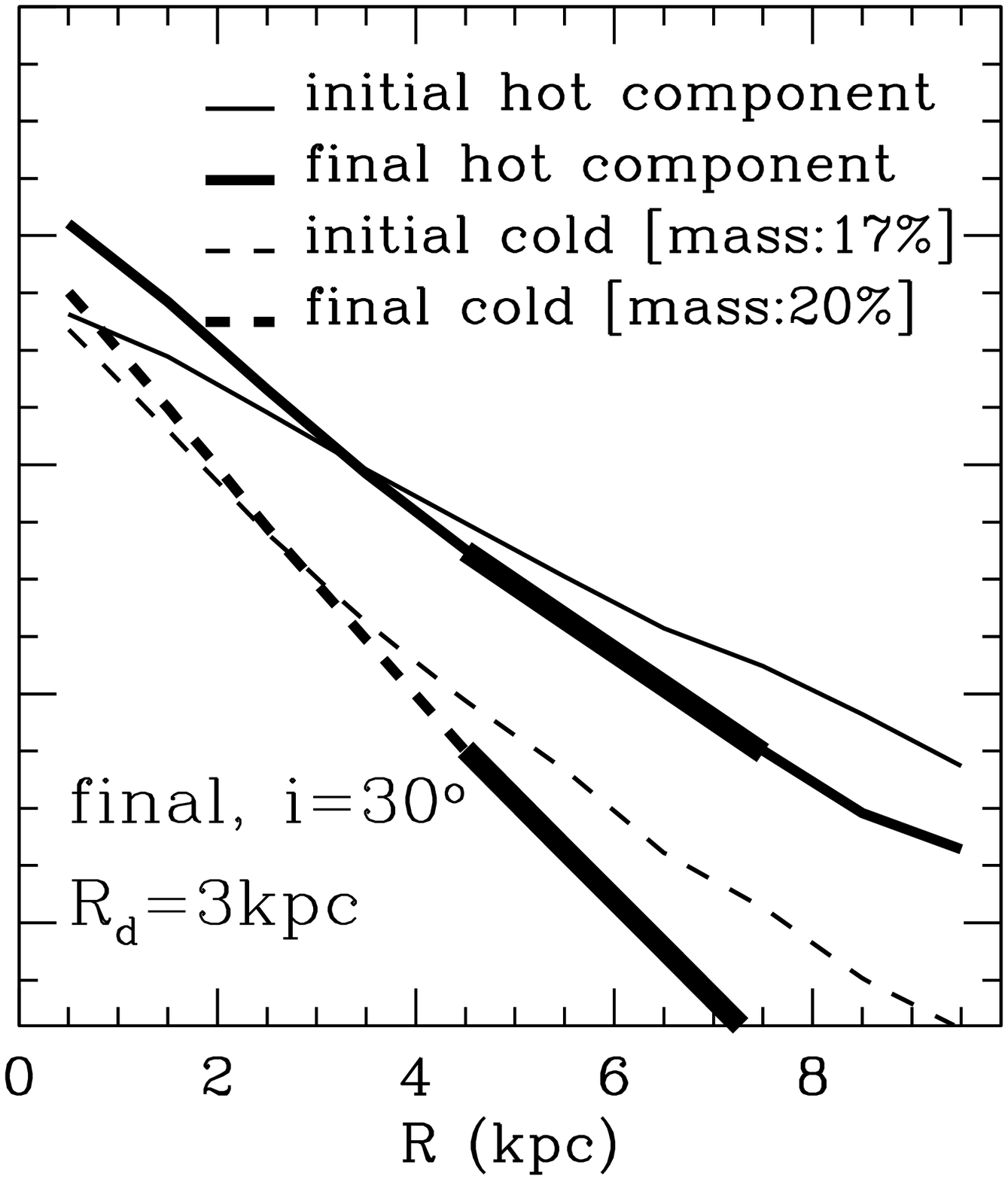}\\
    \includegraphics[width=72mm]{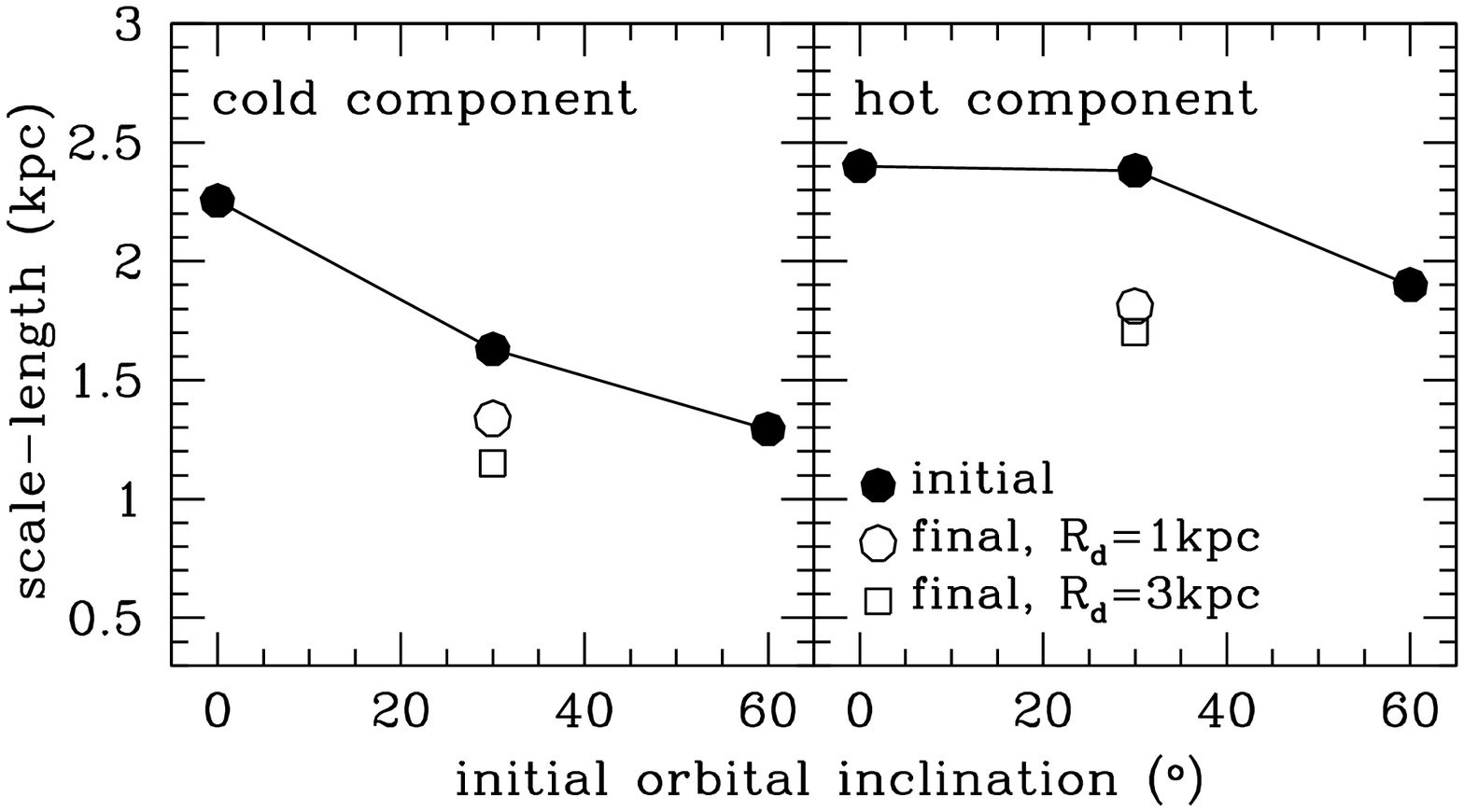}
    \includegraphics[width=72mm]{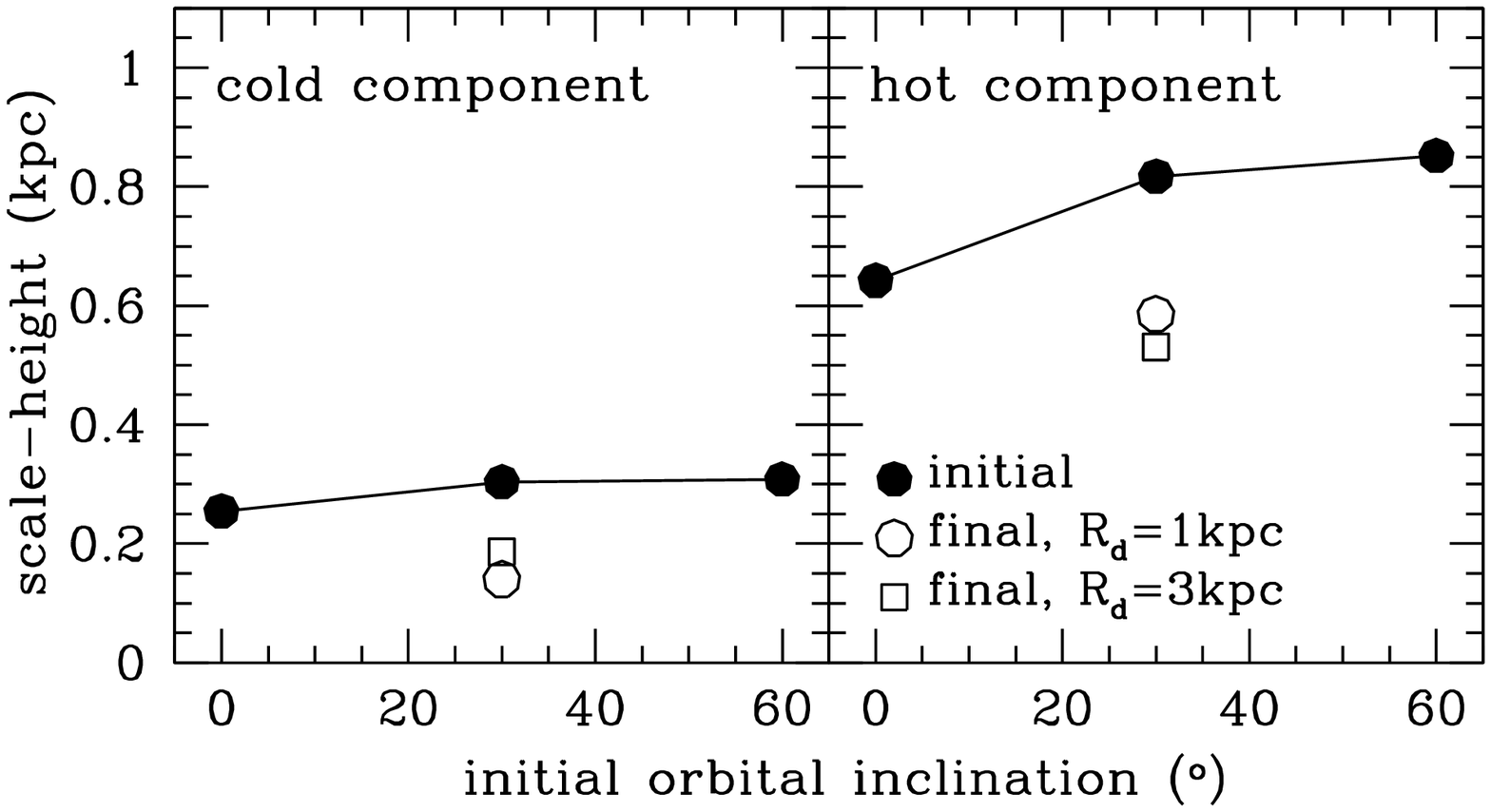}
    \caption{Experiment C3. Final thick-disk structural properties for
      different radial scale-lengths of the growing thin disk:
      $R_d=1$~kpc (left panels) and $R_d=3$~kpc (right panels).  Rows
      1 to 3 as in Figure~\ref{struc-a}. The top right panel shows the
      results when the growing thin disk is included in the
      decomposition analysis.}
    \label{struc-c3}
  \end{center}
\end{figure*}
\begin{figure*}
  \begin{center}
    \includegraphics[width=52mm]{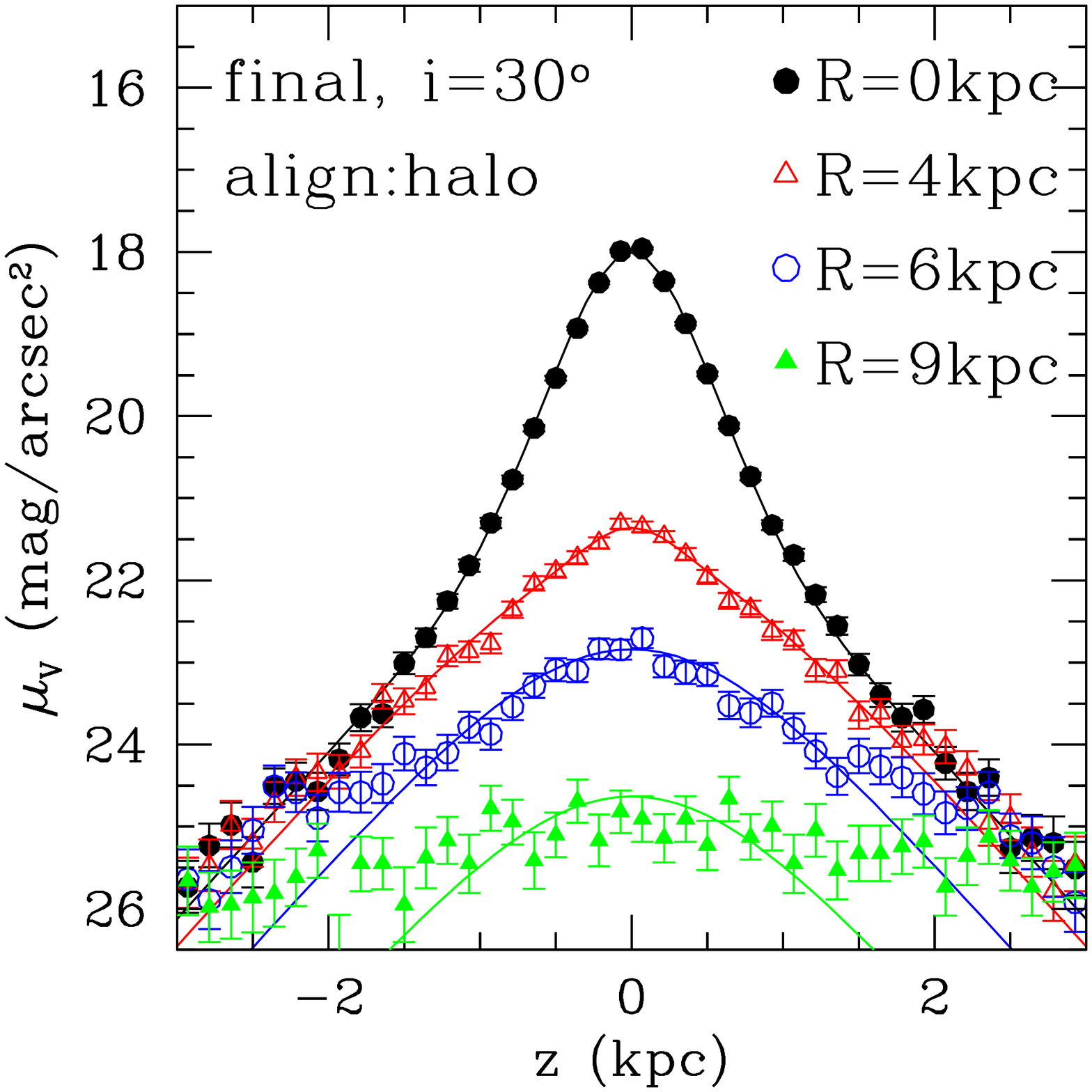}
    \hspace*{-1.2cm}
    \includegraphics[width=52mm]{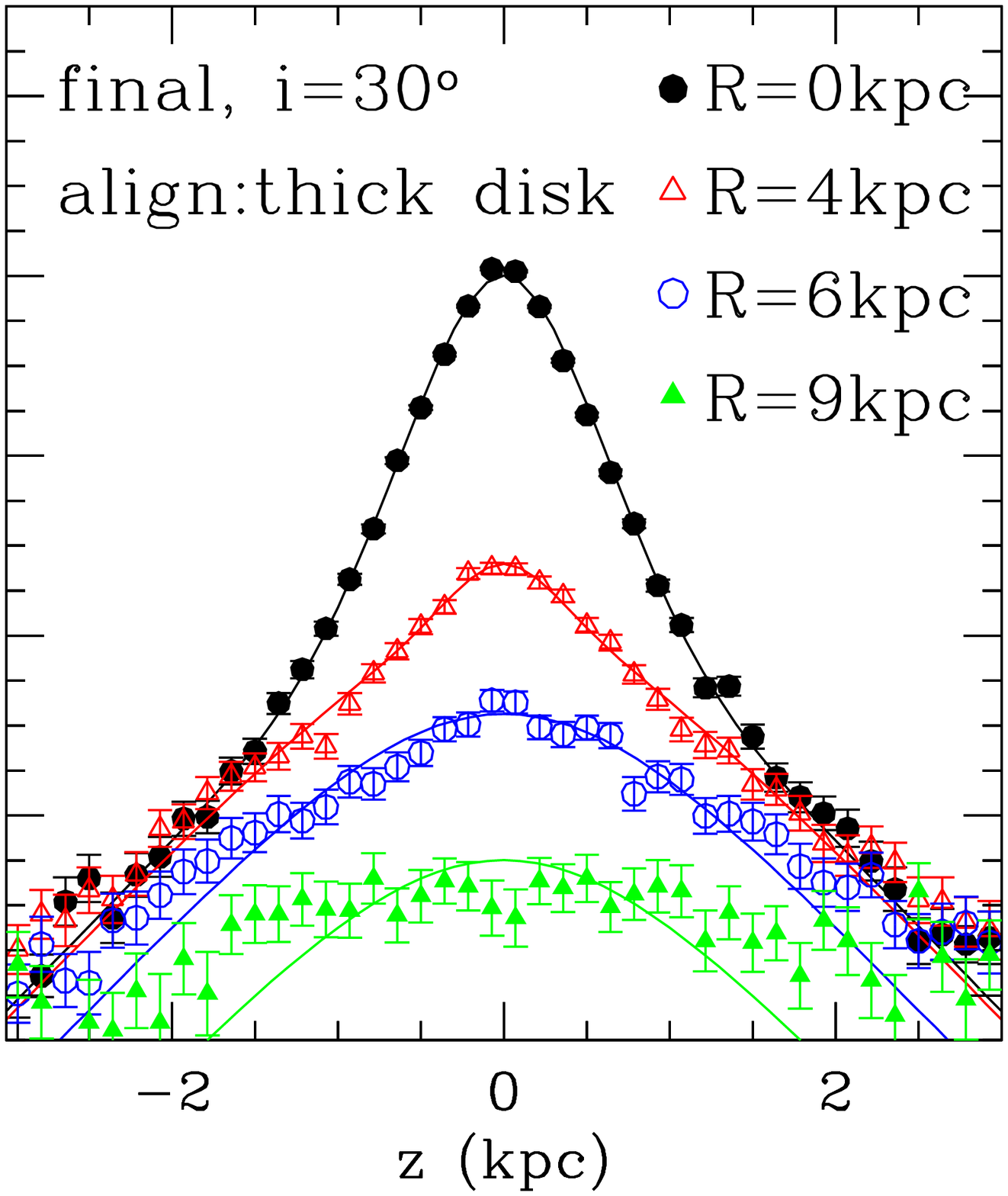}\\
    \includegraphics[width=52mm]{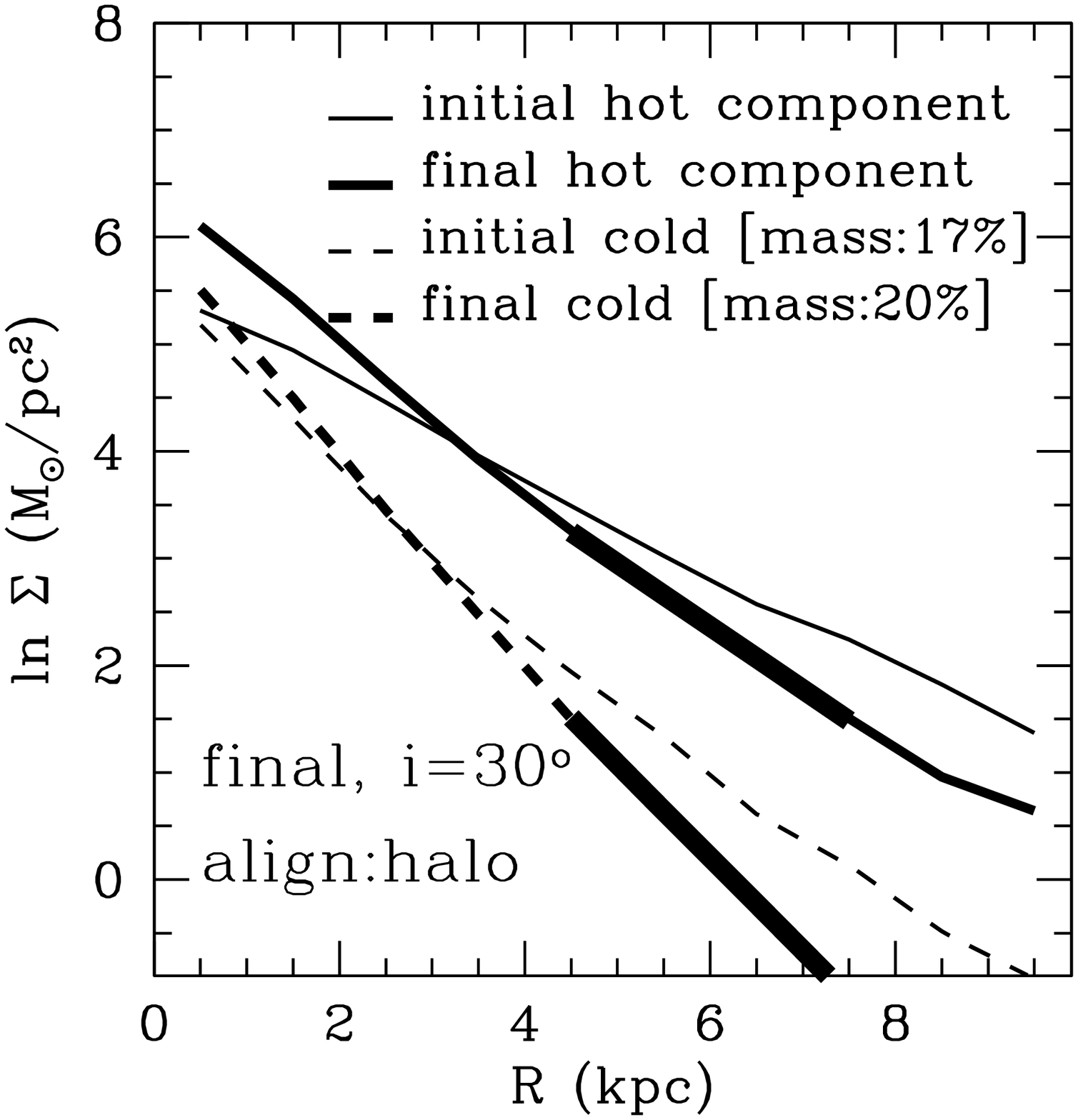}
    \hspace*{-1.2cm}
    \includegraphics[width=52mm]{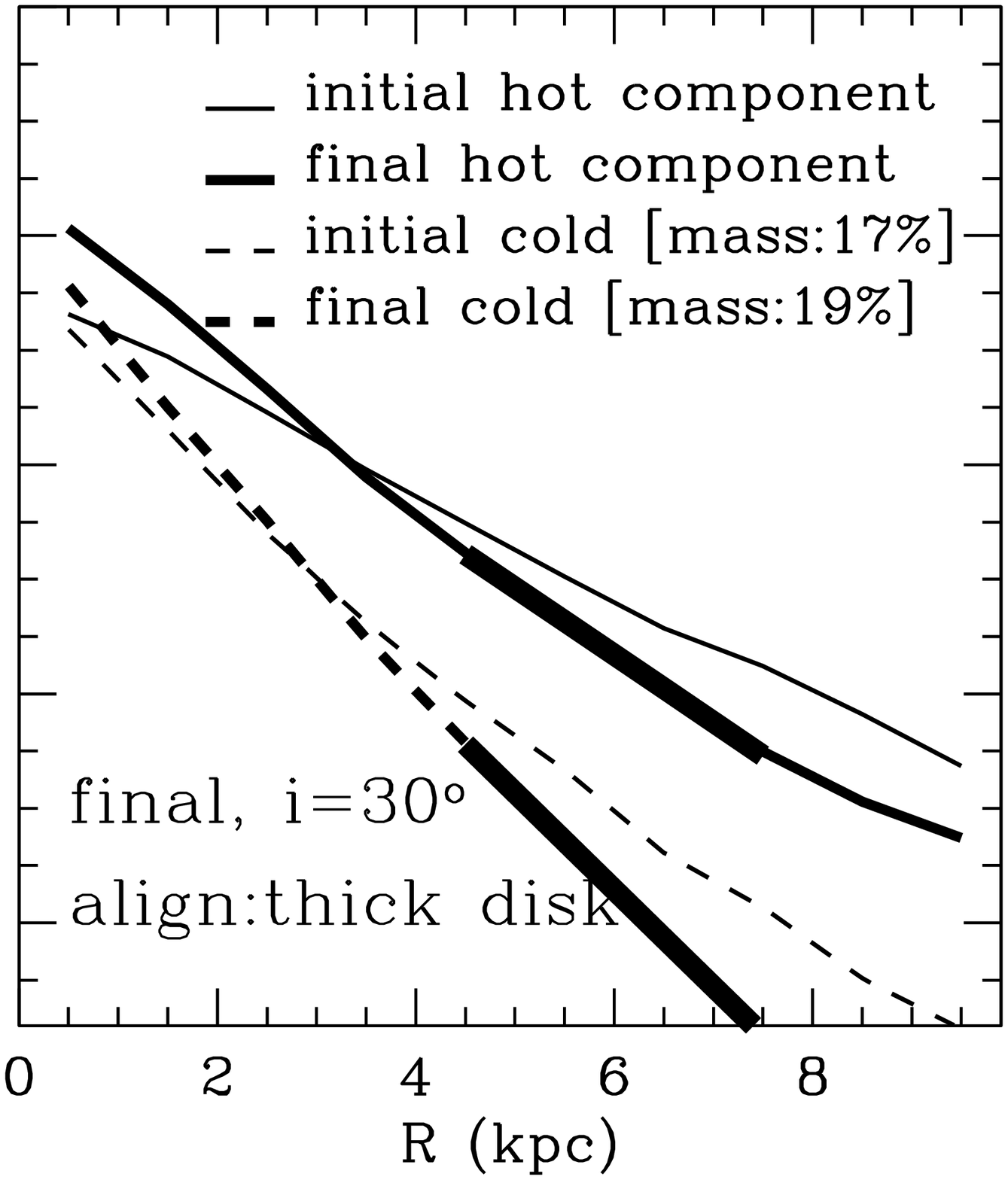}\\
    \includegraphics[width=72mm]{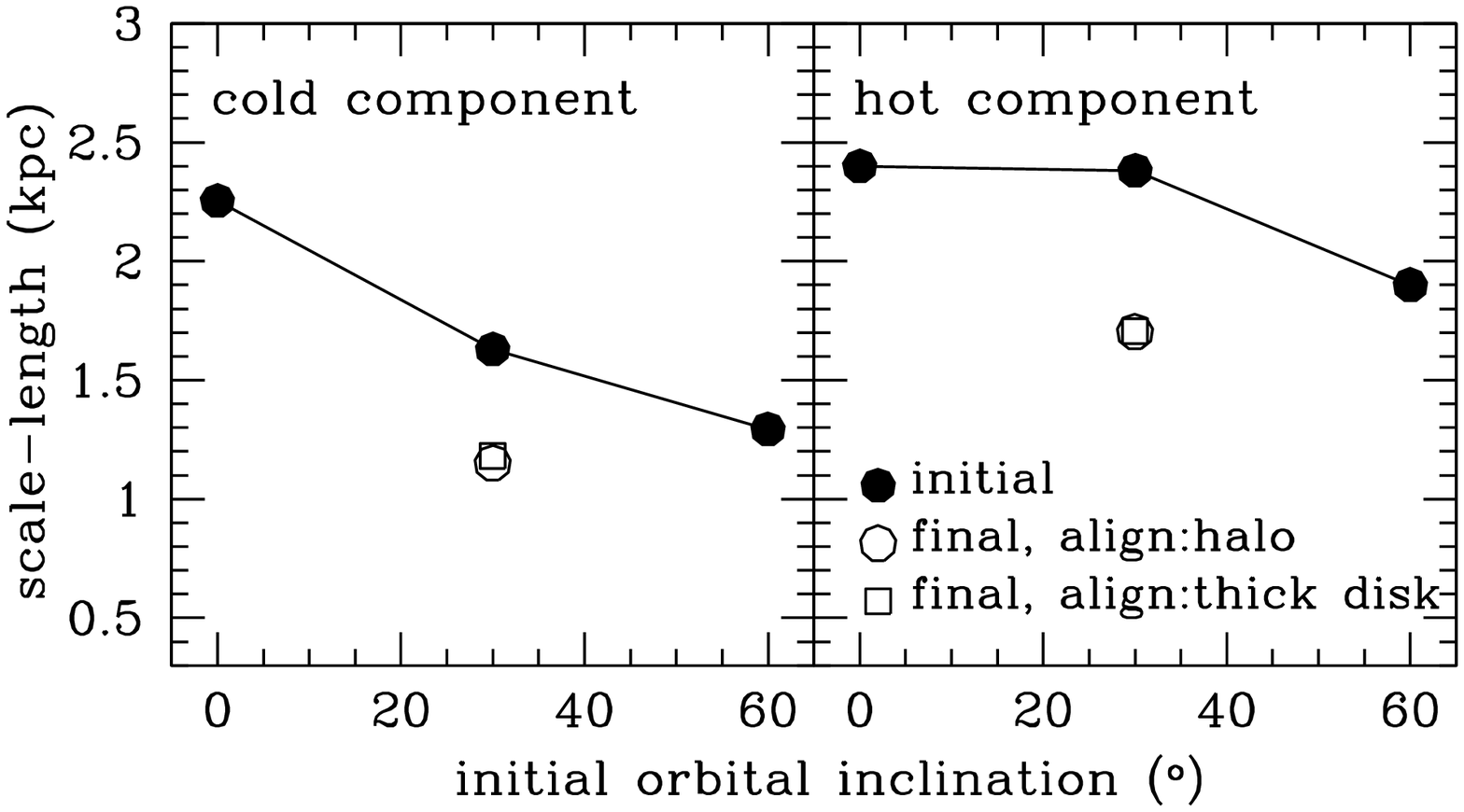}
    \includegraphics[width=72mm]{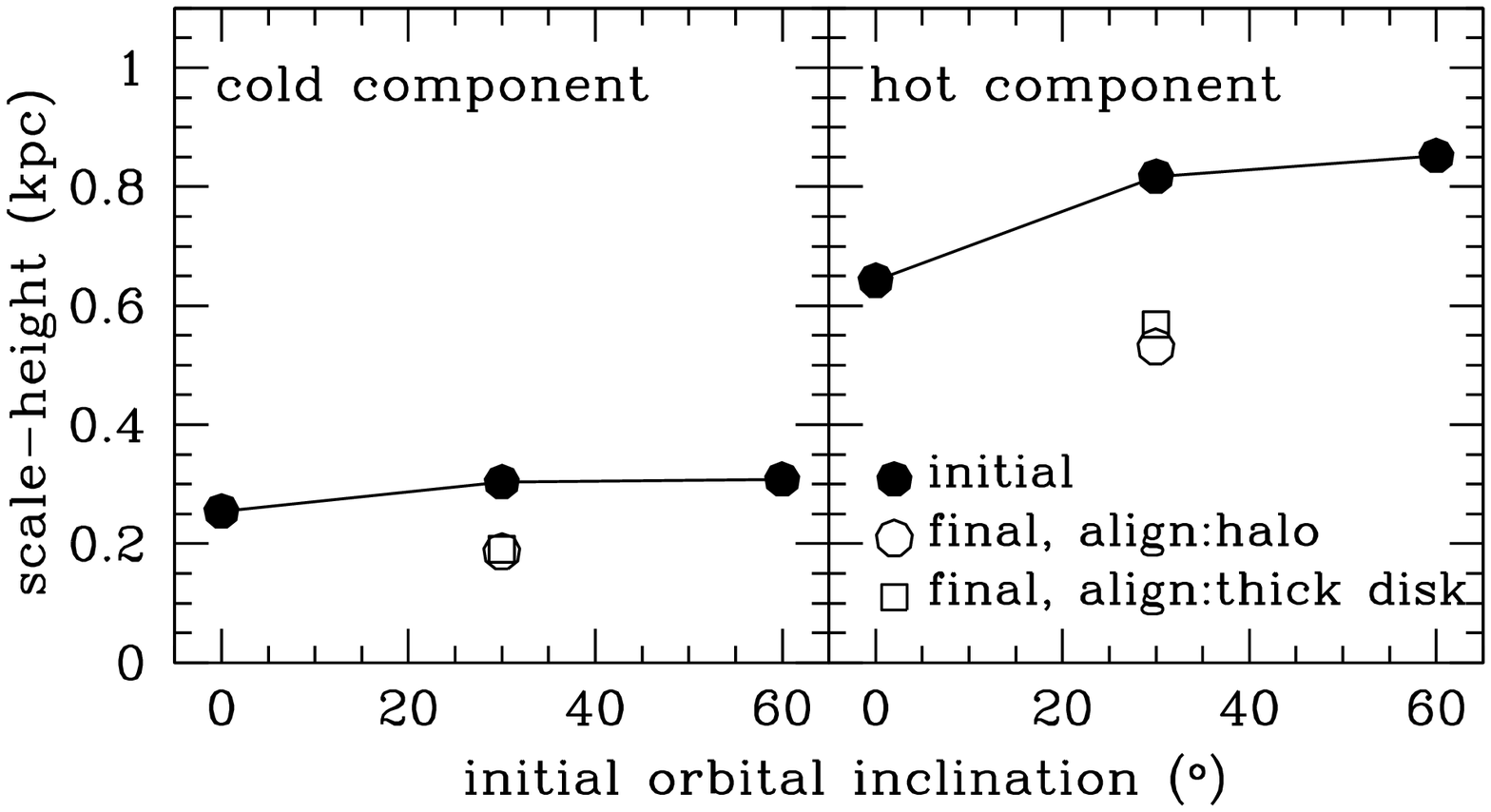}
    \caption{Experiment D. Final thick-disk structural properties for
      different alignments of the total angular momentum vector of the
      growing thin disk: either along the angular momentum of the dark
      matter halo (left panels) or along the angular momentum vector
      of the thick disk (right panels).  Rows 1 to 3 as in
      Figure~\ref{struc-a}.}
    \label{struc-d}
  \end{center}
\end{figure*}
\begin{figure*}
  \begin{center}
    \includegraphics[width=52mm]{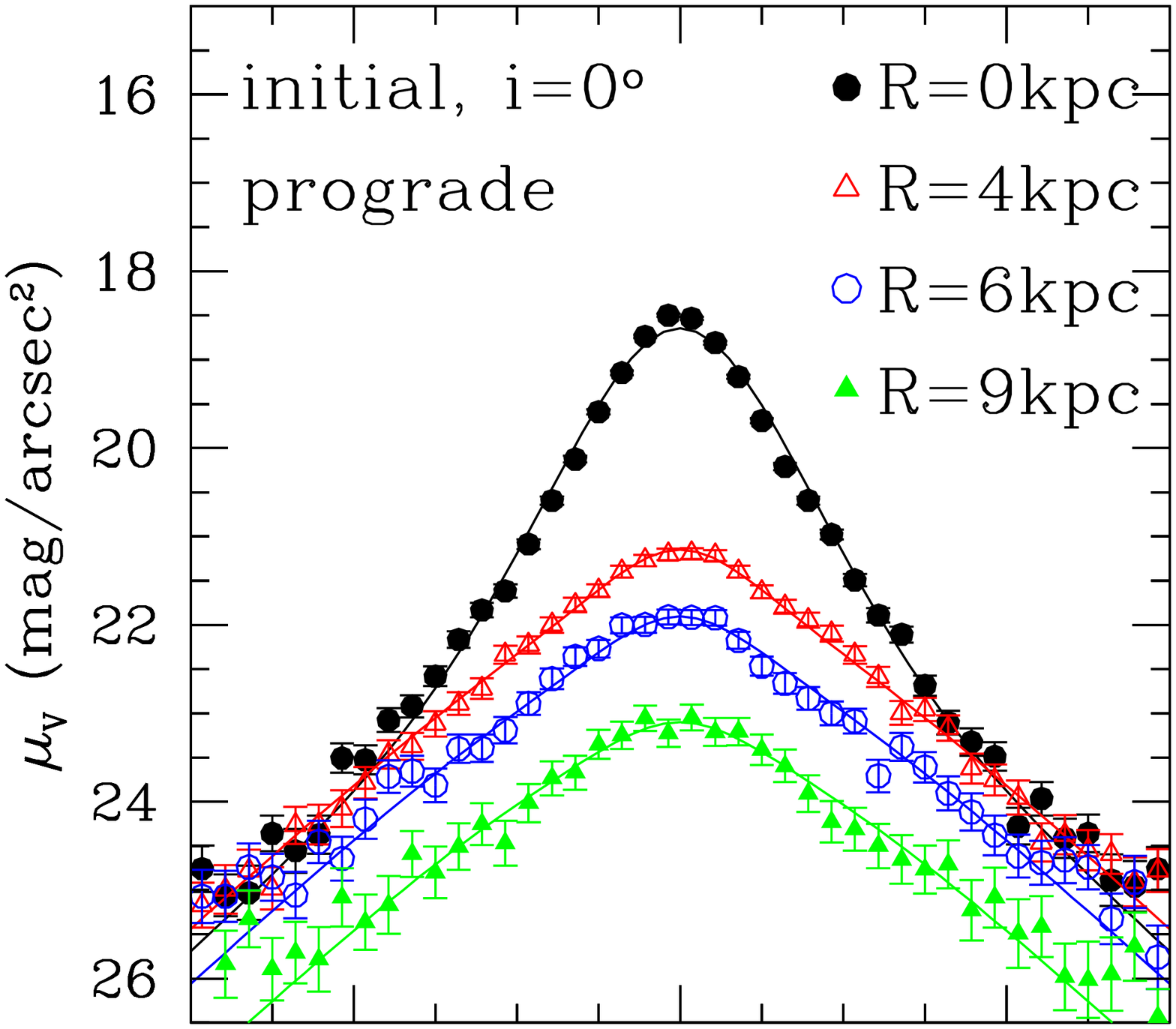}
    \hspace*{-1.2cm}
    \includegraphics[width=52mm]{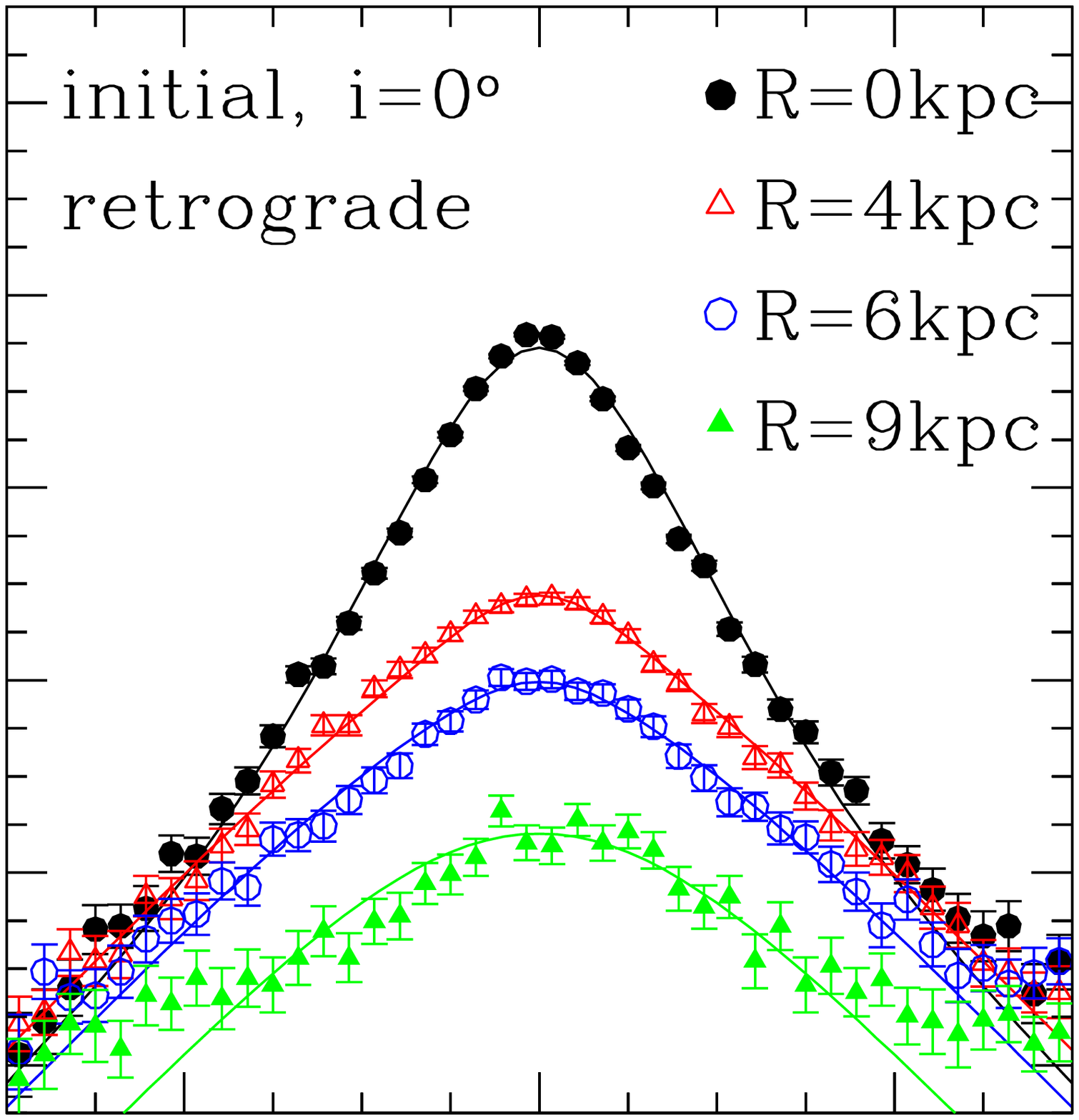}\\
    \vspace*{-0.94cm}
    \includegraphics[width=52mm]{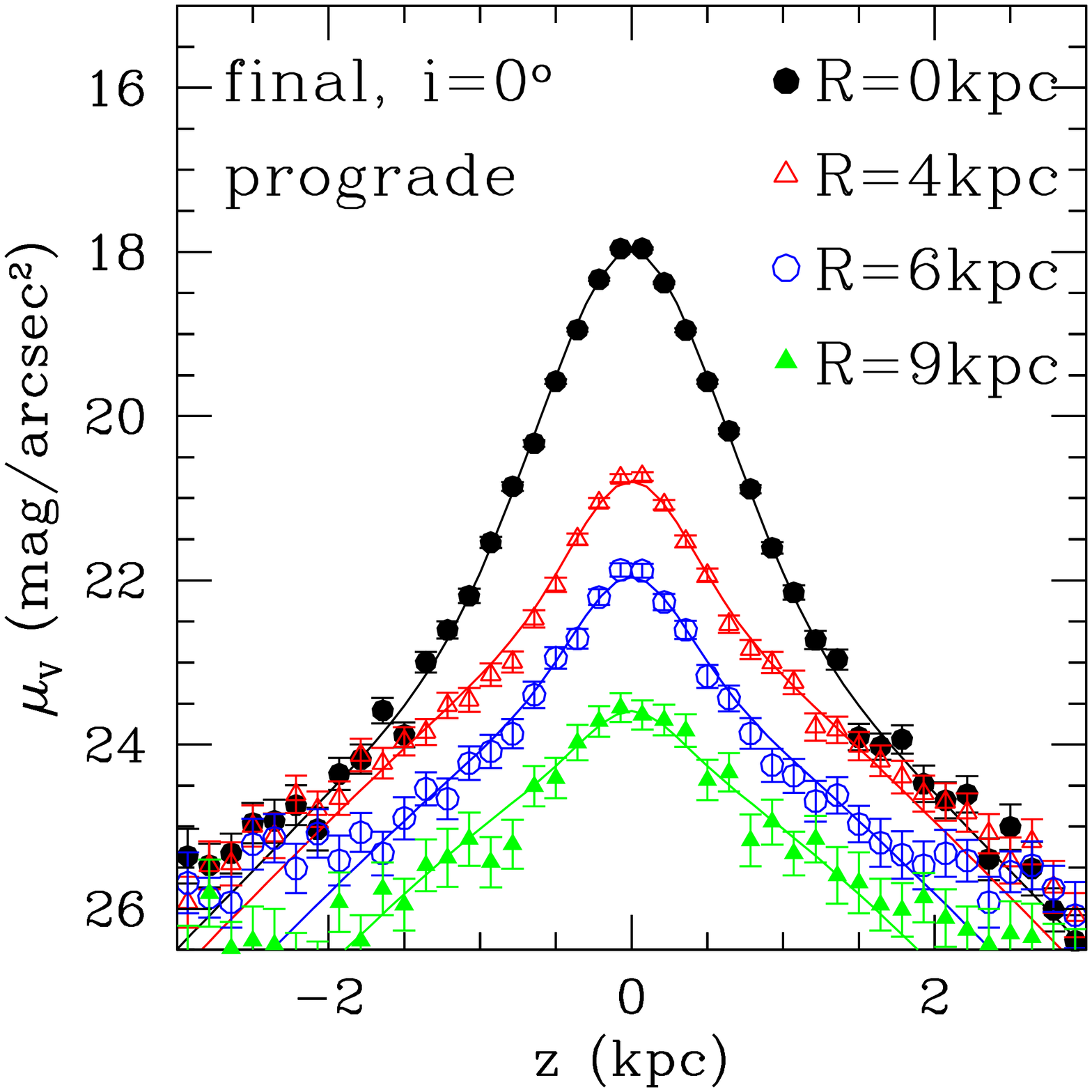}
    \hspace*{-1.2cm}
    \includegraphics[width=52mm]{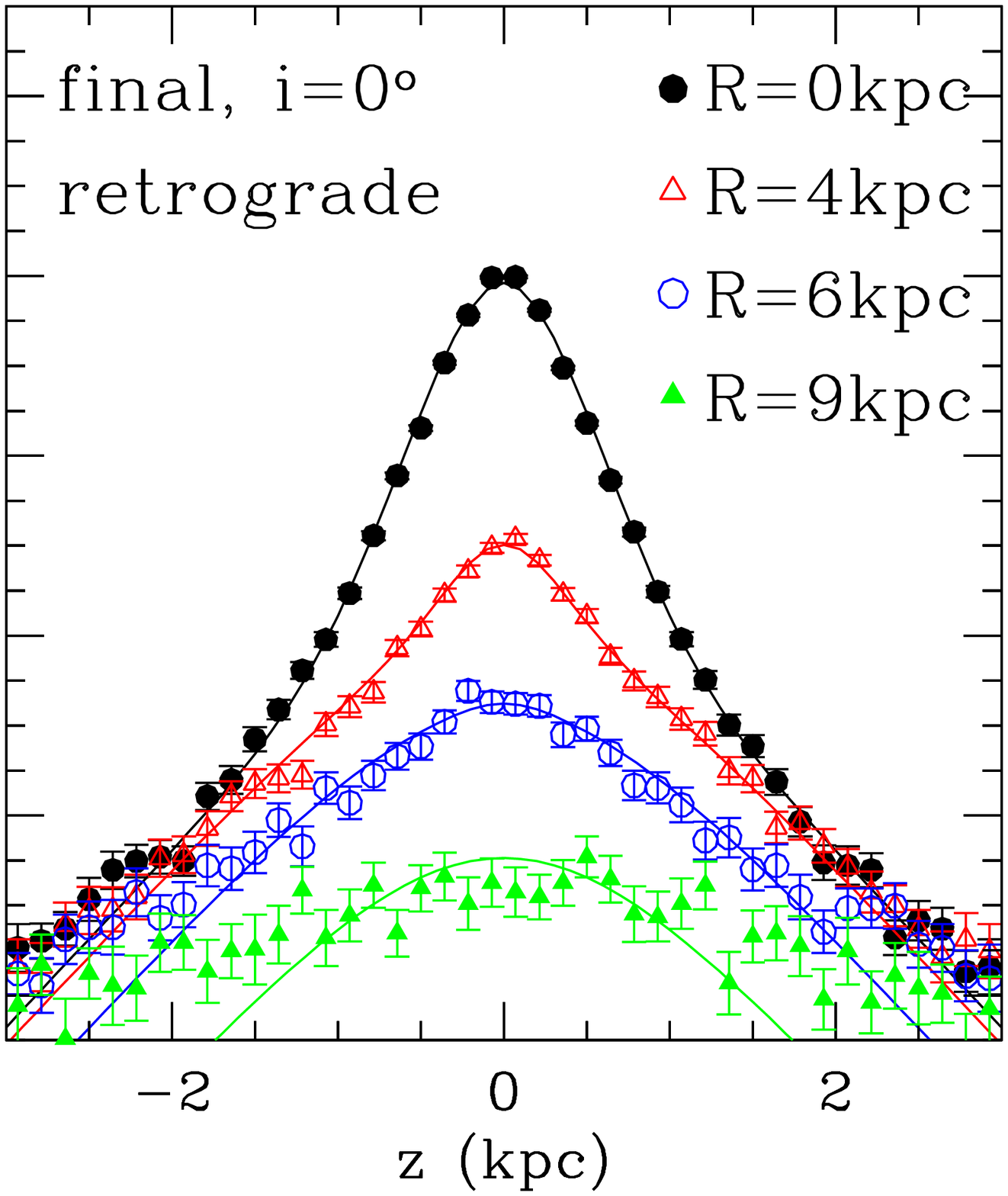}\\
    \includegraphics[width=52mm]{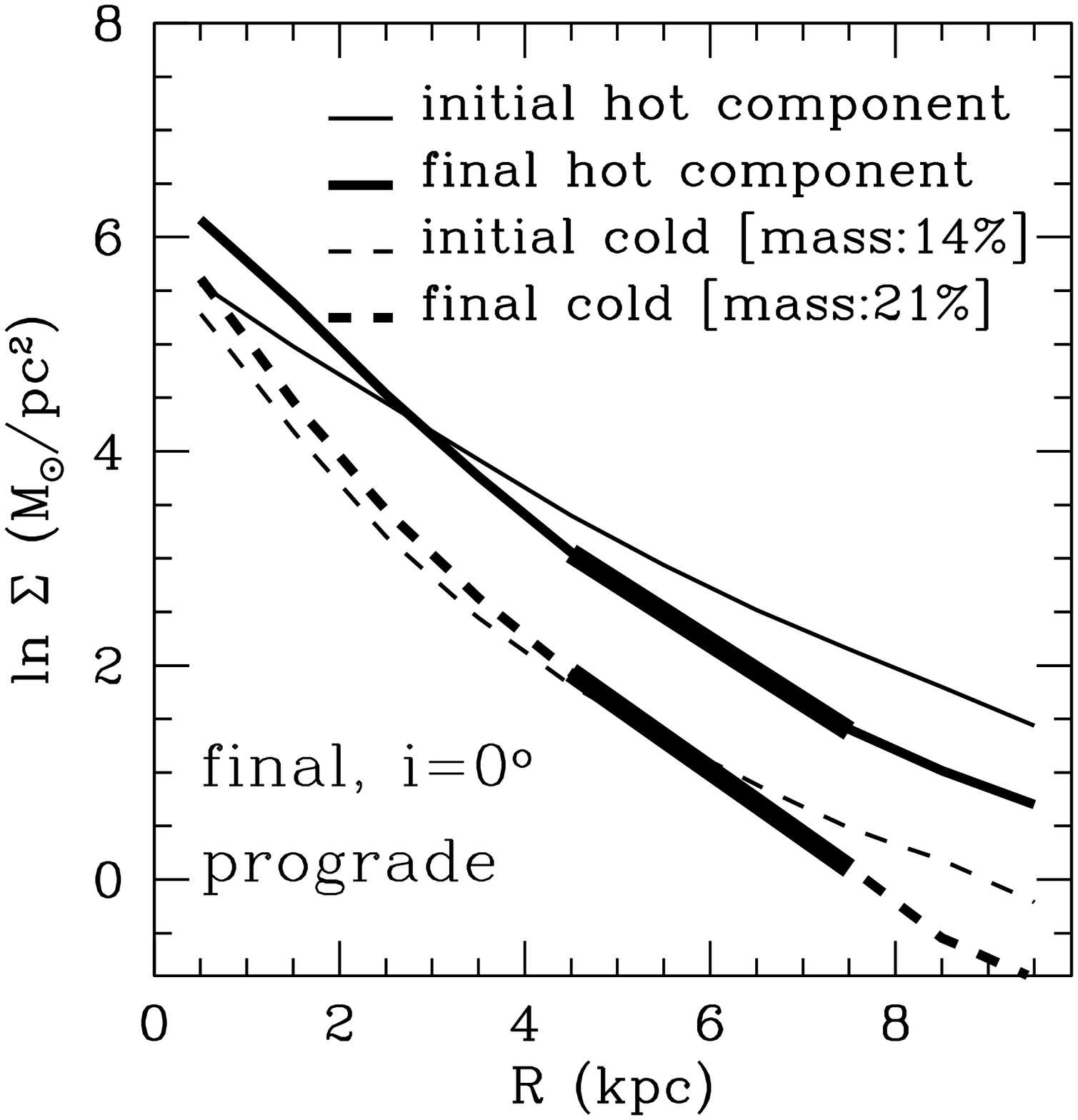}
    \hspace*{-1.2cm}
    \includegraphics[width=52mm]{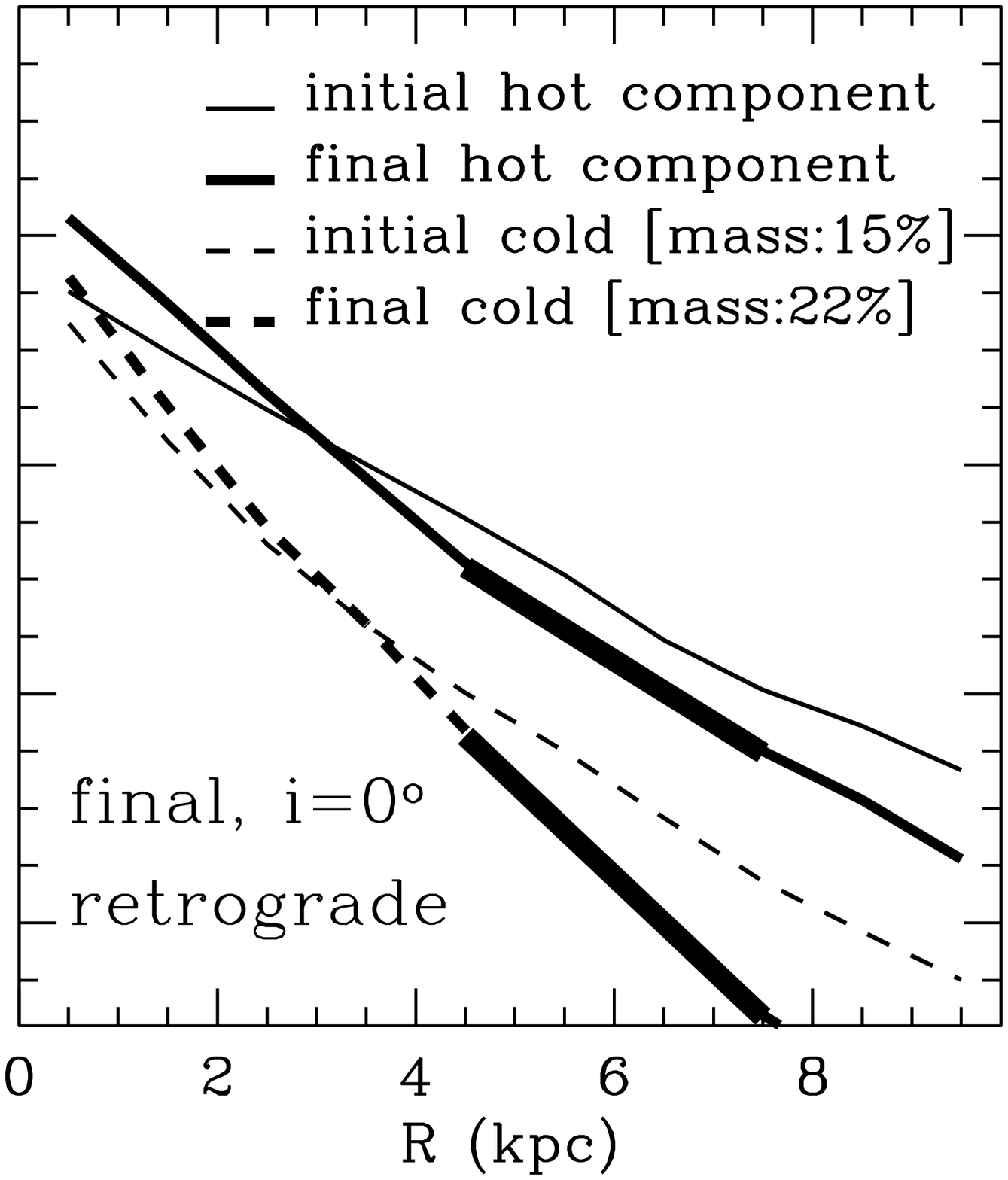}\\
    \includegraphics[width=72mm]{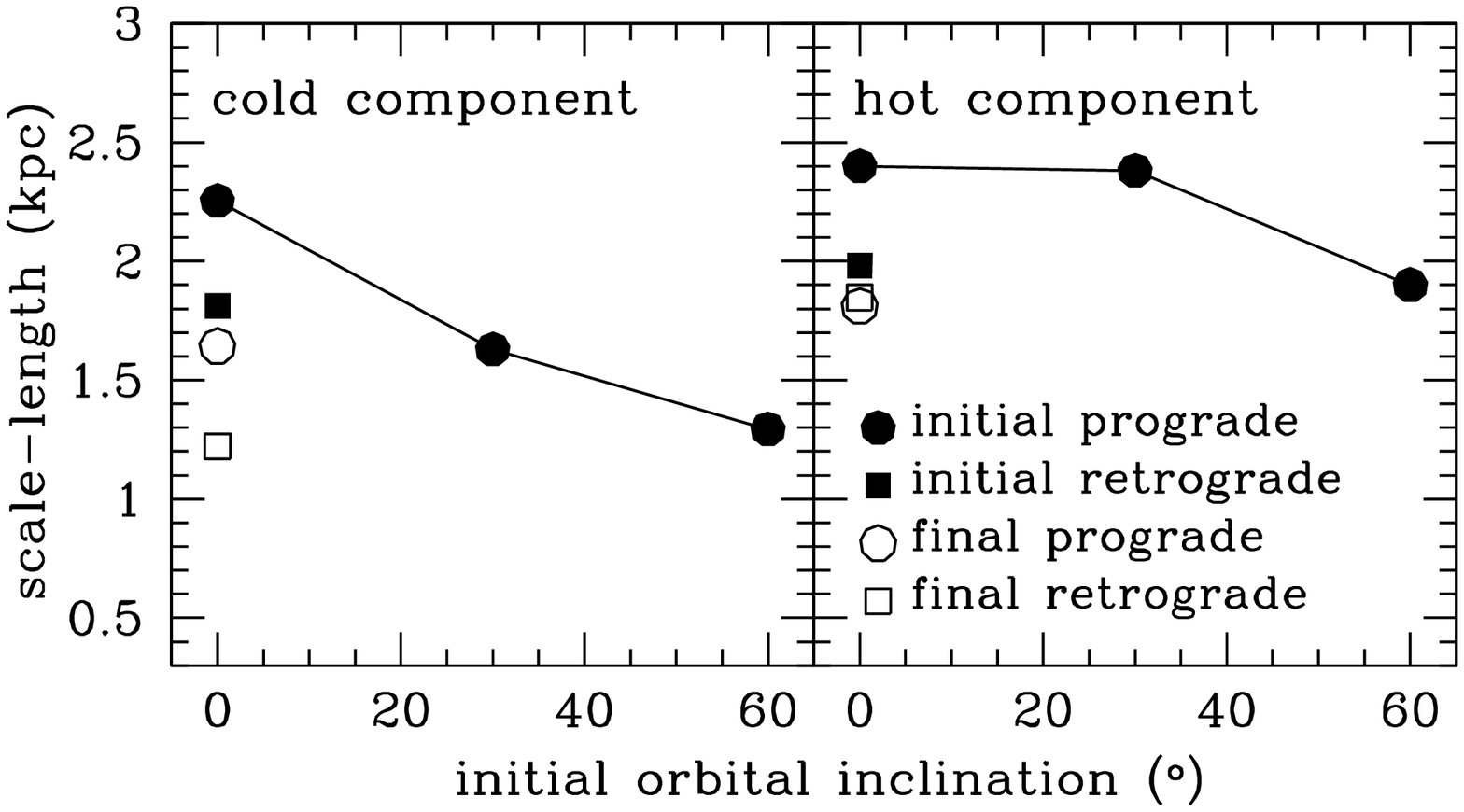}
    \includegraphics[width=72mm]{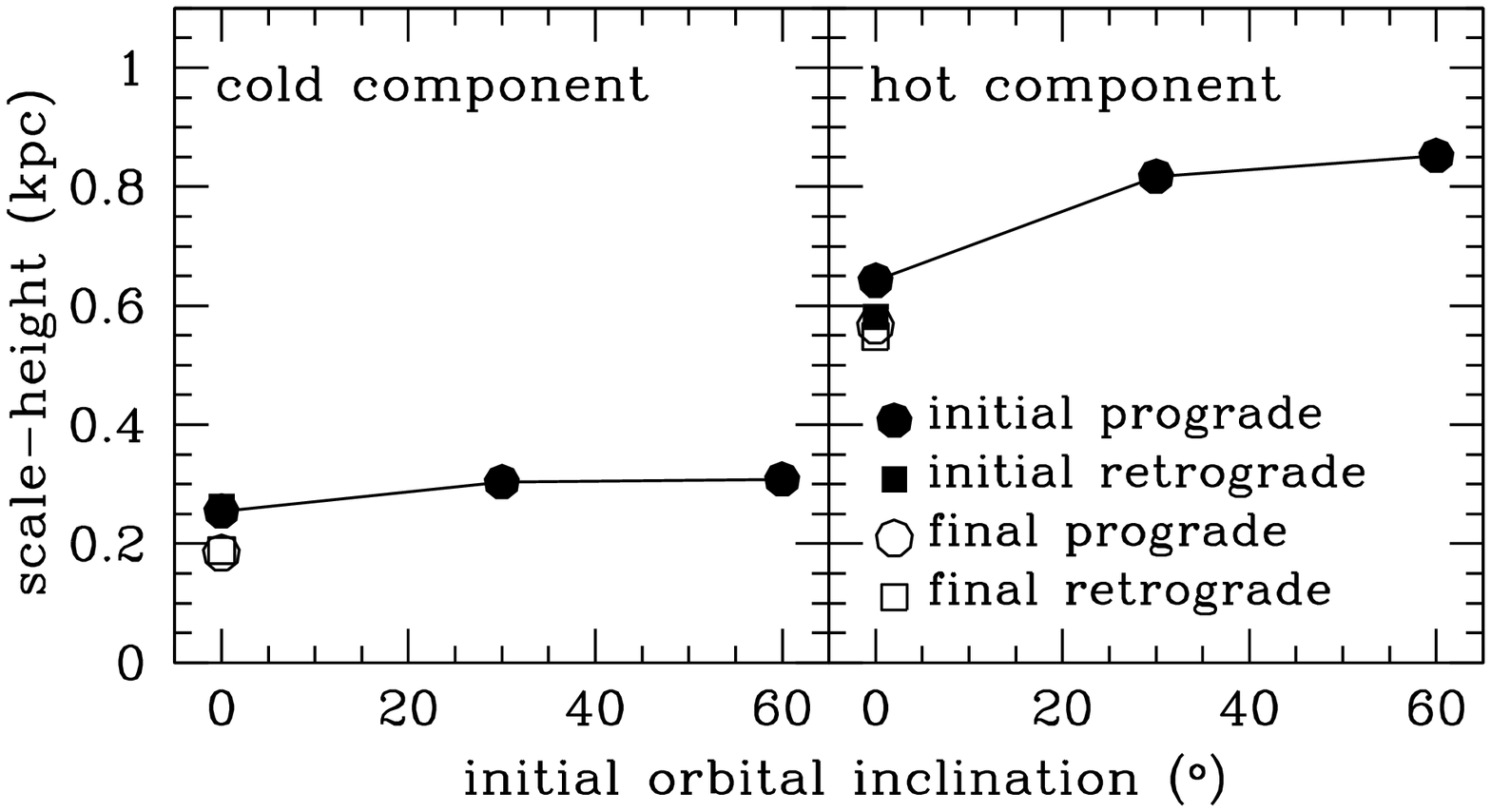}
    \caption{Experiment E. Final thick-disk structural properties for
      different sense of rotation of the initial thick disk: prograde
      (left panels) and retrograde (right panels). Rows 1 to 3 as in
      Figure~\ref{struc-a}.}
    \label{struc-e}
  \end{center}
\end{figure*}
\begin{figure*}
  \begin{center}
    \includegraphics[width=141mm]{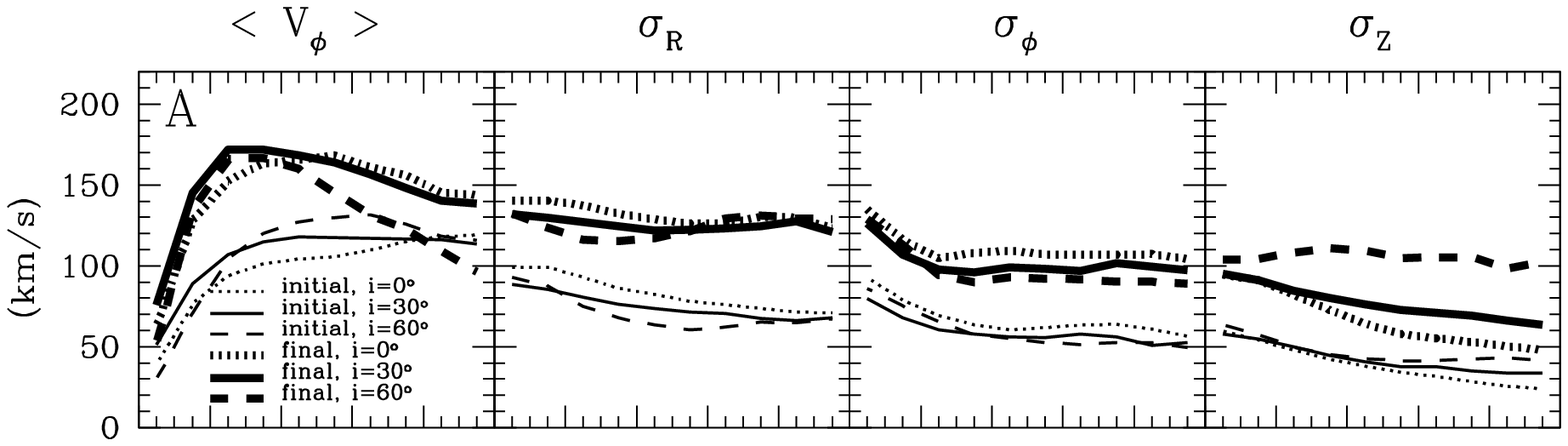}\\
    \vspace*{-1.76cm}
    \includegraphics[width=141mm]{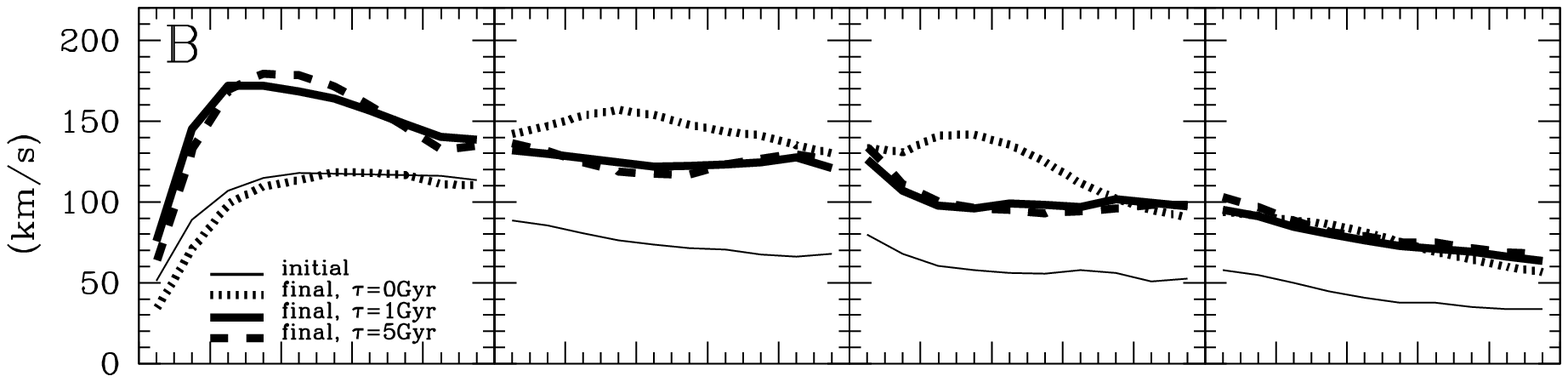}\\
    \vspace*{-1.76cm}
    \includegraphics[width=141mm]{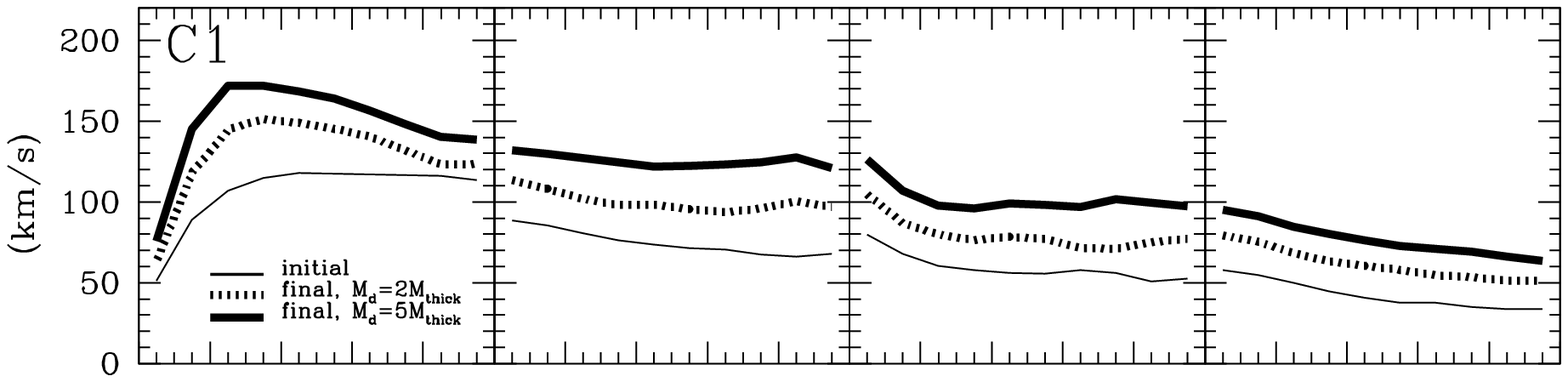}\\
    \vspace*{-1.76cm}
    \includegraphics[width=141mm]{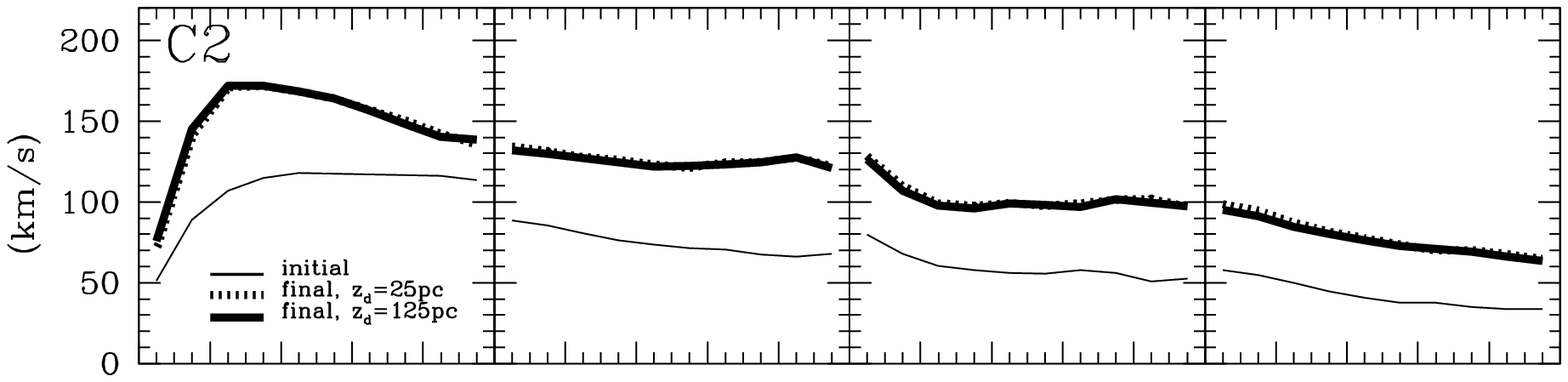}\\
    \vspace*{-1.76cm}
    \includegraphics[width=141mm]{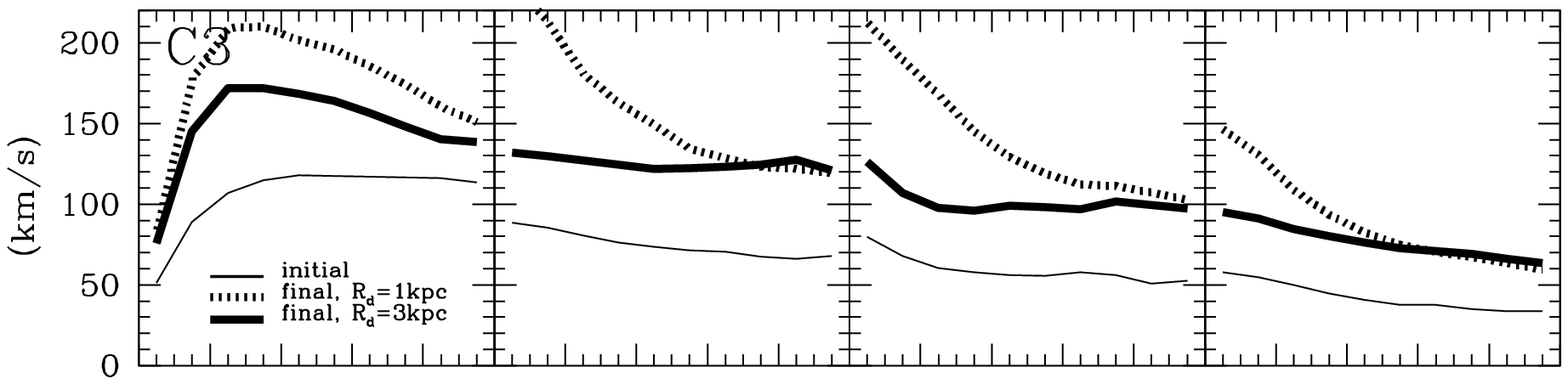}\\
    \vspace*{-1.76cm}
    \includegraphics[width=141mm]{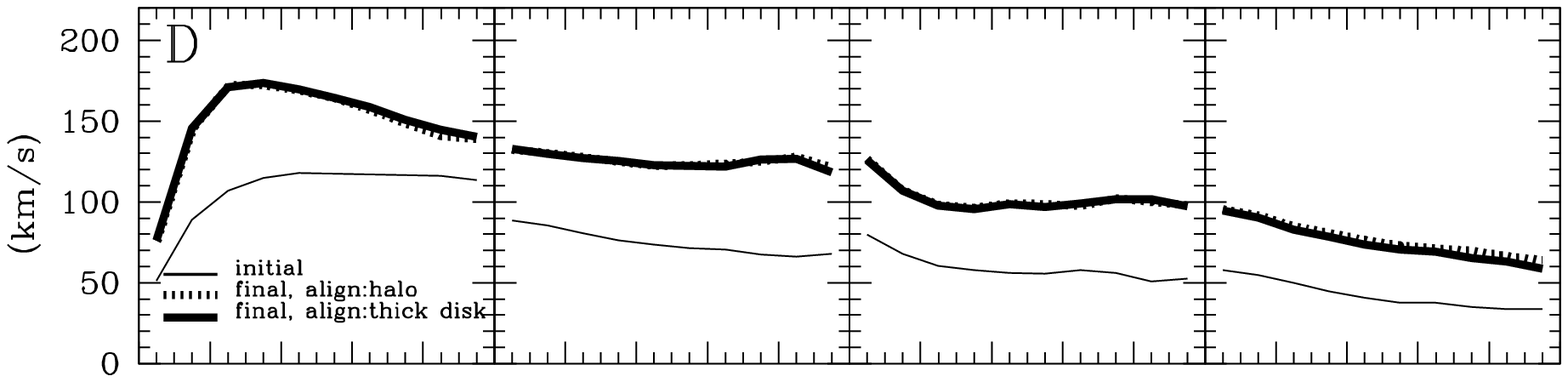}\\
    \vspace*{-1.76cm}
    \includegraphics[width=141mm]{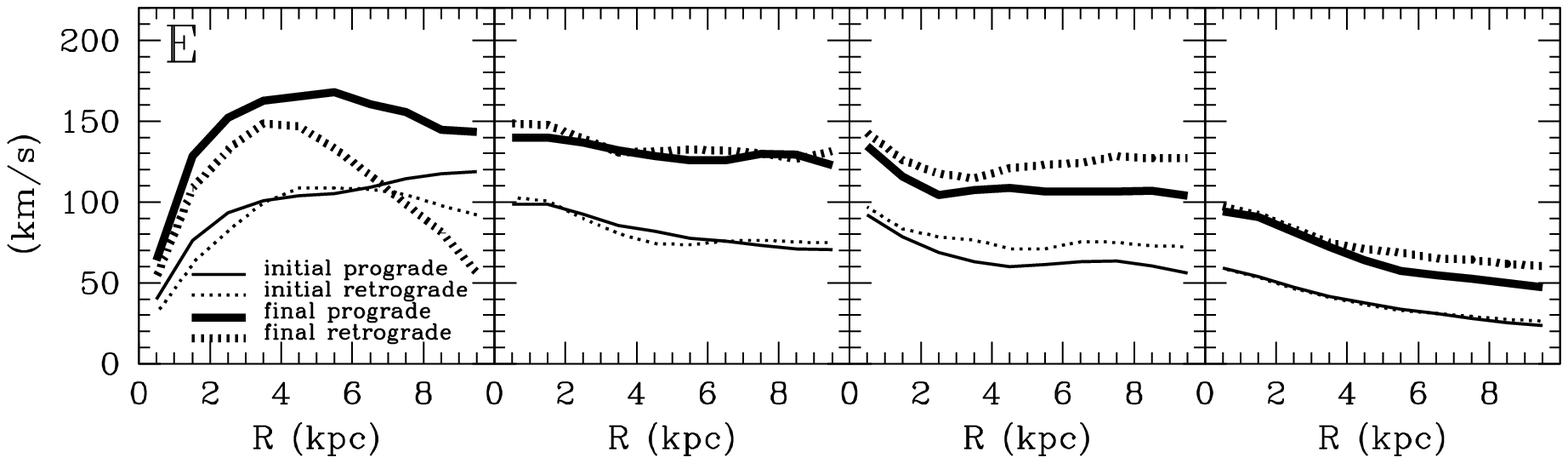}
    \caption{Thick-disk kinematical properties in experiments from A to 
    E. Results are shown for both initial and final thick disks. 
    \label{kin}}
    \end{center}
\end{figure*}

In general, the growth of a new thin disk induces three characteristic
structural changes in all initial thick disks. First, it causes
significant contraction in both the radial and vertical direction.
Figure~\ref{contour-contraction-expB} focuses on experiment A and
shows initial and final surface brightness contours of thick-disk
stars viewed both face-on and edge-on. We stress that no stars from
the growing thin disks are included in this analysis. The fact that
the final contours appear much closer together and significantly less
extended in the outer regions illustrates the contraction experienced
by the disk stars. At the same brightness level, the innermost
contours of the final systems are more extended compared to those of
the initial ones. This indicates that the surface brightness
associated with the central regions of the final systems increases. It
is also important to note that the face-on surface brightness contours
become rounder after the growth of the thin disk. This confirms
results of earlier studies regarding the effects of baryonic
dissipation on the shapes of dark matter halos
\citep[e.g.,][]{dubinski1994,kazantzidis2004}.

Figure~\ref{radial-pos-evol-expB} compares for the same experiment the
initial radial positions of initial heated-disk stars to those after
the growth of the new thin disk. Only stars within $R<15$~kpc and
$|z|<1$~kpc in the final contracted thick disks are included in this
analysis.  The figure shows that the relation between initial and
final radius is remarkably linear in all cases with practically the
same slopes, regardless of the initial orbital inclination of the
infalling satellite. In addition, the structural contraction induces a
significant migration of thick-disks stars inwards. Indeed, stars that
were initially located at $R\sim 14$~kpc end up at $R\sim 8$~kpc after
the growth of the thin disk. 

The angular momenta of individual thick disk particles are well 
conserved after the growth of the new thin disk, as it is shown in 
Figure~\ref{ang-momenta-a1} for experiment A1. In general this is found 
in all experiments, except in B1, as it is expected given its 
non-adiabatic nature. Particles within $R<3$~kpc present a less clear 
conservation of their angular momenta, which is probably due to the 
exchange of angular momentum between halo particles and thick disk 
particles in the inner region.

The second common feature seen in all experiments is that the cold
components present in the initial heated disks
\citepalias[see][]{villalobos-helmi2008} are preserved after the
thin-disk growth. As we discuss in the next section, the mass fraction
associated with these components remains below $\sim 25\%$ of the
total mass of the heated disks in all experiments. In principle, these
thin components could represent old thin-disk populations given their
similar spatial distribution and kinematics.

As discussed in \citetalias{villalobos-helmi2008}, the satellite
accretion events which produce the thick disks create flares in the
outer regions of the primary disk galaxies \citep[see
also][]{kazantzidis2008}. However, in spite of such flares, it was
also found that (radially integrated) vertical surface brightness profiles of 
thick disks are very well represented by two sech$^2$ components. 
Another common characteristic to all experiments described here is the fact 
that the growth of a new thin disk also changes the vertical
structure of thick disks, especially in the outer regions. 
The relevant analysis is presented in Figure~\ref{flare-experB} 
which shows results for run A1.  This figure compares the final 
vertical surface brightness profile (integrated for $R<10$~kpc) of 
the contracted thick disk to that of the initial thick disk evolved 
in isolation (for the same amount of time), and demonstrates that 
they are fairly different. More specifically, the growth of the thin 
disk causes the vertical surface brightness profiles to become 
substantially narrower (a closer inspection shows a significant 
contribution of satellite stars in the region $|z|>1.5$~kpc, especially 
at larger radii). While a two-component sech$^2$ decomposition
provides an accurate description of the vertical structure of the
initial thick disk down to fairly low surface brightnesses, the growth
of the new thin disk highlights the need for a more complicated
functional form to describe the final vertical disk structure at large
heights.

\subsubsection{Experiments from A to E}

Figures~\ref{struc-a} to \ref{struc-e} present results for all
experiments from A to E. In particular, these figures show (1) the
vertical surface brightness profiles of the thick disks at various
projected radii (including the best two-components sech$^2$ fits)
before and after the growth of the new thin disk; (2) the surface
density decomposition (into cold/thinner and hot/thicker components)
of the final thick disks, including the parts of the profiles
considered to estimate the respective scale-lengths and the mass
fraction associated with the cold component; and (3) a comparison
between scale-lengths and scale-heights of thick disks (for cold and
hot components) before and after the growth of the new thin disk.  All
of these properties are computed within $R<10$~kpc and $|z|<4$~kpc.
Lastly, unless otherwise explicitly stated, we do not include stars
from the growing thin disks in the analysis presented below.

Briefly, the decomposition of the vertical surface brightness profiles
is obtained by fitting, at equally spaced projected radii on the 
(properly aligned) thick disks, the function 
\mbox{$L(z) = L_1 \ {\rm sech}^2(z/2z_1) + L_2 \ {\rm sech}^2(z/2z_2)$},
where $L_i$ and $z_i$ are the local brightness on the plane and local 
scale-heights, respectively, and both are free parameters (although in 
general $z_1<z_2$). In these fits, stars from both the disk and the 
(disrupted) satellite are included within the spatial range mentioned 
above. For each thick disk one global scale-height per component is 
finally obtained by averaging the local scale-heights over all radii, 
weighting by the local brightness on the plane. On the other hand, the 
decomposition of the surface density profiles of thick disks is obtained 
by simply considering the region $0 < |z| < 0.5 z_1$ as dominated
by the cold/thinner component, and the region $z_1 < |z| < 4$~kpc as 
dominated by the hot/thicker component. Finally, the scale-lengths of 
each component are computed by applying a linear fit to $\ln \Sigma(R)$, 
avoiding non-axisymmetries associated to both the central regions and 
the very outskirts. For more details, we refer the reader to 
\citetalias{villalobos-helmi2008}. 

Experiment A (Figure~\ref{struc-a}) shows that the final mass fraction 
associated to the cold component is larger than the initial one for 
thick disks formed after an either low or intermediate inclination merger 
($i=0\degr-30\degr$), as opposite to the case when the thick disk is formed
after a high inclination merger ($i=60\degr$), where the final mass fraction
is smaller than the initial one.
It is also found that in an absolute sense the final mass fraction associated 
to the cold component remains within the range $\sim 14-20\%$, as before the 
growth of the new disk. The exception is the $i=60\degr$ case in which the 
mass of the cold remnant may amount up to $50\%$ of that of the initial 
cold component. This fraction however is relatively poorly determined 
because it depends strongly on the region selected to fit the exponential 
profile to the surface density which appears to show a change of slope 
around $ R \sim 4$~kpc. 

Experiment B (Figure~\ref{struc-b}) shows that the thick-disk 
structural evolution does not have a strong dependence on the growth
timescale of the new thin disk for $\tau$=$1$ and $5$~Gyr. In both cases, 
the peak of the final vertical surface brightness profiles becomes narrower 
and is characterized by a central value that is $\sim 1$ magnitude brighter than 
that of the initial configuration. However, at distances $R>4$~kpc, 
the surface brightness levels near the plane ($z=0$~kpc) are lower
compared to the initial thick disk. Both features reflect the radial
and vertical contraction discussed above. Moreover, the initial scale-heights 
of the cold and hot components decrease by $20-35\%$ after the accretion event.
While the scale-lengths have decreased by $\sim 30\%$ for both the cold and 
hot component, the radial surface density profiles are very similar for both 
timescales.

On the other hand, the most dramatic feature of the experiment with 
the instantaneous growth is the ``break'' at $R \sim 3-4$~kpc of the 
originally exponential surface density profiles of both the cold remnant 
and the hot/thicker component.
For $\tau=0$~Gyr the central surface densities reach similar values 
to those of the $\tau=1$~Gyr and $\tau=5$~Gyr cases, while in the 
region $R>4$~kpc they show little evolution compared to the initial 
system. 
  
Experiment C1 shows that a factor of $2.5$ more massive growing thin
disk causes roughly twice more contraction in the initial thick disk
in terms of scale-length and scale-height. Interestingly, the mass
fraction associated with the cold component of the thick disk
increases by only $\sim 3\%$ (this equivalent to a mass increase in
the cold component of roughly $10\%$) (Figure~\ref{struc-c1}).  On the
other hand, the thickness of the growing thin disk (experiment C2)
does not seem to impact significantly the structural evolution of the
thick disks, as shown in Figure~\ref{struc-c2}, where the left and
right panels correspond to $z_d=25$~pc and $z_d=125$~pc, respectively.
For example, in both cases the scale-lengths decreased by $\sim 30\%$,
while the scale-heights decreased by $\sim 40\%$. Finally, variations
in the scale-length of the new disk (experiment C3) lead to a change
in the mass fraction associated with the cold component of the thick
disk. Thin disks with a scale-length 3 times smaller
(Figure~\ref{struc-c3}) reduce the mass of the cold remnant to one
third. This can be expected since a growing disk with a larger radial
extent would be able to attract more mass onto the midplane across the
system. No significant differences are detected in the final
scale-lengths and scale-heights of both components in the contracted
thick disks. However, the estimation of the final scale-length of the
cold remnant in the case of a new disk with $R_d=1$~kpc should be
taken with caution, since it could be affected by the ``break''
feature of the surface density and by the region chosen to derive the
scale-length.

We remind the reader that the previous analysis does not include stars
from the growing thin disks. Therefore, it is important to clarify the
extend to which the structural decomposition of the final thick disks
depends on the inclusion of the growing thin disks. Such an attempt is
presented in Figure~\ref{struc-c3} which compares the final thick-disk
vertical surface brightness profiles in experiment C3 ($R_d=1$~kpc)
with and without the inclusion of the growing thin disk. Not
unexpectedly, when the growing thin disk is included in the
decomposition, the peak of the surface brightness profiles increases
at all radii.  However, the scale-heights of both ``cold'' and
``hot''components are not substantially modified. Specifically, when
the growing thin disk is included, scale-heights change from 140~pc
and 585~pc to 130~pc and 650~pc. We find a similar behavior in the
rest of our experiments for the scale-heights. It is important to note
that other properties can be affected more by the inclusion of the
growing thin disk in the analysis (e.g., scale-lengths). However, the
magnitude of these differences depends sensitively on the criteria
that are used to define the regions dominated by the ``cold'' and
``hot'' component, that is $|z|<0.5 z_{\rm cold}$ and $z_{\rm cold}<
|z| < 3$~kpc, respectively, in terms of the scale-height of the
``cold'' component \citepalias[see][]{villalobos-helmi2008}.

In the numerical simulations discussed in
\citetalias{villalobos-helmi2008}, the angular momenta of the dark
matter halos and thick disks were often slightly misaligned by few
degrees\footnote{Cosmological simulations show that the angular
  momentum of the baryonic component of a galaxy correlates well with
  the angular momentum of the parent dark matter halos, with a typical
  misalignment of $\sim 20\degr$ \citep{sharma2005}.}. In experiment
D, we utilized the thick disk produced by an accretion event with an
initial inclination of $i=30\degr$. The angular momentum vector of the
adopted thick disk is misaligned by $\sim 6\degr$ with respect to the
halo angular momentum.  For this experiment we have allowed the
angular momentum of the growing thin disk to be aligned with the
angular momentum of either the halo or the thick disk.
Figure~\ref{struc-d} shows that no significant differences can be
detected in the structure of the thick disks for either alignment of
the growing disk, which is not surprising given the rather small
misalignment between the initial thick disk and the halo angular
momenta. In principle, the differences in the evolution of the
properties of the thick disk could become larger for a more
significant initial misalignment.

In experiment E (Figure~\ref{struc-e}) the adopted initial thick disks
have been produced by an accretion event with an initial inclination
of $i=0\degr$. We considered two cases where the infalling satellite
was on a prograde and on a retrograde orbit with respect to the
rotation of the pre-existing galactic disk. In both cases, a
significant amount of angular momentum is transferred from the
satellite to the dark matter halo
\citep[see][]{vitvitska2002,hetznecker2006} and much less into the
pre-existing disk. The latter is because by the time the satellite has
reached the disk the orbit is much smaller than the initial one, and
it has lost most of its bound mass. As a result, the halo and the
resulting thick disk rotate in the same sense after the accretion
event in the prograde case, while in the retrograde case they rotate
in opposite directions. We note that our initial dark matter halos are
non-rotating, but the same conclusions would hold for halos with
typical amounts of initial rotation (see below).

The surface density profiles in Figure~\ref{struc-e} show that the
growth of a thin disk on thick disks with opposite sense of of
rotation induces essentially the same increase in the mass fraction
associated with the thin remnant (from $\sim 15\%$ initially to $\sim
22\%$), and that the decrease in the scale-length of the thin remnant
is of similar amplitude ($\sim 30\%$).  The most significant
difference seems to be in the scale-lengths of the thick disk, where
in the prograde case it is induced a decrease of $25\%$, while in the
retrograde one it is induced only a $7\%$ decrease (as for experiment C3
this has some dependency on the region considered to estimate the
scale-lengths). On the other hand, the evolution of the scale-heights
of both the thin remnant and the thick disk do not show a significant
difference between prograde and retrograde cases.

This experiment was motivated by recent observations of
counter-rotating thick disks (with respect to their thin disks)
detected by \citet{yoachim2005}. It has been suggested that such
counter-rotation may pose difficulties for the disk heating scenario
as a viable model for the formation of thick disks. However, an
encounter between a pre-existing thin disk galaxy and a massive
satellite on a retrograde orbit can, in principle, produce a thick
disk and a counter-rotating dark matter halo.
In order to illustrate this we can estimate the total angular momentum
of a halo after such encounter as 
\mbox{$L_{\rm f}^{\rm halo} = L_{\rm i}^{\rm halo} + L_{\rm i}^{\rm sat}$}, 
where the last term on the right is the
initial orbital angular momentum of the infalling satellite.
$L_{\rm i}^{\rm halo}$ can be computed using the halo spin 
parameter, $\lambda$, and its virial properties as
$L_{\rm i}^{\rm halo} = \sqrt{2} \lambda M_{\rm vir} V_{\rm vir} R_{\rm vir}$ 
\citep{bullock2001}. 
On the other hand, $L_{\rm i}^{\rm sat} = M_{\rm sat} R_0 V_0$, in terms
of the satellite mass and its initial orbital radius and velocity.
In the case of experiment E2, $M_{\rm vir}= 5.07 \times 10^{11} \Mo$, 
$V_{\rm vir}=133.87$~km~s$^{-1}$, $ R_{\rm vir}=122.22$~kpc, 
$M_{\rm sat} = 10^{11} \Mo$,  $R_0 = 83.9$~kpc, and 
\mbox{$V_0=-78.8$~km~s$^{-1}$}
(negative sign indicates a retrograde orbit). Thus, assuming a typical 
value for the spin parameter, $\lambda=0.035$, we have
\mbox{$L_{\rm f}^{\rm halo} = - 2.5 \times 10^{14} \Mo$~kpc~km~s$^{-1}$},
where the negative sign shows that the halo is counter-rotating 
with respect to its initial state.

Freshly accreted cold gas from the galactic halo would have its
angular momentum aligned with that of the dark halo and therefore
could form a young thin disk that is counter-rotating with respect to
the older thick disk. In this case, the new disk would also
counter-rotate with respect to any old thin remnant present.

\subsection{Kinematical Evolution of Thick Disks}
\label{kine-evol-sigz-sigr}

\subsubsection{General Features}
\label{kine-evol-gral}

Similarly to the structural evolution, the growth of a thin disk
induces two characteristic changes in thick-disk kinematics that are
common to all experiments. In particular, a growing thin disk causes
an increase in both the mean rotational velocity and the velocity
dispersions of the thick disks. Both effects can be explained by the
fact that the growing thin disk adds mass to the system and by the
subsequent structural contraction described above.

\subsubsection{Experiments from A to E}

%
\begin{figure}
  \begin{center}
    \includegraphics[width=65mm]{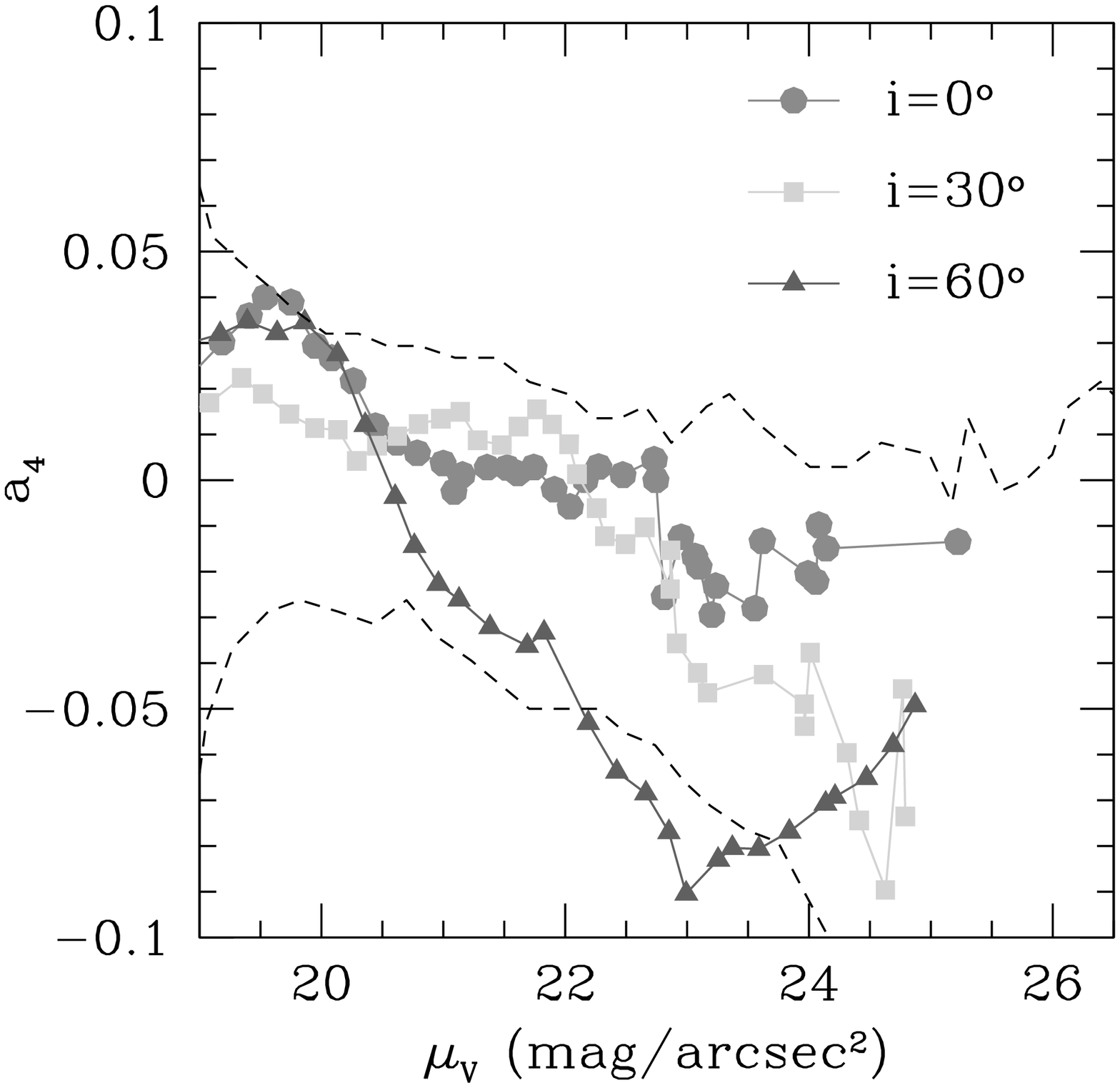}
    \hspace*{0.2cm}
    \includegraphics[width=69mm]{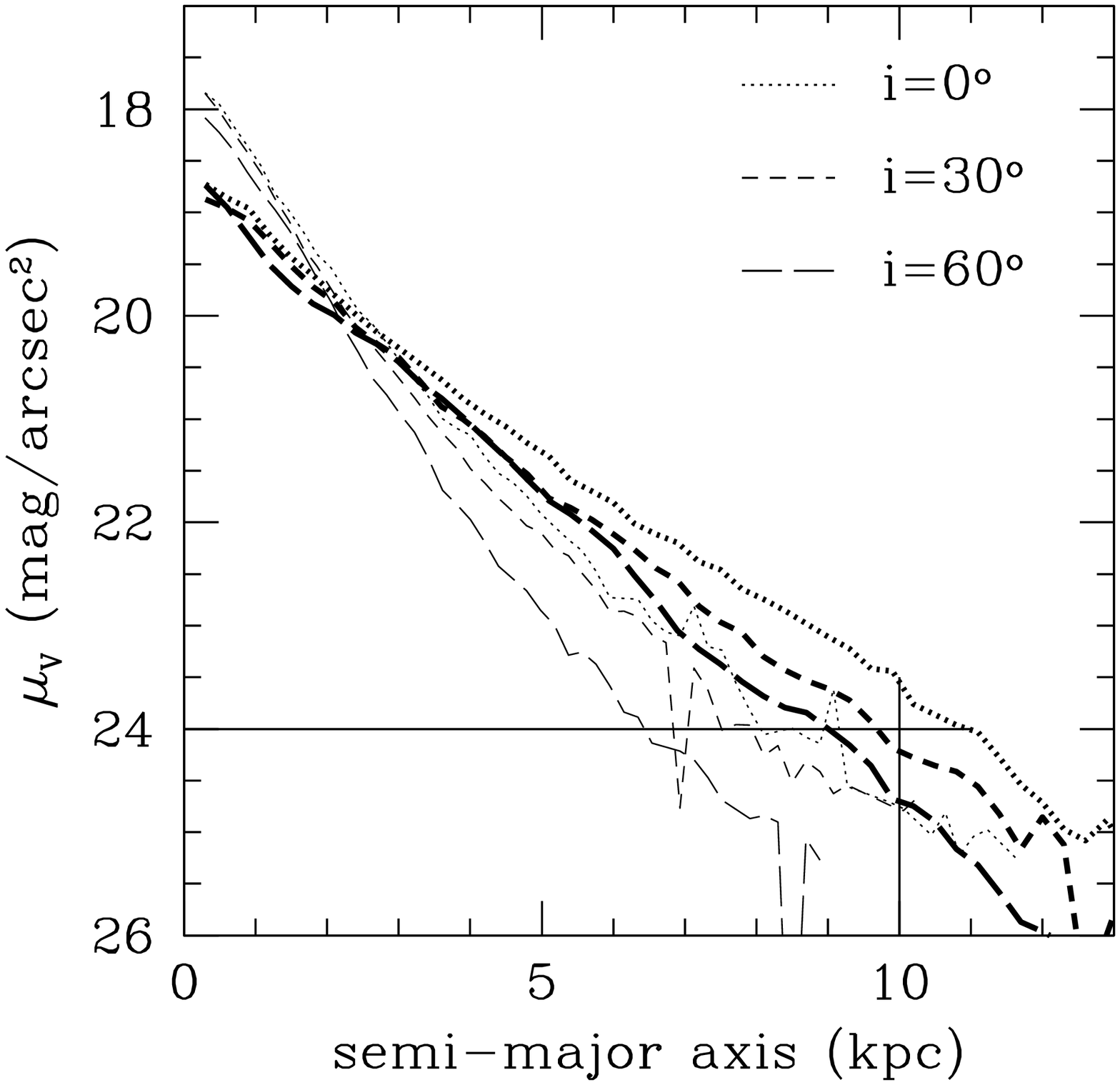}
    \caption{{\it Upper panel}: $a_4$ isophotal shape parameter as a
      function of isophotal surface brightness in the V-band for the
      final thick disks in experiment A. Dashed curves show results
      before the growth of the new disk (see Figure 8 of
      \citetalias{villalobos-helmi2008}) and are included here as a
      reference.  {\it Bottom panel}: V-band surface brightness as a
      function of the isophotal semi-major axis for the same
      experiment.  Thick and thin lines show the profiles before and
      after the thin-disk growth, respectively. The solid horizontal
      line marks the surface brightness limit below which the
      isophotes are consistently boxy. The vertical solid line marks
      the limiting radius $R$ within which the global properties of
      the final thick disks are computed.}
    \label{a4-boxy}
  \end{center}
\end{figure}

Figure~\ref{kin} presents the initial and final global kinematics of
the simulated thick disks as a function of galactocentric radius. We
characterize the thick-disk kinematical response to the growth of a
new thin disk through the evolution of the mean rotational velocity,
$\langle \vphi \rangle$, and velocity ellipsoid ($\sigma_R, \sigma_\phi, \sigma_z$),
where $\sigma_R$, $\sigma_\phi$, and $\sigma_z$ correspond to the
radial, azimuthal, and vertical velocity dispersions, respectively.
These properties are computed within $R<10$~kpc and $|z|<4$~kpc and as
before no stars from the growing thin disks have been included in the
analysis.  It is important to mention that we do not find significant
differences between the kinematics of the global thick disks (i.e.
``cold'' + ``hot'' components) and those from the ``hot'' component
alone in the case of $\sigma_R$ and $\sigma_\phi$. However, $\sigma_z$
and $\vphi$ of the ``hot'' component are larger, as expected, by $\sim
20\kms$ and $\sim 40\kms$, respectively (mainly at smaller radii).

The results of experiment A demonstrate that after the growth of the
new thin disk, the velocity dispersions of the final thick disks
maintain the original trends with the satellite orbital inclination
(that is, the hotter the initial thick disk the hotter the final one).
We note that the mean rotational velocity in the $60\degr$ case shows
a clear decrease at larger radii which is due to the increasing
skewness of the total $\vphi$ distribution towards lower velocities.
This is a consequence of the contribution of the accreted stars, which
are preferentially found in the outskirts and which rotate more slowly
with respect to the stars of the pre-existing thin disk.

The findings of experiment B indicate that for $\tau$=$1$ and $5$~Gyr
the global kinematics of the final thick disks do not depend on the
growth timescale of the new disk. Interestingly, the effect of the
instantaneous growth is more complex. In this case, the mean rotation
of the thick disk is not substantially affected by the growth of the
new disk (except by some decrease in the inner region). This is
consistent with the corresponding surface density profile
(Figure~\ref{struc-b}) which shows little variation with respect to
the initial configuration (except by a distinctive ``break'' at $R\sim
4$~kpc). Although a significant amount of mass has been added to the
system via the growth of the new thin disk, this growth was strongly
non-adiabatic. This fact implies that the eccentricity of the stellar
orbits in the thick disk must have changed significantly, or
equivalently, that the amount of ordered (circular) motion must be
much smaller. Thus in practice, the extra potential energy due to the
new thin disk must have gone into random motions in the plane as can
be seen from the central panels in Figure \ref{kin}, which clearly
show an increase in the velocity dispersions, particularly at
$R\sim4$~kpc.

The results of experiment C1 show that a more massive growing disk
induces a larger increase in both the mean rotation and the velocity
dispersions of the final thick disks. By resorting to the Jeans
equations \citetext{\citealt{binney1987}, Eq.~4.29}, it may be
expected that a thin disk roughly twice as massive causes an increase
of a factor of $2$ in all thick-disk velocity dispersions and in the
mean rotational velocity at a given radius.

The findings of experiment C2 suggest that for our choices of vertical
scale-heights, $z_d=25$ and $125$~pc, the kinematical evolution of
thick-disk stars does not depend essentially on the thickness of the
growing disk.  Of course, this conclusion is valid as long as the mass
of the growing disk is the same as in our case. On the other hand, the
results of experiment C3 highlight that at a fixed mass a more radially 
compact growing disk is responsible for generating a larger increase 
in both the mean rotational velocity and velocity ellipsoid in the inner 
regions of thick disks.

The results of experiment D show that the relative alignment of the
new disk (with respect to either the halo or the initial thick disk
angular momentum) does not seem to affect the kinematical evolution of
thick disks. This could also have been expected from the lack of
structural evolution shown in Figure~\ref{struc-d}.  Note, however,
that the misalignment between the halo and initial thick disk angular
momenta is only $\sim 6\degr$. Such misalignment may be possibly too
small to induce significant differences in both the structure and
kinematics of the final thick disks.

The findings of experiment E demonstrate that the evolution of all
three velocity dispersions is very similar for both prograde and
retrograde thick disks, as expected from Figure~\ref{struc-e}. Note
that for the retrograde case there is a significant decrease in the
mean rotation of the thick disk at $R>4$~kpc. Similarly to experiment
A2, a closer inspection shows that there is significant skewness in
the distribution of $\vphi$ towards lower velocities. This is due to
the contribution of satellite stars at large radii which rotate more
slowly than disk stars. Overall, in the retrograde case the evolution
in the kinematics of the thick disk is found to be very similar 
to that in the prograde case. The most significant
difference is found in $\sigma_{\rm z}$ where the retrograde case
leads to a somewhat higher dispersion at large radii.

\begin{figure*}
  \begin{center}
    \includegraphics[width=140mm]{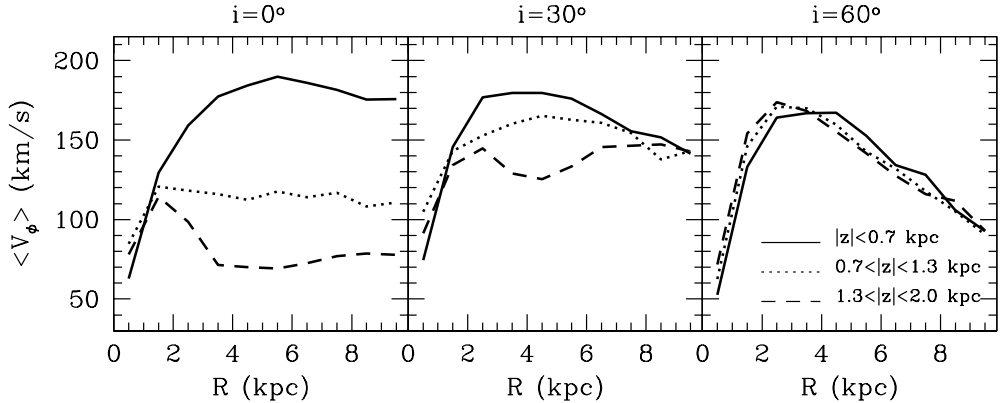}
    \caption{Final mean azimuthal velocity of thick-disk stars as a
      function of galactocentric distance, $R$. Results are presented
      for experiment A at various scale-heights above the disk plane.
      The vertical gradient of the mean azimuthal velocity depends
      sensitively on the orbital inclination of the infalling
      satellite that produced the initial thick disks.}
    \label{vpmean-fn-z-p2p3}
  \end{center}
\end{figure*}
\begin{figure}
  \begin{center}
    \includegraphics[width=70mm]{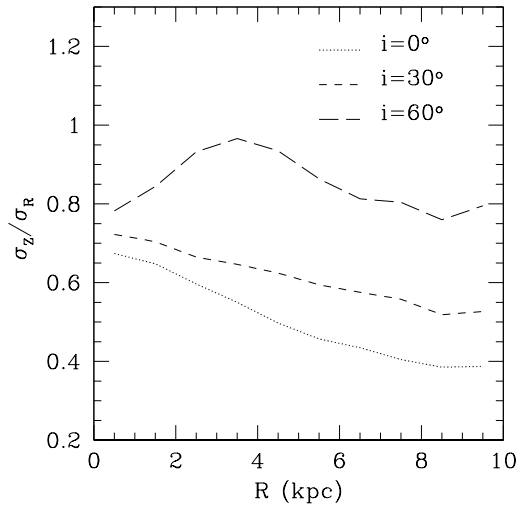}
    \caption{Final velocity dispersion ratio $\sigma_z/\sigma_R$ of
      thick-disk stars as a function of galactocentric distance, $R$.
      Results are shown for experiment A.  Even after the growth of
      the new thin disk, the velocity dispersion ratio
      $\sigma_z/\sigma_R$ depends sensitively on the orbital
      inclination of the infalling satellite.}
    \label{new-sigz-sigr}
  \end{center}
\end{figure}
%

\section{Comparison with Initial Thick Disks}
\label{comparison}

\subsection{Structure}

The results presented in Figure~8 of \citetalias{villalobos-helmi2008}
show that the outer isophotes of the simulated thick disks measured at
surface brightness $\mu_{V} > 25$ \massq, namely $>6$ mag below the
central peak $\mu_{V} \sim 19$ \massq, are consistently more boxy than
the inner ones.  This characteristic feature is present in all the
thick disks simulated in \citetalias{villalobos-helmi2008}, implying
that the detection of such a degree of boxiness in real galaxies could
be used to test thick-disk formation scenarios \citep[see
also][]{bournaud2009}.

Figure~\ref{a4-boxy} shows the $a_4$ parameter as a function of
isophotal surface brightness in the V-band for the final thick disks
of experiment A.  This figure shows that the trend of more boxy
contours at lower surface brightness limits is maintained after the
growth of the new disk. Interestingly, at a given $\mu_{V}$, the
contours are slightly less boxy in comparison to those presented in
Figure~8 of \citetalias{villalobos-helmi2008}. This is likely caused
by the growth of the new (flatter and more massive) thin disk.  Recall
that Figure~\ref{contour-contraction-expB} shows that contours at
lower $\mu_{V}$ are in general closer to the center of the system
after the induced structural contraction. In order to detect the
``boxiness'' in the experiments with growing disks one would need to
reach surface brightness levels beyond $\mu_{V} > 24 - 25$ \massq,
namely $> 6 - 7$ mag below the central peak of the thick disks after
the growth of the new disk, $\mu_{V} \sim 18$ \massq.

The bottom panel of Figure~\ref{a4-boxy} shows for the same experiment
the isophotal surface brightness as a function of the isophotal
semi-major axis before and after the growth of the new disk. The
results in this panel suggest that a possible cause for the
structural change introduced to the vertical surface brightness profiles
in the outskirts of the final thick disks (see Figure~\ref{flare-experB}) 
is the contraction which
brings boxy isophotes within the region $R<10$~kpc.  To this end, the
presence of boxy contours at the outskirts would enhance the
separation between inner contours (that are more packed together) and
outer contours (see Figure~\ref{contour-contraction-expB}). This would
make more noticeable the difference in the profile shown in
Figure~\ref{flare-experB} (right) between $|z|<1$~kpc and $|z|>2$~kpc.
 
The structure of the initial thick disks is in good agreement with
observations of external galaxies and the MW in terms of thick-to-thin
ratios of both scale-lengths and scale-heights (see Section~4.2 of
\citetalias{villalobos-helmi2008}). This conclusion was based on the
assumption that thick-disk structure would {\it not} be strongly
affected by the growth of a new thin disk {\it and} that this thin
disk would follow the same distribution as the cold component of the
thick disk. Overall, the results of the present paper indicate that
thick disks {\it do} respond to the growth of an embedded thin disk.
It is interesting to investigate how these scale ratios compare to
observations after the formation of the new thin disk. In our new
experiments, we find the final thick-to-thin scale-height ratio to be
$\sim 0.6/0.125=4.8$, which is still within the range 2.4-5.3 observed
in the MW and in S0 galaxies \citep[see][]{buser1999,pohlen2004,yoachim2006}.

In what follows we assume that the Galactic thick disk originated from
the vertical dynamical heating of a primordial thin disk by infalling
satellites (\citealt{kazantzidis2008};
\citetalias{villalobos-helmi2008}). It is interesting to use the
results above and attempt to determine the initial properties of the
initial thick disk of the MW, namely those before the structural
contraction induced by the formation of the current thin disk; as well
as the properties of the pre-existing disk, namely those before the
interaction with the satellite. For example, we typically find that
the scale-length of the thick disk after the encounter with the
satellite increases by a factor $f^R_{\rm merger}\sim 1.3 - 1.45$, as
shown in Section~3.4 of \citetalias{villalobos-helmi2008}. On the
other hand, the formation of a factor of $5$ more massive thin-disk
component leads to a decrease in the scale-length of $f^R_{\rm contr}
\sim 0.6$.  This implies that the scale-length of the pre-existing
disk is given by $R_d^{\rm initial} = (f^R_{\rm merger} \times
f^R_{\rm contr})^{-1} \times R_{d,\rm thick}$\footnote{Of course, such
  a relation is only valid under the assumption that the only relevant
  processes are the satellite accretion event and the thin-disk
  growth, and that secular evolution processes are unimportant in
  establishing the galactic structure.} or $R_d^{\rm initial} \sim
1.15 R_{d, \rm thick}$, which is comparable to the present-day value
of the thick disk. This suggests that the radial expansion caused by
the accretion event is nearly completely balanced by the contraction
induced by the growth of the thin disk.  Moreover, using current
estimates for the thick-disk scale-length of $R_{d,\rm thick} = 2.5 -
3$~kpc, we deduce the initial scale-length of the Galactic
pre-existing disk to be $R_d^{\rm initial} \sim 2.9 - 3.5$~kpc.  Of
course, this could represent a considerable problem, since disks are
expected to have been (significantly) smaller in the past
\citep{buitrago2008}.

Similarly, we may compute the initial scale-height of the thick disk
of the MW from $f^z_{\rm merger}\sim 4.7$ and $f^z_{\rm contr} \sim
0.6$ (both for a $30\degr$ inclination encounter).  Thus, after
assuming for the current thick disk a value of $z_{d,\rm
  thick}=1$~kpc, we compute the scale-height of the pre-existing disk
as $z_d^{\rm initial} \sim 340$~pc. Therefore, we estimate the
vertical structure of the pre-existing disk to be similar to that of
the current Galactic thin disk.

Especially relevant in this context is the study of
\citet{elmegreen2006} who investigated thick disks in the Hubble Space
Telescope Ultra Deep Field (UDF). These authors examined whether the
process of adiabatic thick-disk contraction could have determined the
present-day scale height of the thick disk of the MW whose midplane
density ratio of thick-to-thin disk is $12\%$
\citep[e.g.,][]{juric2008}. Their calculations showed that if the
present thick-disk component of the MW (with an estimated exponential
scale-height of 875~pc) began as an equilibrium pure-thick disk at a
young age, and if subsequent accretion of the entire thin disk was
adiabatic, then the initial thick-disk {\it sech$^2$} scale-height had
to be $\sim 3$~kpc (i.e. exponential scale-height $\sim 1.5$~kpc).
Such a value is considerably larger than that observed for young thick
disks in the UDF, where the average scale height is $1.0 \pm 0.4$~kpc.
We note that the magnitude of the contraction induced in the thick
disk by the growth of the new thin disk reported by
\citet{elmegreen2006}, \mbox{$f^z_{\rm contr} \sim 0.6$}, agrees well
with our estimates.

\subsection{Kinematics}

\citetalias{villalobos-helmi2008} showed that when the thick disks are
not decomposed into cold and hot components, their mean rotational
velocities exhibit clear gradients as a function of height (see their
Figure~20). The magnitude of these vertical gradients depends
sensitively on the initial satellite orbital inclination, with larger
gradients corresponding to lower inclinations.

\begin{figure*}
  \begin{center}
    \includegraphics[width=84mm]{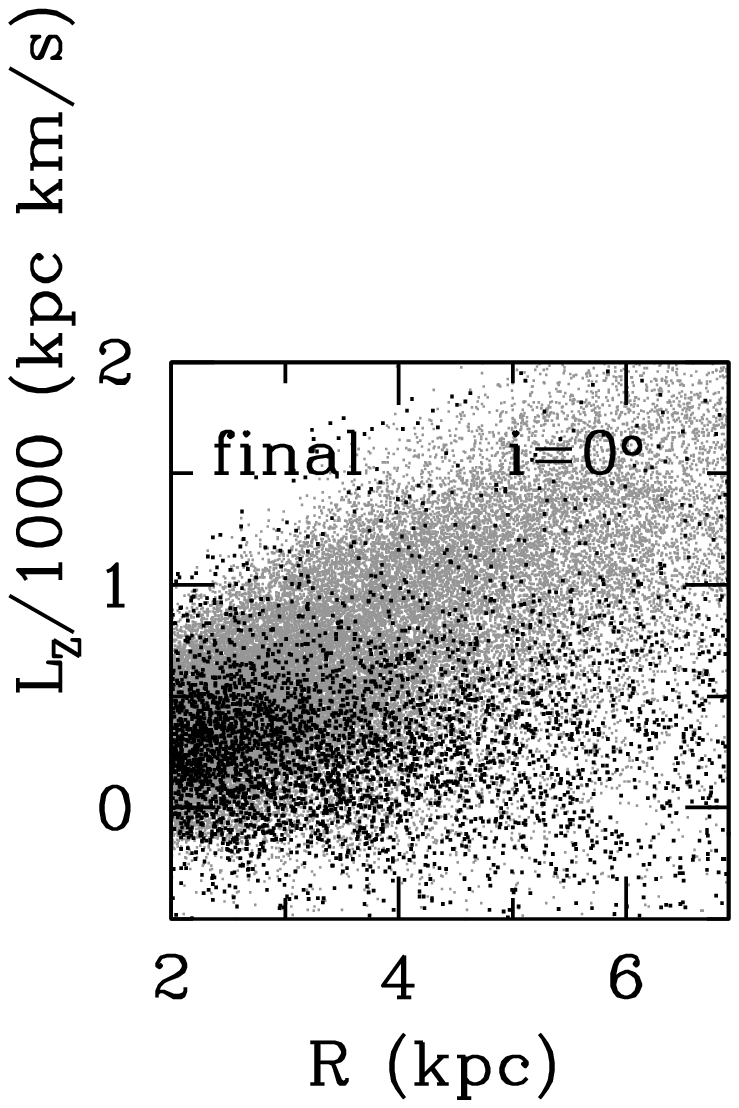}
    \hspace*{-8.54cm}
    \includegraphics[width=84mm]{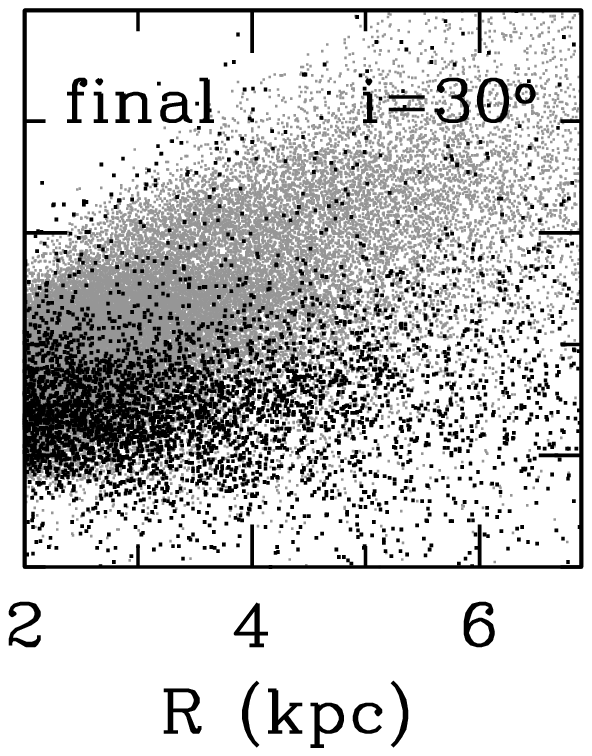}
    \hspace*{-8.54cm}
    \includegraphics[width=84mm]{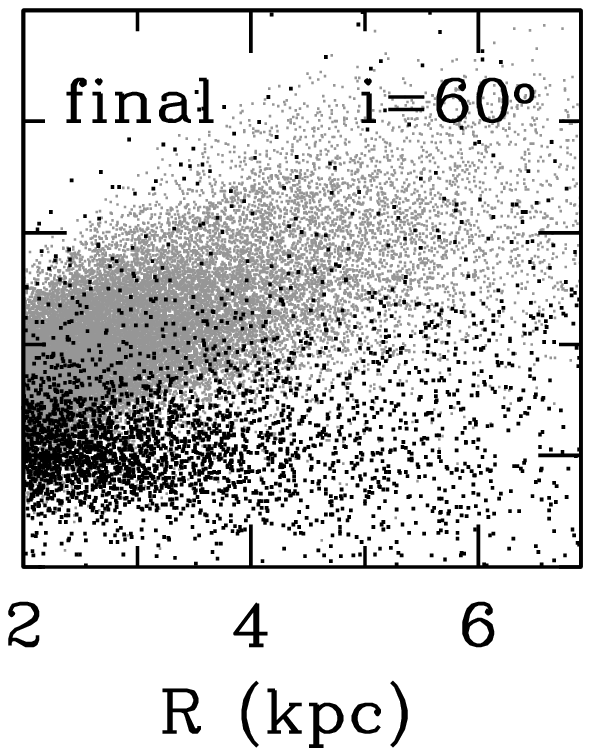}
    \hspace*{-8.54cm}
    \includegraphics[width=84mm]{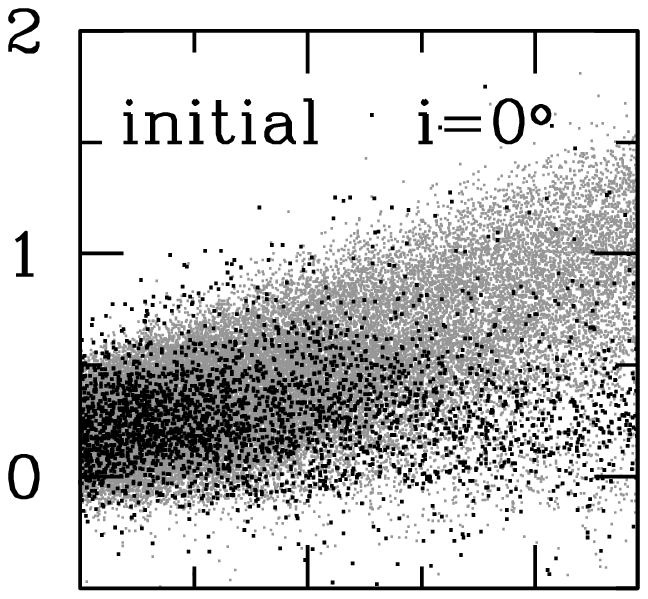}
    \hspace*{-8.54cm}
    \includegraphics[width=84mm]{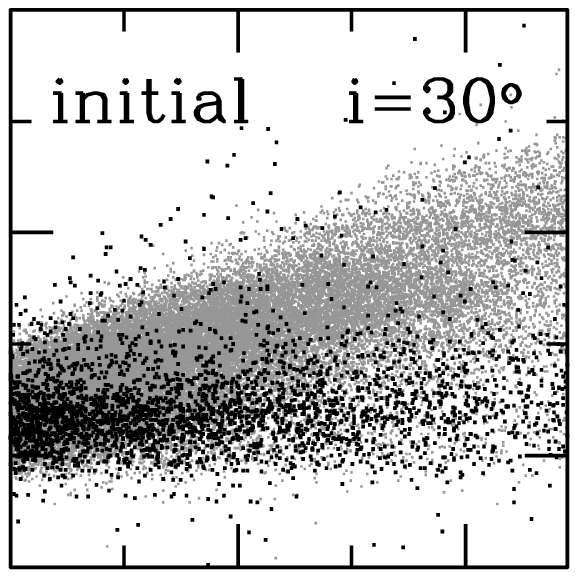}
    \hspace*{-8.54cm}
    \includegraphics[width=84mm]{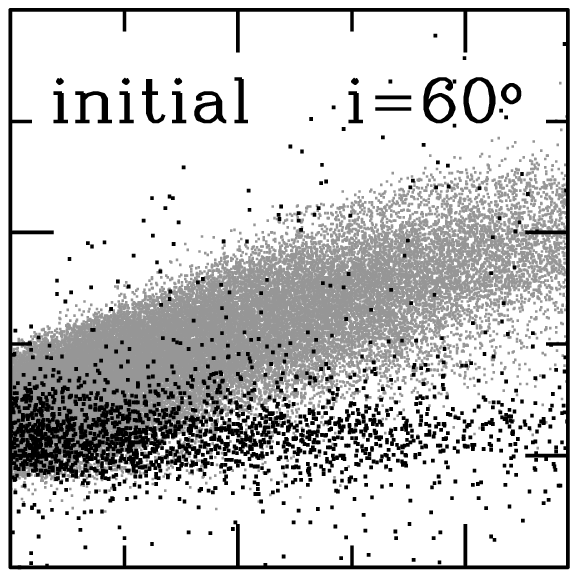}
    \includegraphics[width=84mm]{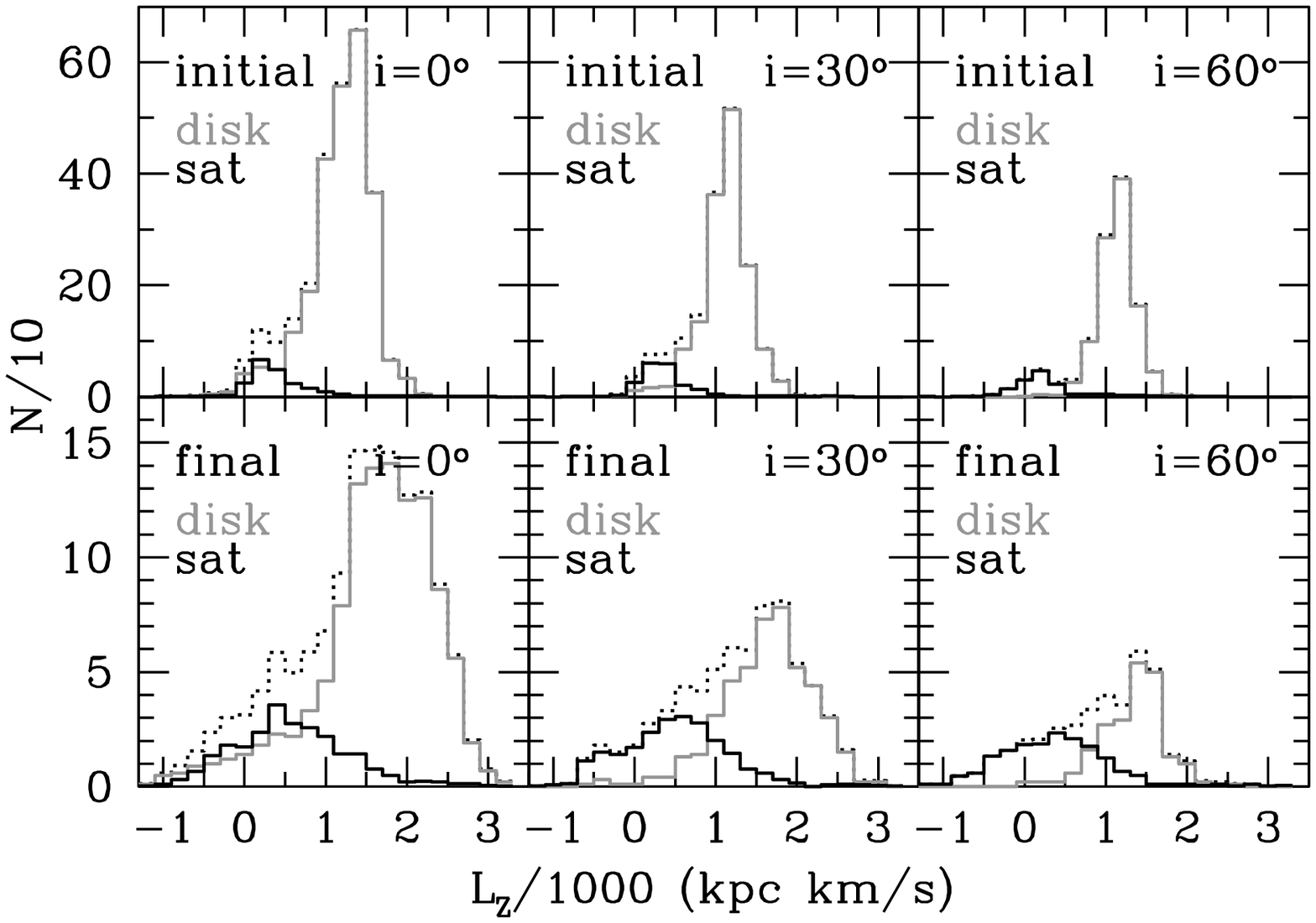}
    \caption{{\it Left panels:} Initial and final scatter plots of
      vertical angular momentum, $\Lz$, versus galactocentric
      distance, $R$, for disk (grey points) and satellite stars (black
      points).  Results are presented for experiment A and for stars
      that were initially located within $2<R<7$~kpc and $|z|<1$~kpc.
      For clarity, only one every five stars is plotted.  {\it Right
        panels:} Initial and final $\Lz$ histograms for disk and
      satellite stars that were initially located within $8<R<9$~kpc
      and $|z|<1$~kpc. Results are presented for experiment A and
      dotted lines show the total distributions.}
    \label{lzevol-pap2pap3}
  \end{center}
\end{figure*}

Figure~\ref{vpmean-fn-z-p2p3} shows the variations in the mean
rotational velocity of the final thick disks in experiment A as a
function of galactocentric distance. Interestingly, after the growth
of the new thin disk the vertical gradients are retained, maintaining
their dependence on satellite orbital inclination.  As concluded in
\citetalias{villalobos-helmi2008}, the possible detection of such
vertical gradients in the thick disk of the MW
\citep[e.g.][]{girard2006,ivezic2008} would suggest a satellite
accretion event with a low or intermediate inclination, {\it if} the
Galactic thick disk was formed via the dynamical heating of a
preexisting thin disk. Note also the progressive decrease of the mean
rotation at outer radii for larger inclinations. As we illustrate
below, this is due to the fact that stars from the heated disk and
those from the satellite have rather different $z$-components of
angular momentum at those radii (see Figure~\ref{lzevol-pap2pap3}).

\citetalias{villalobos-helmi2008} also demonstrated that the velocity
dispersion ratio $\sigma_z/\sigma_R$ is a fairly accurate
discriminator of the initial orbital inclination of the accreted
satellite (see their Figure~21). Recent observations in the solar
neighborhood report $\sigma_z/\sigma_R \sim 0.6$ \citep[e.g.
see][]{layden1996,chiba2001,soubiran2003,alcobe2005,vallenari2006}.
Assuming that the Galactic thick disk was formed according to the disk
heating scenario, such a value may be suggestive of an encounter with
a satellite on a low/intermediate orbital inclination 
\citepalias{villalobos-helmi2008}. Figure~\ref{new-sigz-sigr} shows
the final thick-disk dispersion ratio $\sigma_z/\sigma_R$ as a
function of galactocentric distance in experiment A. It is interesting
that the conclusions previously drawn are still valid after the growth
of the new thin disk. An interaction with a satellite on a
low/intermediate initial inclination is still required to obtain a
value of $\sigma_z/\sigma_R \sim 0.6$. We note that similar results
are valid for the rest of the experiments presented in this study.

For the experiment with initial inclination $30\degr$, we find that
after the satellite accretion event the thick-disk velocity
dispersions increase by factors $(f_{\rm R},f_{\rm \phi},f_{\rm
  z})_{\rm merger} \sim (2,1.7,1.5)$. These values are obtained by
measuring the velocity dispersions at $2.4R_d$ and at $2.4R_{d, \rm
  thick}$ (see Table~1 and Figure~14 of
\citetalias{villalobos-helmi2008}, for ``$z$=1'').  Additionally, the
growth of a factor of $5$ more massive new disk further increases the
velocity dispersions by factors $(f_{\rm R},f_{\rm \phi},f_{\rm
  z})_{\rm contr} \sim (1.8,2,2.2)$, where the final velocity
dispersions were measured at $2.4R_{d, \rm thick}^{\rm final}$ (see
bottom panel of Figure~\ref{struc-b}).  As in the case of computing
the scale-height of the pre-existing disk, we can estimate its initial
velocity dispersions as $(\sigma_{\rm R}^{\rm initial},\sigma_{\rm
  \phi}^{\rm initial},\sigma_{\rm z}^{\rm initial}) \sim
(18,16,12)\kms$. This calculation assumes observed values for the
Galactic thick disk $(65,54,38)\kms$ \citep{layden1996,chiba2001,
soubiran2003,alcobe2005,vallenari2006,veltz2008}. The entire
process of heating and subsequent contraction leads to an increase by
a factor of $\sim 3.4$ in all velocity dispersions for this particular
example.

\begin{figure*}
  \begin{center}
    \includegraphics[width=58mm]{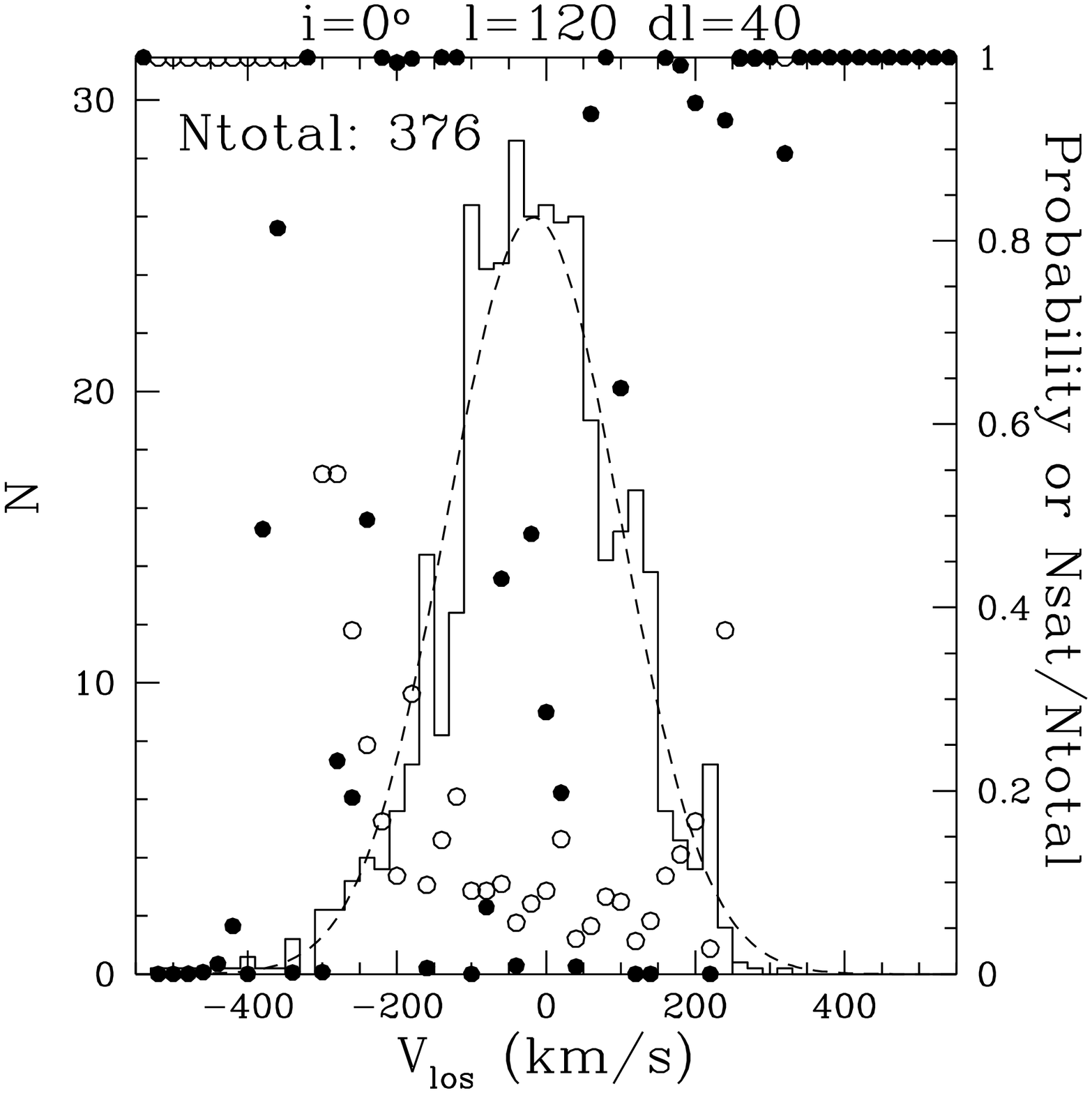}
    \includegraphics[width=58mm]{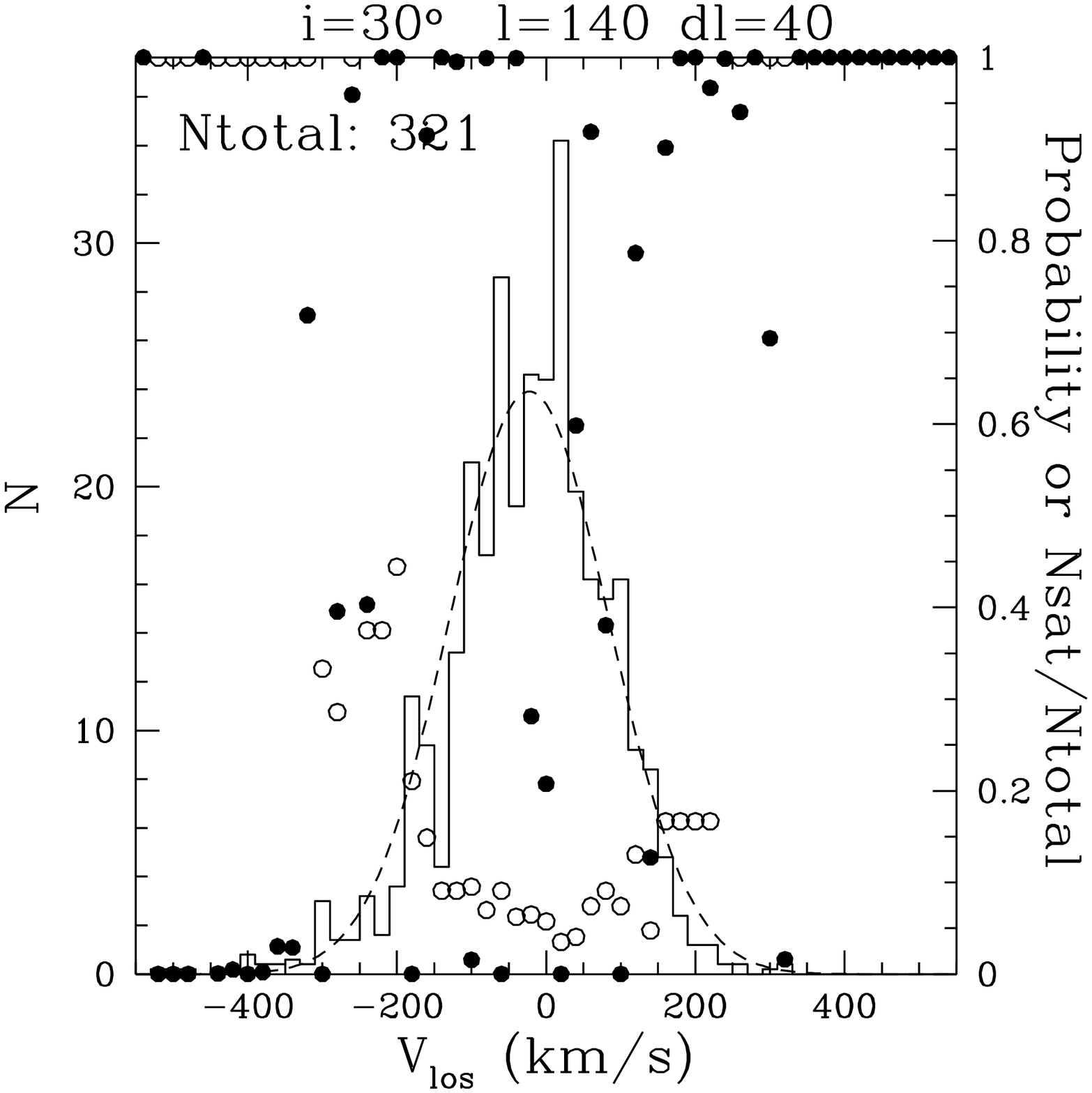}
    \includegraphics[width=58mm]{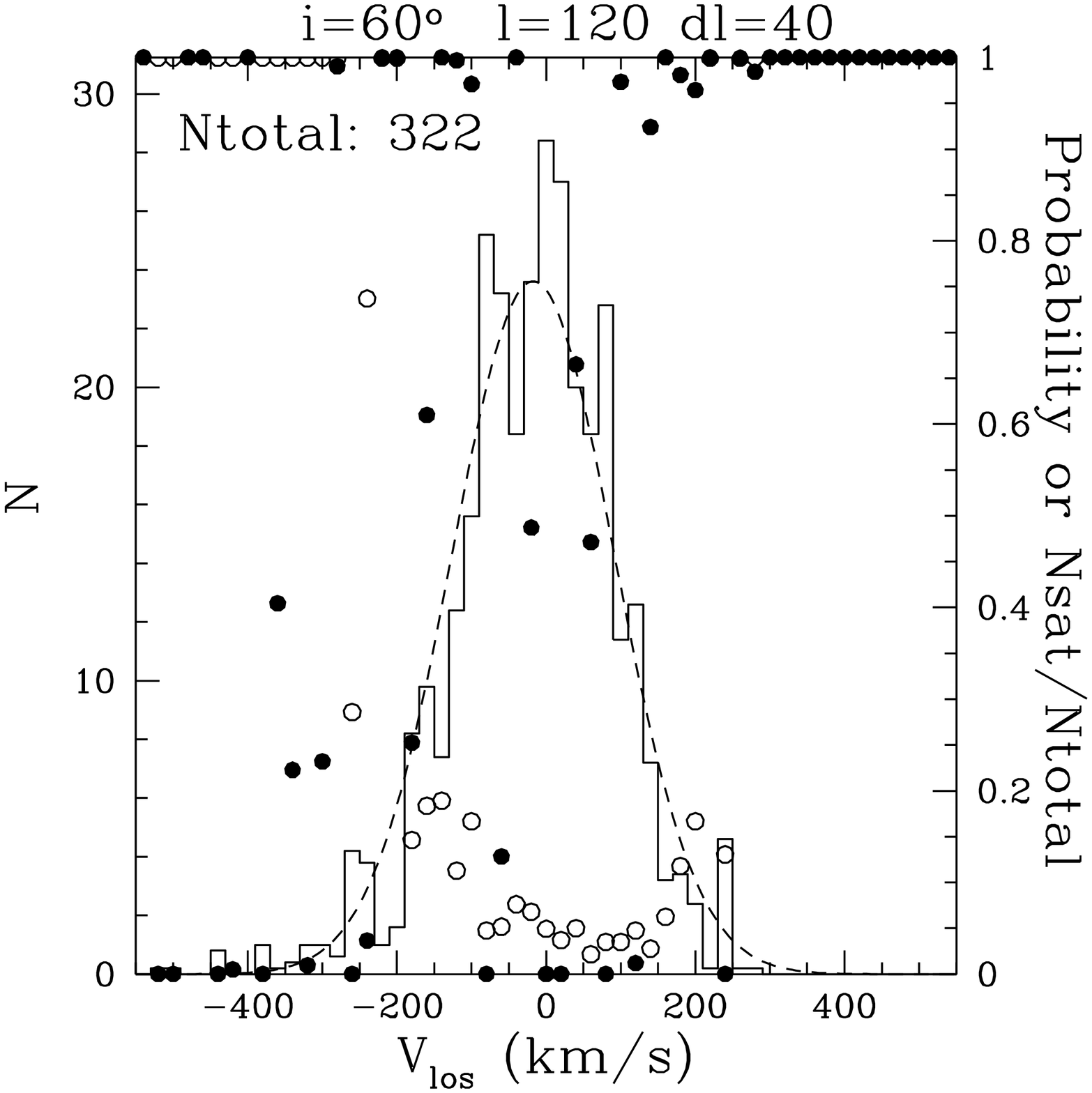}
    \caption{Histograms of final heliocentric line-of-sight velocities
      for thick-disk stars after subtracting the mean rotation of the
      final thick disks. Results are presented for experiment A and
      histograms include disk and satellite stars within a slice
      around $l\sim$140$\degr$, where the contribution of satellite
      stars to the wings is maximal. The width of $l$-slices is
      40$\degr$. Dashed lines present the best-fit Gaussian
      distributions to the histograms.  Open points show the fraction
      of satellite stars in each velocity bin, while filled points
      denote the probability of the observed number of stars compared
      to what is expected from the best-fit Gaussian in each velocity
      bin.}
    \label{ngauss-vlos-cut-l}
  \end{center}
\end{figure*}
%

\subsection{Phase-Space}

\citet[hereafter
\citetalias{villalobos-helmi2009}]{villalobos-helmi2009} showed that
the distribution of the $z$-component of the angular momentum $L_z(R)$
is a good discriminator for separating heated disk and satellite
stars, especially at large radii (see their Figure~6).
Figure~\ref{lzevol-pap2pap3} shows both the initial and final $\Lz$
distributions of disk and satellite stars for experiment A. Stars are
located within $2<R<7$~kpc and $|z|<1$~kpc. Interestingly, the initial
separation between disk and satellite stars is now less clear compared
to that in Figure~6 of \citetalias{villalobos-helmi2009}. This is
likely due to differences in the intrinsic kinematics of the
satellites used in each case (spherical versus disky).  After the
growth of the new disk two effects are expected regarding the $\Lz$
distributions of stars.  First, disk stars are expected to have a
steeper slope in their trend of $\Lz(R)$, while satellite stars should
roughly maintain their almost flat slope.  This is a consequence of
both the radial contraction experienced by thick disks and the
adiabatic invariance of $\Lz$ of each star.  Second, both type of
stars are expected to have a larger $\Lz$ ($=R \vphi$) dispersion at a
given radius $R$. This is because of the increase in the $\vphi$
velocity dispersion measured after the growth of the new disk
(Section~\ref{kine-evol-sigz-sigr}).  Both effects can be observed in
the left panels of Figure~\ref{lzevol-pap2pap3}.  As the slope of disk
stars gets steeper, it is expected that the separation between disk
and satellite stars would become more noticeable. However, the
significant increase in the velocity dispersion appears as the
dominant effect, making the separation even less evident than before
the growth of the new disk.
 
The right panels of Figure~\ref{lzevol-pap2pap3} present the
distributions of disk and satellite stars that were initially located
in the solar neighborhood ($8<R<9$~kpc) and within $|z|<1$~kpc. The
histograms show that while there is no clear distinction between disk
and satellite stars, something that can also be seen in the initial
thick disks (except in the $60\degr$ case), it is possible to identify
a large contribution of satellite stars in the wings of the
distributions at low $\Lz$.  This causes significant asymmetries in
the total distributions. Nevertheless, it is important to note that
the trend in $L_z$ of disk stars as a function of $R$ is maintained
and thus should be observable in the Galactic thick disk if it was
formed via the heating of a pre-existing disk.

\citetalias{villalobos-helmi2009} also showed that the wings of the
distributions of heliocentric line-of-sight velocities, $v_{\rm los}$,
measured in small volumes in the final thick disks, contain mostly
satellite stars. These velocity distributions were found to differ
significantly from Gaussians because of the more heavily populated
wings (see Figure~10 of \citetalias{villalobos-helmi2009}).
Figure~\ref{ngauss-vlos-cut-l} shows a very similar behavior for the
final thick disks of experiment A.  In this case, the volumes are
located somewhat closer to the center, at $\sim 3$~kpc ($1.5$~kpc
radius) to take into account the radial contraction of the systems
after the growth of the new disk.  However, we remind the reader that
the results were not found to depend on either the size or the
location of the volumes.  It is also important to stress that the
dispersions of the \vlos distributions are significantly larger than
those presented in Figure~10 of \citetalias{villalobos-helmi2009}.
Nonetheless, the same tests used in \citetalias{villalobos-helmi2009}
to quantify the statistical significance of features present in the
\vlos distributions can successfully detect the contribution of
satellite stars to the wings in the present study. These statistical
tests measure the likelihood of finding satellite stars at a given
velocity bin of a \vlos distribution by generating a number of random
realizations based on the best Gaussian fit to the distribution.

\section{Summary and Conclusions}
\label{summary-conclusions}

Using a suite of collisionless $N$-body simulations we have examined
the evolution of the structural and kinematical properties of
simulated thick disks induced by the growth of an embedded thin disk.
Our simulation campaign quantifies the importance of various
parameters related to the growing disks that could influence the
response of thick disks, including: (1) the growth timescale; (2) the
final mass; (3) the vertical scale-height; (4) the radial
scale-length; and (5) the sense of rotation of the initial thick disk
with respect to its halo (prograde or retrograde).

Our simulations demonstrate that after the growth of the new thin
disk, thick-disks experience a strong radial as well as vertical
contraction which leads to a significant decrease in their
scale-lengths and scale-heights. This contraction is triggered by the
deepening of the potential well of the system due to the mass growth
associated with the new disk. An important consequence of this
contraction is related to the migration of thick-disk stars from the
outskirts inwards. Stars that were typically at $R\sim15$~kpc before
the growth of the new disk, are found at $R \sim 9$~kpc after the
growth is completed.

Despite the strong contraction, the mass fraction associated with the
kinematically cold component of the final thick disks remains $<25\%$.
This is comparable to the mass present in the remnants of the
satellite accretion events discussed in
\citetalias{villalobos-helmi2008}.  The outskirts of these disks ($R
\sim 3.8 - 5.6 R_{d,\rm thick}$) present boxy contours at very low
surface brightness levels ($>6 - 7$ mag below the central peak).

We find that the thick-disk structural and kinematical evolution is
driven primarily by the total mass and scale-length of the growing
thin disk, with a weak dependence on the characteristics of the
initial thick disk.  Conversely, the thin-disk growth timescale
appears to have a minor influence on the final thick-disk structure
provided that this timescale is sufficiently long for the process to
be considered adiabatic. Similarly, neither the thinness (in terms of
scale-height) nor the alignment of its angular momentum vector (either
to the halo angular momentum or to that of the thick disk) seem to
affect the final properties of the thick disks.  The sense of rotation
of the initial thick disk has an effect on the radial extension of the
cold component of the final thick disk, which is significantly smaller
for retrograde rotation.
 
Kinematically, the growth of a more massive thin disk increases the
mean rotation and the velocity ellipsoid of the initial thick disks.
The increase in the mean rotation is associated with the larger mass
deposited at a given radius after the structural contraction.
Moreover, as a consequence of the contraction and additional mass
accumulation on the midplane of the thick disk, the velocity
dispersion increase is linked to a significant steepening of the
gravitational potential gradient near the disk plane.

We close with a few words of caution and a discussion of fruitful
directions for future work that may lead to more conclusive statements
about the role of a reforming thin disk in establishing the structural
properties of thick disks. We reiterate that we have only modeled the
gravitational potential of a thin disk that slowly grows in mass over
time in the collisionless regime, and we have neglected both the
cosmological context and the complex interplay of effects (e.g., gas
cooling, star formation) relevant to the formation and evolution of
spiral galaxies. Previous studies have been carried out in the context 
of galactic mergers including gas physics \citep{barnes1996,kazantzidis2005,
dimatteo2007,hopkins2008,bournaud2009,callegari2009,moster2009} and 
within a cosmological context \citep{governato2007}, 
therefore we consider that an important future study will be to follow
self-consistently, as the satellites fall in, the growth of the new thin
disk via accretion-induced bursts of star formation, deposition of gas
by the accreting satellites themselves, and smooth gas accretion
through the cooling of hot gas in the galactic halo. While a full
exploration of these contingencies is challenging, we intend to extend
the present investigation in this direction in a forthcoming work.

\acknowledgments

The authors are grateful to the organizers of the Ensenada meeting on
``Galaxy Structure and the Structure of Galaxies'' for providing the
right atmosphere and setting to initiate this project. SK would like
to thank Mandeep Gill and David Weinberg for many stimulating
discussions. SK is supported by the Center for Cosmology and
Astro-Particle Physics (CCAPP) at The Ohio State University. AV and AH
acknowledge financial support from the Netherlands Organization for
Scientific Research (NWO). This work was supported by an allocation of
computing time from the Ohio Supercomputer Center
(http://www.osc.edu). The numerical simulations performed to produce
the thick disks used as initial conditions in this study were
conducted in the Linux cluster at the Center for High Performance
Computing and Visualization (HPC/V) of the University of Groningen in
The Netherlands. This research made use of the NASA Astrophysics Data
System.

\bibliography{bibl2n,bibl2p2,bibl3}

\begin{thebibliography}{85}
\expandafter\ifx\csname natexlab\endcsname\relax\def\natexlab#1{#1}\fi

\bibitem[{{Alcob{\'e}} \& {Cubarsi}(2005)}]{alcobe2005}
{Alcob{\'e}}, S., \& {Cubarsi}, R. 2005, \aap, 442, 929

\bibitem[{{Ardi} {et~al.}(2003){Ardi}, {Tsuchiya}, \& {Burkert}}]{ardi2003}
{Ardi}, E., {Tsuchiya}, T., \& {Burkert}, A. 2003, \apj, 596, 204

\bibitem[{{Bahcall} \& {Soneira}(1980)}]{Bahcall1980}
{Bahcall}, J.~N., \& {Soneira}, R.~M. 1980, \apjs, 44, 73

\bibitem[{{Barnes} \& {White}(1984)}]{barnes1984}
{Barnes}, J., \& {White}, S.~D.~M. 1984, \mnras, 211, 753

\bibitem[{{Barnes} \& {Hernquist}(1996)}]{barnes1996}
{Barnes}, J.~E., \& {Hernquist}, L. 1996, \apj, 471, 115

\bibitem[{{Baugh}(2006)}]{baugh2006}
{Baugh}, C.~M. 2006, Reports on Progress in Physics, 69, 3101

\bibitem[{{Belokurov} {et~al.}(2006)}]{belokurov2006}
{Belokurov}, V., {et~al.} 2006, \apjl, 642, L137

\bibitem[{{Benson}(2005)}]{benson2005}
{Benson}, A.~J. 2005, \mnras, 358, 551

\bibitem[{{Benson} {et~al.}(2004){Benson}, {Lacey}, {Frenk}, {Baugh}, \&
  {Cole}}]{benson2004}
{Benson}, A.~J., {Lacey}, C.~G., {Frenk}, C.~S., {Baugh}, C.~M., \& {Cole}, S.
  2004, \mnras, 351, 1215

\bibitem[{{Binney} \& {Tremaine}(1987)}]{binney1987}
{Binney}, J., \& {Tremaine}, S. 1987, {Galactic dynamics} (Princeton, NJ,
  Princeton University Press, 1987, 755 p.)

\bibitem[{{Blumenthal} {et~al.}(1986){Blumenthal}, {Faber}, {Flores}, \&
  {Primack}}]{blumenthal1986}
{Blumenthal}, G.~R., {Faber}, S.~M., {Flores}, R., \& {Primack}, J.~R. 1986,
  \apj, 301, 27

\bibitem[{{Blumenthal} {et~al.}(1984){Blumenthal}, {Faber}, {Primack}, \&
  {Rees}}]{blumenthal1984}
{Blumenthal}, G.~R., {Faber}, S.~M., {Primack}, J.~R., \& {Rees}, M.~J. 1984,
  \nat, 311, 517

\bibitem[{{Bournaud} {et~al.}(2009){Bournaud}, {Elmegreen}, \&
  {Martig}}]{bournaud2009}
{Bournaud}, F., {Elmegreen}, B.~G., \& {Martig}, M. 2009, \apjl, 707, L1

\bibitem[{{Buitrago} {et~al.}(2008){Buitrago}, {Trujillo}, {Conselice},
  {Bouwens}, {Dickinson}, \& {Yan}}]{buitrago2008}
{Buitrago}, F., {Trujillo}, I., {Conselice}, C.~J., {Bouwens}, R.~J.,
  {Dickinson}, M., \& {Yan}, H. 2008, \apjl, 687, L61

\bibitem[{{Bullock} {et~al.}(2001){Bullock}, {Dekel}, {Kolatt}, {Kravtsov},
  {Klypin}, {Porciani}, \& {Primack}}]{bullock2001}
{Bullock}, J.~S., {Dekel}, A., {Kolatt}, T.~S., {Kravtsov}, A.~V., {Klypin},
  A.~A., {Porciani}, C., \& {Primack}, J.~R. 2001, \apj, 555, 240

\bibitem[{{Buser} {et~al.}(1999){Buser}, {Rong}, \& {Karaali}}]{buser1999}
{Buser}, R., {Rong}, J., \& {Karaali}, S. 1999, \aap, 348, 98

\bibitem[{{Callegari} {et~al.}(2009){Callegari}, {Mayer}, {Kazantzidis},
  {Colpi}, {Governato}, {Quinn}, \& {Wadsley}}]{callegari2009}
{Callegari}, S., {Mayer}, L., {Kazantzidis}, S., {Colpi}, M., {Governato}, F.,
  {Quinn}, T., \& {Wadsley}, J. 2009, \apjl, 696, L89

\bibitem[{{Chiba} \& {Beers}(2001)}]{chiba2001}
{Chiba}, M., \& {Beers}, T.~C. 2001, \apj, 549, 325

\bibitem[{{de Rijcke} {et~al.}(2005){de Rijcke}, {Michielsen}, {Dejonghe},
  {Zeilinger}, \& {Hau}}]{derijcke2005}
{de Rijcke}, S., {Michielsen}, D., {Dejonghe}, H., {Zeilinger}, W.~W., \&
  {Hau}, G.~K.~T. 2005, \aap, 438, 491

\bibitem[{{di Matteo} {et~al.}(2007){di Matteo}, {Combes}, {Melchior}, \&
  {Semelin}}]{dimatteo2007}
{di Matteo}, P., {Combes}, F., {Melchior}, A.-L., \& {Semelin}, B. 2007, \aap,
  468, 61

\bibitem[{{Dubinski}(1994)}]{dubinski1994}
{Dubinski}, J. 1994, \apj, 431, 617

\bibitem[{{Elmegreen} \& {Elmegreen}(2006)}]{elmegreen2006}
{Elmegreen}, B.~G., \& {Elmegreen}, D.~M. 2006, \apj, 650, 644

\bibitem[{{Ferguson} {et~al.}(2002){Ferguson}, {Irwin}, {Ibata}, {Lewis}, \&
  {Tanvir}}]{ferguson2002}
{Ferguson}, A.~M.~N., {Irwin}, M.~J., {Ibata}, R.~A., {Lewis}, G.~F., \&
  {Tanvir}, N.~R. 2002, \aj, 124, 1452

\bibitem[{{Ferguson} {et~al.}(2005){Ferguson}, {Johnson}, {Faria}, {Irwin},
  {Ibata}, {Johnston}, {Lewis}, \& {Tanvir}}]{ferguson2005}
{Ferguson}, A.~M.~N., {Johnson}, R.~A., {Faria}, D.~C., {Irwin}, M.~J.,
  {Ibata}, R.~A., {Johnston}, K.~V., {Lewis}, G.~F., \& {Tanvir}, N.~R. 2005,
  \apjl, 622, L109

\bibitem[{{Font} {et~al.}(2001){Font}, {Navarro}, {Stadel}, \&
  {Quinn}}]{font2001}
{Font}, A.~S., {Navarro}, J.~F., {Stadel}, J., \& {Quinn}, T. 2001, \apjl, 563,
  L1

\bibitem[{{Forbes} {et~al.}(2003){Forbes}, {Beasley}, {Bekki}, {Brodie}, \&
  {Strader}}]{forbes2003}
{Forbes}, D.~A., {Beasley}, M.~A., {Bekki}, K., {Brodie}, J.~P., \& {Strader},
  J. 2003, Science, 301, 1217

\bibitem[{{Gauthier} {et~al.}(2006){Gauthier}, {Dubinski}, \&
  {Widrow}}]{gauthier2006}
{Gauthier}, J.-R., {Dubinski}, J., \& {Widrow}, L.~M. 2006, \apj, 653, 1180

\bibitem[{{Girard} {et~al.}(2006){Girard}, {Korchagin}, {Casetti-Dinescu}, {van
  Altena}, {L{\'o}pez}, \& {Monet}}]{girard2006}
{Girard}, T.~M., {Korchagin}, V.~I., {Casetti-Dinescu}, D.~I., {van Altena},
  W.~F., {L{\'o}pez}, C.~E., \& {Monet}, D.~G. 2006, \aj, 132, 1768

\bibitem[{{Governato} {et~al.}(2007){Governato}, {Willman}, {Mayer}, {Brooks},
  {Stinson}, {Valenzuela}, {Wadsley}, \& {Quinn}}]{governato2007}
{Governato}, F., {Willman}, B., {Mayer}, L., {Brooks}, A., {Stinson}, G.,
  {Valenzuela}, O., {Wadsley}, J., \& {Quinn}, T. 2007, \mnras, 374, 1479

\bibitem[{{Hayashi} \& {Chiba}(2006)}]{hayashi2006}
{Hayashi}, H., \& {Chiba}, M. 2006, \pasj, 58, 835

\bibitem[{{Helmi} {et~al.}(1999){Helmi}, {White}, {de Zeeuw}, \&
  {Zhao}}]{helmi-etal1999}
{Helmi}, A., {White}, S.~D.~M., {de Zeeuw}, P.~T., \& {Zhao}, H. 1999, \nat,
  402, 53

\bibitem[{{Hetznecker} \& {Burkert}(2006)}]{hetznecker2006}
{Hetznecker}, H., \& {Burkert}, A. 2006, \mnras, 370, 1905

\bibitem[{{Hopkins} {et~al.}(2008){Hopkins}, {Hernquist}, {Cox}, {Younger}, \&
  {Besla}}]{hopkins2008}
{Hopkins}, P.~F., {Hernquist}, L., {Cox}, T.~J., {Younger}, J.~D., \& {Besla},
  G. 2008, \apj, 688, 757

\bibitem[{{Huang} \& {Carlberg}(1997)}]{huang1997}
{Huang}, S., \& {Carlberg}, R.~G. 1997, \apj, 480, 503

\bibitem[{{Ibata} {et~al.}(2001{\natexlab{a}}){Ibata}, {Irwin}, {Lewis},
  {Ferguson}, \& {Tanvir}}]{ibata2001b}
{Ibata}, R., {Irwin}, M., {Lewis}, G., {Ferguson}, A.~M.~N., \& {Tanvir}, N.
  2001{\natexlab{a}}, \nat, 412, 49

\bibitem[{{Ibata} {et~al.}(2001{\natexlab{b}}){Ibata}, {Lewis}, {Irwin},
  {Totten}, \& {Quinn}}]{ibata2001a}
{Ibata}, R., {Lewis}, G.~F., {Irwin}, M., {Totten}, E., \& {Quinn}, T.
  2001{\natexlab{b}}, \apj, 551, 294

\bibitem[{{Ibata} {et~al.}(2007){Ibata}, {Martin}, {Irwin}, {Chapman},
  {Ferguson}, {Lewis}, \& {McConnachie}}]{ibata2007}
{Ibata}, R., {Martin}, N.~F., {Irwin}, M., {Chapman}, S., {Ferguson}, A.~M.~N.,
  {Lewis}, G.~F., \& {McConnachie}, A.~W. 2007, \apj, 671, 1591

\bibitem[{{Ibata} {et~al.}(1994){Ibata}, {Gilmore}, \& {Irwin}}]{ibata1994}
{Ibata}, R.~A., {Gilmore}, G., \& {Irwin}, M.~J. 1994, \nat, 370, 194

\bibitem[{{Ivezi{\'c}} {et~al.}(2008)}]{ivezic2008}
{Ivezi{\'c}}, {\v Z}., {et~al.} 2008, \apj, 684, 287

\bibitem[{{Juri{\'c}} {et~al.}(2008)}]{juric2008}
{Juri{\'c}}, M., {et~al.} 2008, \apj, 673, 864

\bibitem[{{Kalirai} {et~al.}(2006){Kalirai}, {Guhathakurta}, {Gilbert},
  {Reitzel}, {Majewski}, {Rich}, \& {Cooper}}]{kalirai2006}
{Kalirai}, J.~S., {Guhathakurta}, P., {Gilbert}, K.~M., {Reitzel}, D.~B.,
  {Majewski}, S.~R., {Rich}, R.~M., \& {Cooper}, M.~C. 2006, \apj, 641, 268

\bibitem[{{Kazantzidis} {et~al.}(2008){Kazantzidis}, {Bullock}, {Zentner},
  {Kravtsov}, \& {Moustakas}}]{kazantzidis2008}
{Kazantzidis}, S., {Bullock}, J.~S., {Zentner}, A.~R., {Kravtsov}, A.~V., \&
  {Moustakas}, L.~A. 2008, \apj, 688, 254

\bibitem[{{Kazantzidis} {et~al.}(2004){Kazantzidis}, {Kravtsov}, {Zentner},
  {Allgood}, {Nagai}, \& {Moore}}]{kazantzidis2004}
{Kazantzidis}, S., {Kravtsov}, A.~V., {Zentner}, A.~R., {Allgood}, B., {Nagai},
  D., \& {Moore}, B. 2004, \apjl, 611, L73

\bibitem[{{Kazantzidis} {et~al.}(2009){Kazantzidis}, {Zentner}, {Kravtsov},
  {Bullock}, \& {Debattista}}]{kazantzidis2009}
{Kazantzidis}, S., {Zentner}, A.~R., {Kravtsov}, A.~V., {Bullock}, J.~S., \&
  {Debattista}, V.~P. 2009, \apj, 700, 1896

\bibitem[{{Kazantzidis} {et~al.}(2005){Kazantzidis}, {Mayer}, {Colpi}, {Madau},
  {Debattista}, {Wadsley}, {Stadel}, {Quinn}, \& {Moore}}]{kazantzidis2005}
{Kazantzidis}, S., {et~al.} 2005, \apjl, 623, L67

\bibitem[{{Khochfar} \& {Burkert}(2006)}]{khochfar2006}
{Khochfar}, S., \& {Burkert}, A. 2006, \aap, 445, 403

\bibitem[{{Layden} {et~al.}(1996){Layden}, {Hanson}, {Hawley}, {Klemola}, \&
  {Hanley}}]{layden1996}
{Layden}, A.~C., {Hanson}, R.~B., {Hawley}, S.~L., {Klemola}, A.~R., \&
  {Hanley}, C.~J. 1996, \aj, 112, 2110

\bibitem[{{Majewski} {et~al.}(2003){Majewski}, {Skrutskie}, {Weinberg}, \&
  {Ostheimer}}]{majewski2003}
{Majewski}, S.~R., {Skrutskie}, M.~F., {Weinberg}, M.~D., \& {Ostheimer}, J.~C.
  2003, \apj, 599, 1082

\bibitem[{{Malin} \& {Hadley}(1997)}]{malin_hadley1997}
{Malin}, D., \& {Hadley}, B. 1997, Publications of the Astronomical Society of
  Australia, 14, 52

\bibitem[{{Mart{\'{\i}}nez-Delgado} {et~al.}(2005){Mart{\'{\i}}nez-Delgado},
  {Butler}, {Rix}, {Franco}, {Pe{\~n}arrubia}, {Alfaro}, \&
  {Dinescu}}]{martinez2005}
{Mart{\'{\i}}nez-Delgado}, D., {Butler}, D.~J., {Rix}, H.-W., {Franco}, V.~I.,
  {Pe{\~n}arrubia}, J., {Alfaro}, E.~J., \& {Dinescu}, D.~I. 2005, \apj, 633,
  205

\bibitem[{{Matthews}(2000)}]{matthews2000}
{Matthews}, L.~D. 2000, \aj, 120, 1764

\bibitem[{{Mayer} {et~al.}(2008){Mayer}, {Governato}, \&
  {Kaufmann}}]{mayer2008}
{Mayer}, L., {Governato}, F., \& {Kaufmann}, T. 2008, Invited Review in
  "Advanced Science Letters" (astro-ph/0801.3845)

\bibitem[{{Mo} {et~al.}(1998){Mo}, {Mao}, \& {White}}]{mo1998}
{Mo}, H.~J., {Mao}, S., \& {White}, S.~D.~M. 1998, \mnras, 295, 319

\bibitem[{{Moster} {et~al.}(2009){Moster}, {Maccio'}, {Somerville},
  {Johansson}, \& {Naab}}]{moster2009}
{Moster}, B.~P., {Maccio'}, A.~V., {Somerville}, R.~S., {Johansson}, P.~H., \&
  {Naab}, T. 2009, MNRAS submitted (astro-ph/0906.0764)

\bibitem[{{Naab} \& {Ostriker}(2006)}]{naab_ostriker2006}
{Naab}, T., \& {Ostriker}, J.~P. 2006, \mnras, 366, 899

\bibitem[{{Navarro} {et~al.}(1997){Navarro}, {Frenk}, \& {White}}]{navarro1997}
{Navarro}, J.~F., {Frenk}, C.~S., \& {White}, S.~D.~M. 1997, \apj, 490, 493

\bibitem[{{Newberg} {et~al.}(2002)}]{newberg2002}
{Newberg}, H.~J., {et~al.} 2002, \apj, 569, 245

\bibitem[{{Peng} {et~al.}(2002){Peng}, {Ford}, {Freeman}, \&
  {White}}]{peng2002}
{Peng}, E.~W., {Ford}, H.~C., {Freeman}, K.~C., \& {White}, R.~L. 2002, \aj,
  124, 3144

\bibitem[{{Pohlen} {et~al.}(2004){Pohlen}, {Balcells}, {L{\"u}tticke}, \&
  {Dettmar}}]{pohlen2004}
{Pohlen}, M., {Balcells}, M., {L{\"u}tticke}, R., \& {Dettmar}, R.-J. 2004,
  \aap, 422, 465

\bibitem[{{Purcell} {et~al.}(2009){Purcell}, {Kazantzidis}, \&
  {Bullock}}]{purcell2009}
{Purcell}, C.~W., {Kazantzidis}, S., \& {Bullock}, J.~S. 2009, \apjl, 694, L98

\bibitem[{{Quinn} \& {Goodman}(1986)}]{quinn1986}
{Quinn}, P.~J., \& {Goodman}, J. 1986, \apj, 309, 472

\bibitem[{{Quinn} {et~al.}(1993){Quinn}, {Hernquist}, \&
  {Fullagar}}]{quinn1993}
{Quinn}, P.~J., {Hernquist}, L., \& {Fullagar}, D.~P. 1993, \apj, 403, 74

\bibitem[{{Read} {et~al.}(2008){Read}, {Lake}, {Agertz}, \&
  {Debattista}}]{read2008}
{Read}, J.~I., {Lake}, G., {Agertz}, O., \& {Debattista}, V.~P. 2008, ArXiv
  e-prints, 803

\bibitem[{{Reid} \& {Majewski}(1993)}]{reid1993}
{Reid}, N., \& {Majewski}, S.~R. 1993, \apj, 409, 635

\bibitem[{{Sellwood} {et~al.}(1998){Sellwood}, {Nelson}, \&
  {Tremaine}}]{sellwood1998}
{Sellwood}, J.~A., {Nelson}, R.~W., \& {Tremaine}, S. 1998, \apj, 506, 590

\bibitem[{{Shang} {et~al.}(1998)}]{shang1998}
{Shang}, E., {et~al.} 1998, \apjl, 504, L23

\bibitem[{{Sharma} \& {Steinmetz}(2005)}]{sharma2005}
{Sharma}, S., \& {Steinmetz}, M. 2005, \apj, 628, 21

\bibitem[{{Soubiran} {et~al.}(2003){Soubiran}, {Bienaym{\'e}}, \&
  {Siebert}}]{soubiran2003}
{Soubiran}, C., {Bienaym{\'e}}, O., \& {Siebert}, A. 2003, \aap, 398, 141

\bibitem[{{Spitzer}(1942)}]{spitzer1942}
{Spitzer}, L.~J. 1942, \apj, 95, 329

\bibitem[{{Stadel}(2001)}]{stadel2001}
{Stadel}, J.~G. 2001, PhD thesis

\bibitem[{{Tormen}(1997)}]{tormen1997}
{Tormen}, G. 1997, \mnras, 290, 411

\bibitem[{{T\'oth} \& {Ostriker}(1992)}]{toth1992}
{T\'oth}, G., \& {Ostriker}, J.~P. 1992, \apj, 389, 5

\bibitem[{{Vallenari} {et~al.}(2006){Vallenari}, {Pasetto}, {Bertelli},
  {Chiosi}, {Spagna}, \& {Lattanzi}}]{vallenari2006}
{Vallenari}, A., {Pasetto}, S., {Bertelli}, G., {Chiosi}, C., {Spagna}, A., \&
  {Lattanzi}, M. 2006, \aap, 451, 125

\bibitem[{{Vel\'azquez} \& {White}(1999)}]{velazquez1999}
{Vel\'azquez}, H., \& {White}, S.~D.~M. 1999, \mnras, 304, 254

\bibitem[{{Veltz} {et~al.}(2008){Veltz}, {Bienaym{\'e}}, {Freeman}, \& {and
  collaborators}.}]{veltz2008}
{Veltz}, L., {Bienaym{\'e}}, O., {Freeman}, K.~C., \& {and collaborators}.
  2008, \aap, 480, 753

\bibitem[{{Villalobos} \& {Helmi}(2008)}]{villalobos-helmi2008}
{Villalobos}, {\'A}., \& {Helmi}, A. 2008, \mnras, 391, 1806

\bibitem[{{Villalobos} \& {Helmi}(2009)}]{villalobos-helmi2009}
---. 2009, \mnras, 399, 166

\bibitem[{{Vitvitska} {et~al.}(2002){Vitvitska}, {Klypin}, {Kravtsov},
  {Wechsler}, {Primack}, \& {Bullock}}]{vitvitska2002}
{Vitvitska}, M., {Klypin}, A.~A., {Kravtsov}, A.~V., {Wechsler}, R.~H.,
  {Primack}, J.~R., \& {Bullock}, J.~S. 2002, \apj, 581, 799

\bibitem[{{Wainscoat} {et~al.}(1989){Wainscoat}, {Freeman}, \&
  {Hyland}}]{wainscoat1989}
{Wainscoat}, R.~J., {Freeman}, K.~C., \& {Hyland}, A.~R. 1989, \apj, 337, 163

\bibitem[{{Walker} {et~al.}(1996){Walker}, {Mihos}, \&
  {Hernquist}}]{walker1996}
{Walker}, I.~R., {Mihos}, J.~C., \& {Hernquist}, L. 1996, \apj, 460, 121

\bibitem[{{White} \& {Rees}(1978)}]{white_rees1978}
{White}, S.~D.~M., \& {Rees}, M.~J. 1978, \mnras, 183, 341

\bibitem[{{Yanny} {et~al.}(2000)}]{yanny2000}
{Yanny}, B., {et~al.} 2000, \apj, 540, 825

\bibitem[{{Yoachim} \& {Dalcanton}(2005)}]{yoachim2005}
{Yoachim}, P., \& {Dalcanton}, J.~J. 2005, \apj, 624, 701

\bibitem[{{Yoachim} \& {Dalcanton}(2006)}]{yoachim2006}
---. 2006, \aj, 131, 226

\bibitem[{Zeldovich {et~al.}(1980)Zeldovich, Klypin, Khlopov, \&
  Chechetkin}]{zeldovich1980}
Zeldovich, Y.~B., Klypin, A.~A., Khlopov, M.~Y., \& Chechetkin, V.~M. 1980,
  Sov. J. Nucl. Phys., 31, 664

\end{thebibliography}

\end{document}